\documentclass[letterpaper]{article}
\usepackage[margin=22mm]{geometry}
\usepackage{moreverb,url, dsfont, soul, dsfont,amsmath,tikz}

\usepackage[round]{natbib}
\usepackage[colorlinks, bookmarksopen, bookmarksnumbered, citecolor=red, urlcolor=red]{hyperref}
\usepackage{verbatim}
\usepackage{xr}
\usepackage{xr-hyper}
\usepackage{xcolor}
\hypersetup{colorlinks = true, linkcolor={red}, citecolor={blue}, urlcolor={red}}
\AtBeginDocument{\hypersetup{pdfborder={0 0 0}}}
\usepackage{moreverb, url, soul, subcaption, threeparttable, multicol, multirow, lscape, makecell, bm, bbm, soul}
\usepackage{enumitem}
\usepackage[title]{appendix}
\usepackage{mathtools, color, xcolor}
\usepackage{tabularx, ltablex, floatrow, inputenc}
\usepackage[]{graphicx}
\usepackage{color, soul, threeparttable, multirow, multicol, setspace}
\usepackage{amsmath, amssymb, amsfonts, caption, tikz, booktabs, colortbl, dcolumn}
\usepackage{algorithmic, algorithm}

\providecommand{\keywords}[1]
{   \small	
	\textbf{{Keywords}} #1}

\title{\bf\Large Variance Estimation for Weighted Average Treatment Effects} 

\author{
    Huiyue Li$^{1,\dagger}$, Yi Liu$^{2,\dagger}$, Yunji Zhou$^3$, Jiajun Liu$^1$, Dezhao Fu$^4$, and Roland A. Matsouaka$^{1,5}$\footnote{Corresponding author: Roland A. Matsouaka; \tt{roland.matsouaka@duke.edu}}  \\
    \small $^1$Department of Biostatistics and Bioinformatics, Duke University, Durham, North Carolina\\
    \small $^2$Department of Statistics, North Carolina State University, Raleigh, North Carolina\\
    \small $^3$Department of Biostatistics, University of Washington, Seattle, Washington\\
    \small $^4$Department of Biostatistics and Bioinformatics, George Washington University, Washington, DC\\
    \small $^5$Program for Comparative Effectiveness Methodology, Duke Clinical Research Institute,  Durham, North Carolina \\
    \small $^\dagger$These authors made equal efforts and contributions.
}
\date{} 

\newcommand{\Ex}{\mathbb{E}}

\newcommand{\mc}{\mathcal}
\newcommand{\bd}{\boldsymbol}
\newcommand{\mb}{\mathbf}
\newcommand{\bigCI}{\mathrel{\text{\scalebox{1}{$\perp\mkern-10mu\perp$}}}}

\newtheorem{remark}{Remark}
\newtheorem{assumption}{Assumption}

\begin{document}
\maketitle


\begin{abstract}
    Common variance estimation methods for weighted average treatment effects (WATEs) in observational studies include nonparametric bootstrap and model-based, closed-form sandwich variance estimation. However, the computational cost of bootstrap increases with the size of the data at hand. Besides, some replicates may exhibit random violations of the positivity assumption even when the original data do not. Sandwich variance estimation relies on regularity conditions that may be structurally violated. Moreover, the sandwich variance estimation is model-dependent on the propensity score model, the outcome model, or both; thus it does not have a unified closed-form expression. Recent studies have explored the use of wild bootstrap to estimate the variance of the average treatment effect on the treated (ATT). This technique adopts a one-dimensional, nonparametric, and computationally efficient resampling strategy. In this article, we propose a ``post-weighting'' bootstrap approach as an alternative to the conventional bootstrap, which helps avoid random positivity violations in replicates and improves computational efficiency. We also generalize the wild bootstrap algorithm from ATT to the broader class of WATEs by providing new justification for correctly accounting for sampling variability from multiple sources under different weighting functions. We evaluate the performance of all four methods through extensive simulation studies and demonstrate their application using data from the National Health and Nutrition Examination Survey (NHANES). Our findings offer several practical recommendations for the variance estimation of WATE estimators. 
\end{abstract} \hspace{10pt}




\section{Introduction}\label{sec:intro}

A common causal estimand for measuring the effect of a treatment, an exposure, or an intervention is the average treatment effect (ATE). The inferential target of ATE is the overall population of participants where the data are sampled from. 
To identify ATE from observational data, we rely on a number of assumptions, including the essential positivity assumption \citep{rosenbaum1983central, imbens2015causal}. The positivity assumption requires, for each participant in the sample, a non-zero propensity score $P(Z=z\mid \mb X)$, i.e., the probability to receive any of the treatment options $z$ available ($z=1$ if treated and $z=0$ otherwise), given the set of baseline covariates $\mb X$ that is sufficient to eliminate confounding. 

When the positivity assumption is violated, it is oftentimes more sensible to shift the target of inference to the treatment effect on a sub-population (i.e., a weighted population), using a pre-specified weight (or tilting) function $g(\mb x)$. This provides a more general framework to assess treatment effects, with the ATE being a special case with the tilting function $g(\mb x) =1.$ When the positivity assumption is violated, we may for instance focus on the subpopulation of participants whose PSs fall inside the interval $[0.1,0.9]$ \citep{crump2006moving, crump2009dealing}, i.e., the subpopulation $\mc O(\mb X)=\{\mb X:0.1\leq e(\mb X)\leq 0.9\}$, based on the tilting function $g(\mb x) = I(0.1\leq e(\mb x)\leq 0.9)$, where $e(\mb X)=P(Z=1\mid \mb X)$ is the PS (i.e., the probability of receiving treatment $Z=1$, given the baseline covariates).  Another example is the subpopulation of participants from whom there is clinical equipoise, using the tilting function $g(\mb x)=e(\mb x)\{1-e(\mb x)\}$ (the overlap weight function \citep{li2018balancing}), which weights each observation according to how close it to the ``clinical equipoise'' \citep{thomas2020overlap, li2019addressing, matsouaka2024causal}, More generally, in this paper, we will consider the average causal effect defined on a weighted population is called the \textit{weighted average treatment effect (WATE).} Examples of WATEs include the average treatment effect on the population defined by overlap weights (ATO) \citep{li2018balancing, crump2006moving, hirano2003efficient}, matching weights (ATM) \citep{li2013weighting}, trapezoidal weights (ATTZ) \citep{mao2019propensity}, entropy weights (ATEN) \citep{zhou2020propensity}, and beta weights (ATB) \citep{matsouaka2024causal}.

There are different estimators of the WATE, among which the augmented estimator is a popular approach \citep{li2021propensity, matsouaka2024causal}. The augmented estimator involves modeling on both treatment and outcomes. It has been shown to be a ``nearly double-robust'' estimator by a number of simulation and empirical studies,  that is, when either the outcomes or treatment model is correctly specified, it is nearly unbiased to estimating the WATE \citep{matsouaka2024causal, matsouaka2024overlap, matsouaka2023variance, li2021propensity, mao2019propensity, kostouraki2024variance}. In addition, when both models are correctly (or fully) specified, it achieves the semiparametric efficiency bound \citep{hirano2003efficient}. In this paper, we compare a number of variance estimation methods for the augmented estimator of WATE under different model specifications. 

To quantify the inherent uncertainty in estimating WATE, particularly to derive the variance of an augmented estimator, we need to consider three different sources of variability, i.e., those due to estimating the WATE estimand itself, the treatment model, and the outcome models \citep{lunceford2004stratification, williamson2012variance, li2013weighting, zou2016variance, mao2020flexible,  matsouaka2023variance, abadie2016matching, mao2018propensity, matsouaka2024overlap, austin2022bootstrap, shook2024double}. Under some regularity conditions, standard bootstrap (based on resampling the data at hand) and close-form sandwich (based on large-sample normal approximation) estimators have shown to be two valid methods that successfully considered these variabilities \citep{lunceford2004stratification, matsouaka2023variance, yang2018asymptotic}, but they also have some potential and practical limitations (see Section \ref{sec:var-est}). 

As alternatives, we {proposed} two additional bootstrap methods, the ``weight-then-resample'' or ``post-weighting'' bootstrap (denoted by BOOT II) and the wild bootstrap (WB). With BOOT II, we first use the original sample to estimate the propensity score (PS) for each participant. Then, from the original sample, we resample the data with replacement to estimate the WATE, using the readily available PSs (rather than re-estimating them) to compute the corresponding WATE weights, while still re-estimating the outcome models as needed. This method is similar to the ``simple bootstrap'' proposed by \cite{austin2014use}; it has the advantage of circumventing random violations of the positivity assumption and providing stable variance estimation 
\citep{westreich2010invited,matsouaka2023variance,matsouaka2024overlap,zhou2020propensity}. 
The WB, proposed for matching estimators \citep{bodory2020finite,otsu2017bootstrap} has also been successfully leveraged and shown to have good performance for variance estimation on doubly robust estimators for ATT and average treatment effect on the controls (ATC) estimators \citep{matsouaka2023variance}. It consists of perturbing the estimated influence function (IF) of a given estimator of an WATE, using judiciously chosen independent random variables. The variance of the WATE is then calculated as the mean square of the perturbed IFs, evaluated across the perturbed samples. 

We are interested in applying and evaluating these bootstrap variance estimators for the WATE.  Nevertheless, the extension of WB is not straightforward (as explained in Sections \ref{sec:var-est} and \ref{sec:sim}), mainly due to different roles of the PS plays across various WATE estimands. When the PS is ancillary; for instance, when estimating ATE and ATT in \cite{hahn1998role}, the WB works well. However, when it is not ancillary (e.g., for ATO, ATM, ATEN), the uncertainty quantification becomes more challenging. In a recent discussion about the role PS plays in the variance estimation of different WATEs, \cite{kostouraki2024variance} emphasized that the performance of variance estimators depends on the choice of estimands. At the end of this paper, we provide several practical recommendations on the optimal choice of variance estimation approach for each estimand based on our explorations. 

The remainder of this paper is organized as follows. In Section \ref{sec:setup}, we review the potential outcomes framework and present the augmented estimator for WATE. In Section \ref{sec:var-est}, we introduce the variance estimation methods we consider while highlighting their pros and cons.  We conduct an extensive Monte Carlo simulation study, in Section \ref{sec:sim}, to compare the performance of these methods, varying degrees of PS overlap, treatment effect heterogeneity, and model specifications. In Section \ref{sec:data}, we apply the variance estimation methods to analyze the impact of fish consumption on blood mercury levels in participants, using data from the National Health and Nutrition Examination Survey (NHANES) 2013–2014. Finally, we conclude the paper in Section \ref{sec:conclusion} with some comments and remarks. 

\section{Notation and Definitions}\label{sec:setup}

\subsection{Potential outcomes framework}\label{subsec:PO}

Let $Z\in\{0,1\}$ denote a binary treatment assignment (1 for treated and 0 for control), $Y$ the observed outcome of interest, and $\mb X=(X_1, \ldots, X_p)'\in\mc X\subset\mathbb R^p$  a vector of $p$ baseline covariates, $p>0$. The observed data $\mc O=\{\mb O_i=(Z_i, \mb X_i, Y_i): i=1,\dots, N \}$ represent an independently identically distributed (i.i.d.)  sample of $N$ participants in the study. Following the potential outcome framework \citep{neyman1923applications, rubin1974estimating}, we assume that each participant has two potential outcomes $Y(z)$, $z=0,1$, where $Y(z)$ corresponds to the outcome that would have been observed, probably contrary to the fact, had a participant received the treatment $Z=z$. Thus, assuming consistency, we have the relationship of the observed outcome $Y$ and potential outcomes as $Y=Y(Z)=ZY(1)+(1-Z)Y(0)$. We also assume there is only one version of the treatment and that the potential outcome of a participant does not depend on the treatment(s) received by the other participants, i.e., the stable-unit treatment value assumption (SUTVA). Finally, as previously indicated, we define the PS by $e(\mb x)=P(Z=1\mid \mb X=\mb x)$, which is a subject-specific probability for receiving the treated given the baseline covariates value $\mb X=\mb x$. 

Moreover, we also make the two following assumptions for the identification of causal effects \citep{matsouaka2024causal, zhou2020propensity}:

\begin{assumption}[Unconfoundedness]\label{assp:unconfound}
$Y(z)\bigCI Z\mid \mb X$,  for  $z=0,1$.
\end{assumption}
\begin{assumption}[Positivity]\label{assp:positivity}
There are two positive constants $c_1$ and $c_2$ such that $0<c_1\leq e(\mb X)\leq c_2<1$ with probability 1, for all $\mb X\in\mc X$.
\end{assumption}

\subsection{Weighted average treatment effect and corresponding augmented estimator}\label{subsec:WATE}

Given the potential outcomes framework and assumptions in Section \ref{subsec:PO}, we can define a number of causal effects. For instance, the individual treatment effect (ITE) defined by $Y(1)-Y(0)$, the conditional average treatment effect (CATE) given by $\tau(\mb x)=\Ex\{Y(1)-Y(0)\mid \mb X=\mb x\}$, and the average treatment effect (ATE), which is defined as $\tau=\Ex\{Y(1)-Y(0)\}=\Ex\{\tau(\mb X)\}$. Furthermore, as mentioned in Section \ref{sec:intro}, when we want to shift the target of inference to a weighted population, we measure the weighted ATE (WATE)  defined as  

\begin{align}\label{eq:wate}
	\tau_{g} = \frac{\displaystyle \mathbb{E}\{g(\mb X)\tau(\mb X)\}}{\displaystyle \mathbb{E}\{g(\mb X)\}} 
\end{align} 
where 
$g(\mb x)$ is a \textit{tilting function} characterizing the weights to the overall population. 

Examples of tilting function include those for PS trimming $g(\mb x)=I(\alpha\leq e(\mb x)\leq 1-\alpha)$ for an $\alpha\in(0,0.5)$ with $I(\cdot)$ the indicator function \citep{crump2006moving,crump2009dealing}; overlap weights $g(\mb x)=e(\mb x)\{1-e(\mb x)\}$ \citep{li2018balancing}; matching weights $g(\mb x)=\min\{e(\mb x), 1-e(\mb x)\}$ \citep{li2013weighting} and entropy weights $g(\mb x)=-e(\mb x)\log\{e(\mb x)\}-(1-e(\mb x))\log\{1-e(\mb x)\}$ \citep{zhou2020propensity, matsouaka2024overlap}. Interested readers are referred to Table 1 in \cite{matsouaka2024causal} for a more comprehensive summary. 

A common estimator for WATE is the following augmented estimator \citep{mao2019propensity}
\begin{align}\label{eq:aug-wate}
	\displaystyle
	\widehat\tau_{g}^{\text{aug}} = & \sum_{i=1}^{N}\frac{\widehat g(\mb X_i)\{\widehat m_1(\mb X_i) - \widehat m_0(\mb X_i)\}}{\displaystyle\sum_{i=1}^{N}\widehat g(\mb X_i)} + 
	 \sum_{i=1}^{N}\frac{Z_i\widehat  \omega_1(\mb X_i)\{Y_i-\widehat m_1(\mb X_i)\}}{\displaystyle\sum_{i=1}^{N}Z_i\widehat \omega_1(\mb X_i)}-
	\sum_{i=1}^{N}\frac{(1-Z_i)\widehat \omega_0(\mb X_i) \{Y_i-\widehat m_0(\mb X_i)\}}{\displaystyle\sum_{i=1}^{N}(1-Z_i)\widehat \omega_0(\mb X_i)} 
\end{align} where $\widehat m_z(\mb X) = \widehat{\mathbb{E}}\{Y(z)\mid\mb X\}$, by some postulated outcome regression (OR) model for $Y(z)$, and $\widehat \omega_z(\mb x) =\widehat g(\mb x) \widehat e(\mb x)^{-z}(1-\widehat e(\mb x))^{z-1}, z=0,1$. 

When estimating ATE, ATT and ATC, the  estimator $\widehat\tau_{g}^{\text{aug}} $ is doubly-robust in the sense that, whenever one of the PS or OR model is correctly specified, it is consistent. Furthermore, for other tilting functions $g(\mb x)$, we have the following slightly different results: {when the PS model is correctly specified}, the estimator $\widehat\tau_{g}^{\text{aug}}$ is consistent regardless of the model specifications of $Y(0)$ and $Y(1)$; however, when the PS model is misspecified, $\widehat\tau_{g}^{\text{aug}}$ is only consistent to an estimand where the true PS is replaced by the probability limit of the misspecified PS model, regardless the specifications of OR models \citep{mao2019propensity, matsouaka2024causal}. Nevertheless, many empirical studies have shown that, if the two OR models are correctly specified, regardless the specification of PS model, $\widehat\tau_g^{\text{aug}}$ still has a small bias 
 \citep{matsouaka2024overlap, li2021propensity}. 

\section{Variance Estimation Methods}\label{sec:var-est}
\subsection{Sandwich variance estimation (SAND)}\label{subsec:sandwich}

The sandwich variance estimator is constructed based on the M-estimation theory \citep{van2000asymptotic, lunceford2004stratification}, using an unbiased estimating equation as follows \citep{stefanski2002calculus} 
$$
\sum_{i=1}^{N}\psi_{\bd\theta}(Z_i, \mb X_i, Y_i)=\bd 0,
$$ 
for some parameter ${\bd\theta}$ related to $\tau_{g}^{\text{aug}}$ in the form of $\tau_{g}^{\text{aug}}=\mb c'{\bd\theta}$, for some constant vector $\mb c$  (see Appendix \ref{subapx:sandwich} for the details related to the derivation of the variance). This estimating equation considers the variability in estimating both nuisance functions and the estimand. The solution to the equation, denoted by $\widehat{\bd\theta}$, is called the M-estimator. The pseudo-truth of ${\bd\theta}$, denoted by ${\bd\theta}^*$, is the solution of $\mathbb{E}\{\psi_{{\bd\theta}^*}(Z_i, \mb X_i, Y_i)\}=\bd 0$. We have $\sqrt{N}(\widehat{{\bd\theta}}-{{\bd\theta}^*})\to_d\mathcal N({\bd 0}, \bd\Sigma({\bd\theta}^*))$ as $N\to\infty$ under some regularity conditions \citep{van2000asymptotic}, where
$$
{\bd\Sigma}({\bd\theta}^*) = \mb A({\bd\theta}^*)^{-1}\mb B({\bd\theta}^*)\{\mb A({\bd\theta}^*)^{-1}\}',
$$ 
which is  estimated by plugging in $\widehat{\bd\theta}$ in lieu of ${\bd\theta}^*$ to obtain $\widehat{\bd\Sigma}(\widehat{\bd\theta})=\mb A_N(\widehat{\bd\theta})^{-1}\mb B_N(\widehat{\bd\theta})\{\mb A_N(\widehat{\bd\theta})'\}^{-1}$, where  $\mb A({\bd\theta}), \mb B({\bd\theta}),$ $ \mb A_N(\widehat{\bd\theta})$ and $\mb B_N(\widehat{\bd\theta})$ are given by 
\begin{align*}
	\mb A_N(\widehat{{\bd\theta}}) & =-\frac{1}{N}\displaystyle\sum_{i=1}^{N}\frac{\partial \psi_{\widehat{\bd\theta}}(Z_i, \mb X_i, Y_i)}{\partial{{\bd\theta}'}}, \quad\mb B_N(\widehat{{\bd\theta}})= \frac{1}{N}\displaystyle\sum_{i=1}^{N}\psi_{\widehat{\bd\theta}}(Z_i, \mb X_i, Y_i)\psi_{\widehat{\bd\theta}}(Z_i, \mb X_i, Y_i)',
	\\
	\mb A({\bd\theta}^*)&=-\mathbb{E}\left\{\dfrac{\partial\psi_{{\bd\theta}^*}(Z, \mb X, Y)}{\partial{\bd\theta}'}\right\},\quad 
	\mb B({\bd\theta}^*)=\mathbb{E}\left\{ \psi_{{\bd\theta}^*}(Z, \mb X, Y)\psi_{{\bd\theta}^*}(Z, \mb X, Y)'\right\},
\end{align*} 
such that $\mb A_N(\widehat{\bd\theta})\to_p \mb A({\bd\theta}^*)$  and $\mb B_N(\widehat{\bd\theta})\to_p \mb B({\bd\theta}^*)$, as $N\to\infty$. The variance of $\widehat \tau^{\text{aug}}$ is determined as $\mb c'\widehat{\bd\Sigma}(\widehat{\bd\theta})\mb c$. The above random variable $\widehat{\bd\Sigma}(\widehat{\bd\theta})$ is referred to as the 
\textit{sandwich variance estimator} for ${\bd\theta}$. 

Even though the sandwich variance estimator is theoretically robust under certain conditions, there are however some disadvantages in using it: 
\begin{itemize}
    \item The sandwich variance estimator is not as flexible as the nonparametric resampling methods. The estimating equation must be solved on the case-by-case basis as it depends on the postulated PS and outcome models. Thus, the sandwich variance estimator needs to be specially derived for different (parametric) models. Unfortunately, the number of models to choose from in some user-friendly software (e.g., PSW \citep{mao2018package} and PSweight \citep{zhou2020psweight} packages in R) is limited. When machine learning methods \citep{van2007super, mccaffrey2013tutorial} are used to predict the nuisance functions, some of these packages ignore the uncertainty associated in the estimating equation for nuisance functions, assuming the machine learning methods predict them well. 
    \item The sandwich variance is generally sensitive to model misspecifications, sometimes for a mild misspecification \citep{freedman2006so, kauermann2001note} and requires more regularity conditions (see Appendix \ref{subapx:sandwich}) than bootstrap methods. When there is multi-collinearity, limited sample size, or poor overlap,  the sandwich variance estimator can be sharply inflated or even not obtainable \citep{matsouaka2024overlap, pan2002small,matsouaka2023variance}. In Appendix \ref{apx:additional-sim}, we also provide an example where the sandwich variance estimator for WATE does not work. 
\end{itemize}

\subsection{Bootstrap methods (BOOT)}\label{subsec:bootstraps}
The bootstrap is a nonparametric resampling method for assigning measures of precision to a sample estimator \citep{efron1986bootstrap}. In the standard bootstrap procedure, we resample data with replacement to produce multiple bootstrap samples of the same size as the original sample and calculate multiple point estimates from each sample.
The standard deviation of these point estimates can then be the estimate of the model-based standard error of the sample estimator under some regularity conditions, e.g., asymptotic linearity of the point estimator
\citep{shao2012jackknife, van2000asymptotic}. Since the augmented estimator is asymptotically linear, when the nuisance functions belong to some small classes of functions, e.g., Donsker class \citep{shorack2009empirical, pollard1990empirical}, it enables the use of the standard bootstrap for variance estimation. 

\textbf{Bootstrap I (BOOT I).}   As in any standard bootstrap method, we resample the original data $(\mb X,Z,Y)$ with replacement to get a bootstrap sample. Then, we estimate the PSs and calculate the corresponding weights, fit the outcome models (if we use augmented estimators), and finally estimate the WATE. We repeat this process in $R$ independent bootstrap replicates and obtain $R$ point estimates. The variance estimate is calculated as the empirical variance of the  $R$ corresponding WATE point estimates. 

One of the key {assumptions of PS methods is the positivity assumption \citep{westreich2010invited,zhou2020propensity, matsouaka2024causal}}. {Unfortunately,} the use of the above BOOT I may lead to  {violations of the positivity assumption: some of the resampled bootstrap replicates may drastically change the within  proportion of treated participants or increase the number of  participants at the fringe of the PS spectrum}  and thus lead to extreme PS weights \citep{matsouaka2024overlap}. Due to the presence of extreme PS weights, BOOT I may lead to the unstable standard error of the WATE. We therefore {propose} the following Bootstrap II procedure. 

\textbf{Bootstrap II (BOOT II).} Unlike BOOT I, with BOOT II, {we propose a  ``weight-then-resample" approach}. We first obtain the needed estimated PS $\widehat e(\mb x)$ using the original sample, then we resample $(\mb X,Z,Y,\widehat e(\mb X))$ all together. The rest of the algorithm is the same as BOOT I: we calculate the desired $(\widehat g,\widehat\omega_0, \widehat\omega_1)$  and also fit the necessary outcome models $m_1$ and $m_0$ within each bootstrap replicate. The steps for the BOOT II are specified in the Algorithm \ref{algo:bootII}. We also refer to this ``weight-then-resample" approach as the ``post-weighting bootstrap.''

{The rationale and theoretical justification for BOOT II are based on the following consideration. Although BOOT II appears to bypass the need to account for variability in the estimation of the PS, \citet{hahn1998role} showed that the asymptotic variance of the augmented estimator for the ATE remains unchanged whether the true PS is known or estimated. This result follows from two key facts: (i) the augmented estimator is derived from the efficient influence function, and (ii) the semiparametric efficiency bound for the ATE is unaffected by whether the PS is known. This finding provides theoretical support for the use of BOOT II in estimating variance for the ATE. Motivated by Hahn's result, we further investigate whether such a post-weighting bootstrap procedure can also yield valid variance estimates for other WATEs. }

{As a remark, in BOOT II, resampling without re-estimating the PS does not affect the consistency of the augmented estimator \eqref{eq:aug-wate}, since the point estimate is computed using the full, original sample; the resampling procedure is used solely to estimate its variance. At the same time, re-estimating the PS or not in the bootstrap samples does not affect the consistency of the bootstrap estimates; that is, the consistency of each bootstrap estimate aligns with that of the original sample estimate, as the model specifications for the PS and outcomes remain unchanged across bootstrap replicates. }

\begin{algorithm*}
  \caption{Post-weighting bootstrap for variance estimation of WATE} \label{algo:bootII}
  \begin{algorithmic}[1]
  \STATE {\bfseries Input:} {Observational data $\mc O = \{(Z_i, \mb X_i, Y_i), i=1,\dots, N\}$; tilting function $g(\mb x)$ of the given WATE; number of bootstrap replicates $R$; confidence level $\alpha\in(0,0.5)$. } 
  
  \STATE {\bfseries Output:} {Standard error estimate $\widehat\Sigma_g^{1/2}$ and a $100\times(1-\alpha)\%$ confidence interval $\widehat C_\alpha(\mc O)$. }

  \STATE Fit a PS working model by $\widehat e(\mb X)$, e.g., using a logistic regression model $\widehat e(\mb X)=e(\mb X;\widehat{\bd\beta})$. 

  \STATE Derive the predicted PS $\widehat e(\mb X_i)$. 

  \STATE Augment data $\mc O$ by $\mc O^{\text{aug}} = \{(Z_i, \mb X_i, Y_i,  \widehat e(\mb X_i), i=1,\dots, N\}$

  \FOR{$r=1,\dots, R$}

  \STATE Resample $\mc O^{\text{aug}}$ with replacement by size $N$, which gives a bootstrap data $\mc O^{\text{aug},(r)}$ with some observations might be repeated.
\STATE Calculate the tilting function $\widehat g(\mb X_i)^{(r)}$ and weights $(\widehat\omega_0^{(r)}(\mb X_i), \widehat\omega_1^{(r)}(\mb X_i))$ for $i=1,\dots, N$. 
  \STATE Fit two working outcome models $\widehat m_0^{(r)}(\mb X)$ and $\widehat m_1^{(r)}(\mb X)$ for outcomes, respectively, on $\mc O^{\text{aug},(r)}\cap\{Z_i=0\}$ and $\mc O^{\text{aug},(r)}\cap\{Z_i=1\}$, for $i=1,\dots,N$, and predict for all data. 

  \STATE Plug $\mc O^{\text{aug},(r)}\cup\{\widehat\omega_0^{(r)}(\mb X_i), \widehat\omega_1^{(r)}(\mb X_i), \widehat m_0^{(r)}(\mb X_i), \widehat m_0^{(r)}(\mb X_i), i=1,\dots, N\}$ in \eqref{eq:aug-wate} and get the augmented estimator $\widehat\tau_g^{(r)}$. 
  
  \ENDFOR
  
  \STATE \textbf{Return: } 
  The standard error estimate: $\widehat\Sigma^{1/2}_g = \dfrac{\text{IQR}(\widehat\tau_g^{(1)},\dots, \widehat\tau_g^{(R)})}{z_{0.75}-z_{0.25}}$ or the standard deviation of $\{\widehat\tau_g^{(1)},\dots, \widehat\tau_g^{(R)}\}$; the $100\times(1-\alpha)\%$ confidence interval $\widehat C_\alpha(\mc O) = \widehat\tau_g^{\text{aug}}\pm\dfrac{z_{(1-\alpha/2)}}{\sqrt{N}}\widehat\Sigma^{1/2}_g$, where $z_{0.75}-z_{0.25}=1.349$ is the interquartile range of the standard normal distribution, and $\widehat\tau_g^{\text{aug}}$ is by \eqref{eq:aug-wate}.  
  \end{algorithmic}
\end{algorithm*}

There are two additional good reasons that BOOT II {may offer advantages} over BOOT I. First, BOOT I is less favorable when the dataset is large. If we have $p$ covariates and a sample size of $N$, it can be shown that estimating PS incurs an $O(Np^2)$  complexity \citep{sengupta2016subsampled}. For massive data, when both $N$ and $p$ can be large, the BOOT II avoids fitting PS in all $R$ replications, thus reducing the computing time. Second, random violation of the positivity assumption \citep{westreich2010invited} can occur in standard bootstrap samples. For example, when $P(Z=1)$, the proportion of the treated participants, is  small (or large) in the original sample and, depending on whether the sample size is small or not, some of the standard bootstrap samples might not contain or just contain very few treated (or control) participants. In this case, a good number of PS estimates will be equal or distributed around 0 (resp. 1), which easily incurs extreme weights. Even if extreme weights already exist in the original sample by the way of BOOT II, weights from within some of the standard bootstrap samples of BOOT I might be more extreme.  

It is worth mentioning that, in a similarly related context, the order of bootstrap and PS matching was investigated by \cite{austin2014use}. They found that matching before bootstrapping results in estimates of standard error closer to the empirical standard deviation. Hence, as a parallel, we are also interested in the efficiency of estimating the PS (and thus the corresponding weights) before resampling with replacement (as with standard bootstrap), i.e., whether it leads to better variance estimation than running standard bootstrap---where the weights are estimated after resampling (as with the above BOOT I). In a sense, our work will extend the application of the match-then-resample and the findings of \cite{austin2014use} in the context of propensity weighting methods.

\subsection{Wild bootstrap (WB)}\label{subsec:IF-WB}

In the WB algorithm, sampling variability is introduced via multiplying an independent random variable by the part of an estimator that incurs randomness (i.e., perturbation); this occurs, for instance, when we perturb the residual in the linear regression setting \citep{wu1986jackknife, matsouaka2023variance}.

The estimator $\widehat\tau_g^{\text{aug}}$ given in \eqref{eq:aug-wate} is regular and asymptotically linear  \citep{tsiatis2007semiparametric}, i.e., $\sqrt{N}(\widehat \tau_g^{\text{aug}}-\tau_g)=\displaystyle \frac{1}{\sqrt{N}}\displaystyle \sum_{i=1}^{N}\phi_g(Z_i, \mb X_i, Y_i) + o_p(1)$, where $\phi_g(Z, \mb X, Y)$ is the influence function (IF) of $\widehat\tau_g^{\text{aug}}$, where $\mathbb{E}\{\phi_g(Z, \mb X, Y)\}=0$ and $\mathbb{V}\{\phi_g(Z, \mb X, Y)\} = \mathbb{E}\{\phi_g(Z, \mb X, Y)^2\} = \Sigma_g$. Following \cite{matsouaka2023variance}, we consider the WB by perturbing the estimated IF using independent random variables. 

The established IF of estimating $\tau_g$ is given by \citep{hirano2003efficient, crump2006moving} 
\begin{align}\label{eq:IF-wate}
	\phi_g(\mb X,Z,Y) & = \frac{g(\mb X)}{\mu_g}\left\{F(\mb X,Z,Y) + \tau(\mb X)-\tau_g\right\} + \frac{1}{\mu_g}\psi_g(\mb X,Z,Y),
\end{align}
where
$$
F(\mb X,Z,Y) = \frac{Z}{e(\mb X)}\left\{Y-m_1(\mb X)\right\} - \frac{1-Z}{1-e(\mb X)}\left\{Y-m_0(\mb X)\right\}, 
$$
$m_z(\mb X)=\mathbb{E}\{Y(z)\mid \mb X\},z=0,1$, $\mu_g=\mathbb{E}\{g(\mb X)\}$, and $\tau(\mb X) = \mathbb{E}\{Y(1)-Y(0)\mid \mb X\}$. 

Furthermore, we considered comparing two choices of $\psi_g(\mb X,Z,Y)$ in \eqref{eq:IF-wate}, {resulting in two different WB approaches (specified by WB I and WB II below)}:
{
\begin{align}
   & \textbf{WB I}: \psi_g(\mb X,Z,Y) = 0  \text{ \citep{hirano2003efficient}, ~~and}\label{eq:eif-I} \\
   & \textbf{WB II}: \psi_g(\mb X,Z,Y) = \dfrac{\partial g(\mb X)}{\partial e(\mb X)}\{\tau(\mb X)-\tau_g\}\{Z-e(\mb X)\} \text{ \citep{crump2006moving}.} \label{eq:eif-II}
\end{align}
We considered these two IF choices for the following reason because they originate from different literature and are derived under different assumptions. The IF in WB I, from \cite{hirano2003efficient}, assumes the PS is known in the observed data structure, whereas the IF in \cite{crump2006moving} (used in WB II) assumes the PS is unknown and must be modeled and estimated. This distinction is critical for practical implementation, as the PS is typically unknown in observational studies. Incorporating both IFs into the WB framework allows us to explore the role of the PS across different WATE estimands, offering valuable insights into accurately quantifying uncertainty in WATE estimation. 
}

\begin{remark}\label{rmk:bound}
    {It is easy to verify that using option I of the $\psi_g(Z,\mb X, Y)$ term, the equation \eqref{eq:eif-I}, in the IF corresponds to the form of the augmented estimator \eqref{eq:aug-wate}.  
    However, the }empirical version of $\mathbb{E}\{\phi_g(Z, \mb X, Y)^2\}$ {by using this $\psi_g(Z,\mb X, Y)$ in \eqref{eq:eif-I} is generally not useful} for variance estimation, because in practice, the corresponding estimated IF, $\widehat\phi_g$, is obtained by plugging in estimated PS and outcomes; hence, the uncertainty in estimating nuisance functions is not accounted for. Technically, we need to further project the $\phi_g(Z, \mb X, Y)$ onto the orthogonal complement of the nuisance tangent space $\Lambda^\perp$. Thus, the actual variance quantification should be $\mathbb{E}\{\Pi(\phi_g(Z, \mb X, Y)\mid\Lambda^\perp)^2\}$, where  $\Pi(\phi_g(Z, \mb X, Y)\mid\Lambda^\perp)$ denote this projection of  $\phi_g(Z, \mb X, Y)$  onto the nuisance tangent space $\Lambda^\perp$. It is worth mentioning that $\mathbb{E}\{\Pi(\phi_g(Z, \mb X, Y)\mid\Lambda^\perp)^2\}$  is smaller than $\mathbb{E}\{\phi_g(Z, \mb X, Y)^2\}$ by Pythagorean theorem \citep{tsiatis2007semiparametric}. However, the projection $\Pi(\phi_g(Z, \mb X, Y)\mid\Lambda^\perp)$  is generally hard to express and is not model-free, which is why we seek other flexible alternative variance estimation methods for help{, such as by evaluating whether the WB perturbation can be helpful}.
\end{remark}

We specify the WB algorithm for WATE estimators as follows. 
Consider $\widehat \phi_g(Z, \mb X, Y)$ an estimator of  $\phi_g(Z, \mb X, Y)$, where we replace $\tau_g$ by its estimator $\widehat \tau_g^{\text{aug}}$. 
We generate random vectors $\bd\xi = (\xi_1, \dots, \xi_N)$ of independent identically distributed (i.i.d.) random variables---with mean 0 (or 1), variance 1, and $\mathbb{E}(\xi_i^3)<\infty $, where $\xi$ does not depend on the observed data $\{(Z_i, \mb X_i, Y_i): i=1\dots, N \}$---and calculate a perturbed estimator $\widehat \tau_g^* = \widehat \tau_g^{\text{aug}} + \displaystyle \frac{1}{N}\sum_{i=1}^{N}\xi_i\widehat \phi_g(Z_i, \mb X_i, Y_i)$. 

Assuming some regularity conditions,  
$$\widehat \tau_g^*-\tau_g=\displaystyle \frac{1}{N}\displaystyle \sum_{i=1}^{N}(1+\xi_i)\widehat \phi_g(Z_i, \mb X_i, Y_i) + o_p(N^{-1/2}),$$
$$\sqrt{N}(\widehat \tau_g^*-\tau_g)\to_d\mc N(0, \Sigma_g^*), \qquad \Sigma_g^* = \mathbb{E}\{(1+\xi)^2\phi_g^2\}, \text{ as } N\to\infty.$$ 

Moreover, the empirical distribution of $ \displaystyle \frac{1}{\sqrt{N}}\sum_{i=1}^{N}\xi_i\widehat \phi_g(Z_i, \mb X_i, Y_i)$ approximates the sampling distribution of $\sqrt{N}(\widehat\tau_g^{\text{aug}}-\tau_g).$ This result establishes the validity of the WB procedure in the following Algorithm \ref{algo:wildboot} based on \cite{chernozhukov2018sorted}.

\begin{algorithm*}
  \caption{Wild bootstrap (WB) for variance estimation of WATE} \label{algo:wildboot}
  \begin{algorithmic}[1]
  \STATE {\bfseries Input:} {Observational data $\mc O = \{(Z_i, \mb X_i, Y_i), i=1,\dots, N\}$; tilting function $g(\mb x)$ of the given WATE; number of WB replicates $R$; confidence level $\alpha\in(0,0.5)$; {estimated influence function $\widehat\phi_g(Z, \mb X, Y)$, based on the chosen $\psi_g$ from \eqref{eq:eif-I} or \eqref{eq:eif-II} (corresponding to WB I or WB II, respectively); independent random variables $\xi_i$ for perturbation, e.g., $\xi_i$ follows the exponential EXP$(1)$ or the Rademacher distribution (defined by the probability mass function $P(\xi_i=1)=P(\xi_i=-1)=1/2$).}  }
  
  \STATE {\bfseries Output:} {Standard error estimate $\widehat\Sigma_g^{1/2}$ and a $100\times(1-\alpha)\%$ confidence interval $\widehat C_\alpha(\mc O)$. }

  \FOR{$r=1,\dots, R$}

  \STATE Generate a random vector $\bd\xi=(\xi_1,\dots,\xi_N)'$ with i.i.d. components {following the chosen distribution for perturbation. }
  
  \STATE Compute $\widehat \Delta_g^{(r)} = \displaystyle\frac{1}{\sqrt{N}}\sum_{i=1}^{N}\xi_i\widehat \phi_g(Z_i, \mb X_i, Y_i)$. 
  \ENDFOR
  
  \STATE \textbf{Return: } 
  The standard error estimate by WB: $\widehat\Sigma^{1/2}_g = \dfrac{\text{IQR}(\widehat\Delta_g^{(1)},\dots, \widehat\Delta_g^{(R)})}{z_{0.75}-z_{0.25}}$ or the standard deviation of $\{\widehat\Delta_g^{(1)},\dots, \widehat\Delta_g^{(R)}\}$; the $100\times(1-\alpha)\%$ confidence interval $\widehat C_\alpha(\mc O) = \widehat\tau_g^{\text{aug}}\pm\dfrac{z_{(1-\alpha/2)}}{\sqrt{N}}\widehat\Sigma^{1/2}_g$, where $z_{0.75}-z_{0.25}=1.349$ is the interquartile range of the standard normal distribution, and $\widehat\tau_g^{\text{aug}}$ is by \eqref{eq:aug-wate}.  
  \end{algorithmic}
\end{algorithm*}


{
\begin{remark}\label{rmk:WBvar}
    How are the three sources of variability in estimating a WATE accounted for in WB? The WB method consists of two key components: (i) the estimated IF of the augmented estimator and (ii) the perturbation procedure using i.i.d. variables $\xi$'s. The IF characterizes the infinitesimal impact of a single data point on the estimator \citep{van2000asymptotic}, capturing variability that arises from both the outcome model and PS estimation. The perturbation procedure mimics sampling variability by resampling the empirical IF terms. Together, these components account for uncertainty due to the outcome models, PS estimation, and finite sample variation in a unified manner. For the augmented estimator, the IF is not model-free, as it depends on the specific parametric or nonparametric specifications of the PS and outcome models. In this work, we leverage two candidate IFs proposed in the literature---equations \eqref{eq:eif-I} and \eqref{eq:eif-II} \citep{hirano2003efficient, crump2006moving}---and combine them with the WB perturbation to assess their performance and provide our findings and practical recommendations. 
\end{remark}
}

\section{Numerical Experiments}\label{sec:sim}

\subsection{Data generating process}\label{subsec:sim-DGP}

We assessed the performance of all variance estimation methods using Monte Carlo simulations.  We considered a  data generating process (DGP) similar to that of  \cite{matsouaka2023variance}. We used the covariate vector $\mb X=(1, X_1,\dots,X_7)'$, for $N$ independent observations, such that $X_{4}\sim \text{Bern}(0.5), X_{3}\sim \text{Bern}(0.2X_{4}+0.4)$, $(X_1,X_2)'\sim \mc N(\bd\mu,\bd\Sigma)$, $X_{5} = X_{1}^2, X_{6} = X_{1}X_{2}$, $X_{7} = X_{2}^2$, where $\bd\mu = \left(-X_3+X_4+0.5X_3X_4, X_3-X_4+X_3X_4\right)'$  and  $
    \bd\Sigma = X_3\begin{pmatrix}
    1&0.5\\0.5&1
    \end{pmatrix} + 
    (1-X_3)\begin{pmatrix}
    2&0.25\\0.25&2
    \end{pmatrix}
    $. 

Then, we simulated the potential outcomes $\allowdisplaybreaks Y(z)=0.5+X_{1}+0.6X_{2}+2.2X_{3}-1.2X_{4}+(X_{1}+X_{2})^2 + z\delta(\mb X) +\varepsilon(z)$, with $\varepsilon(z)\sim \mc N(0,1)$ as an i.i.d. random error, for $z=0,1$. We consider both $\delta(\mb X)=4$ for a homogeneous treatment effect and $\delta(\mb X)=4+3(X_{1}+X_{2})^2+X_{1}X_{3}$ for a heterogeneous treatment effect.
Next, we generated the treatment 
$Z\sim \text{Bern}\left(\{1+\exp(-\mb X'\bd\beta)\}^{-1}\right)$ from a logistic regression model, where $\bd\beta$ is the regression coefficients of the PS model. {The observed outcome $Y$ for each participant is then generated by $Y=ZY(1)+(1-Z)Y(0)$. }

\begin{figure}[H]
    \centering
  \includegraphics[trim=10 5 10 30, clip, width=0.5\textwidth]{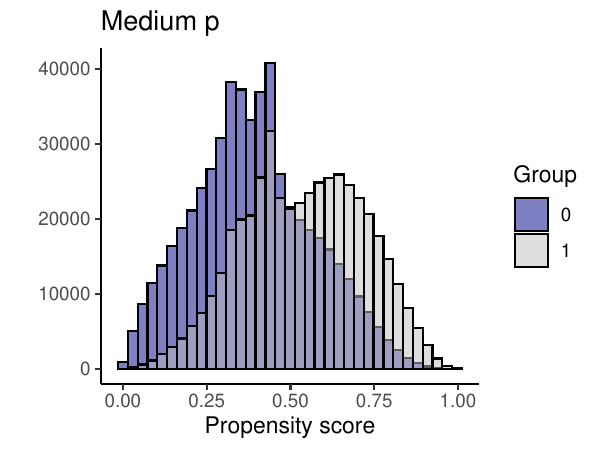}
\caption{PS distribution for Model 2 (proportion of treated participants $\approx 45.9\%$)}\label{fig:sim-ps-mod2}
\end{figure}

{We varied the value of the  above $\bd\beta$ parameter to generate 5 PS models to assess our methods under different practical scenarios. }We chose $\allowdisplaybreaks\bd\beta=(\beta_0, 0.3, 0.4, 0.4, 0.4, -0.1, -0.1, 0.1)$, with $\beta_0 = -2.17, -0.78$ and $0.98$ respectively, for the first 3 models. {These three models represent practical scenarios encompassing different proportions of treated participants $p=P(Z=1)$, where we label them ``Models 1--3,'' corresponding to small ($p=0.20$), medium ($p=0.45$) and large ($p=0.80$) proportions of treated in possible real-world data, respectively. In addition, the first three models show moderate to high overlaps in the true PS distributions of the two treatment groups (see Figure \ref{fig:sim-ps-dist1} in Appendix \ref{apx:additional-sim}). For the fourth model, we let $\allowdisplaybreaks\bd\beta = (0.2, 1.0, -0.9, -0.9, 0.9, 0.15, 0.15, -0.2)$, which creates a case of poor overlap for the PS distributions of the two treatment groups (see Figure \ref{fig:sim-ps-dist2} in Appendix \ref{apx:additional-sim}), accompanied by extreme IPW weights, and we label this case ``Model 4.'' This model has $p=0.49$ (a medium $p$ scenario). Finally, we explored a scenario where the SAND method may not work by utilizing the $\bd\beta$ of Model 3 ($p=0.80$) but under a small sample size ($N=50$), which we labeled ``Model 5.'' } Figure \ref{fig:sim-ps-mod2}, shows the distribution of PSs under Model 2 {(the medium $p$ with a moderate to high overlap)}, using a randomly generated sample of size $N_{\text{super}}=10^6$.  

{Let us now clarify why the chosen DGP is appropriate for evaluating both the proposed and competing methods. First, covariates $X_1$--$X_4$ act as confounders, with $X_5$--$X_7$ representing their higher-order terms, mimicking complex settings often found in data from observational studies. Potential outcomes $Y(z)$ and treatment assignment $Z$ are generated as functions of these covariates, along with independent random error terms $\varepsilon(z)$ for $z=0,1$. Thus, this setup fits the structure of  observational data and satisfies Assumption \ref{assp:unconfound} (unconfoundedness), as both outcomes and treatment depend on a common set of covariates. The observed outcome also follows the consistency assumption, where $Y = ZY(1) + (1 - Z)Y(0)$ for all participants. Moreover, by varying the PS model parameters $\bd\beta$, we can alter the distribution of $Z$, allowing us to simulate a range of practical scenarios through different PS distributions, including scenarios where there are violations of the positivity assumption. }

 To evaluate the performance of different variance estimation methods under repeated sampling, we generated data of sample size $N = 1000$ for Models 1--4 and $N=50$ for Model 5, with $M = 2000$ Monte Carlo simulation replicates. We
considered the augmented ATE, ATT, ATO, ATM, and ATEN as estimators of interest in our simulation; the true values of all the five estimands under heterogeneous treatment effects
can be found in Table \ref{tab:truth}. For BOOT I, BOOT II and all WB's, we considered $R = 200$ repeated (or perturbed) samples.

\subsection{{Competing methods, evaluation criteria, and guide to interpreting results}}\label{subsec:sim-measure}

{
We considered several competing variance estimation methods in our simulation study. For clarity and conciseness, we use the following acronyms throughout Sections \ref{subsec:sim-measure} and \ref{subsec:sim-results}: 
\begin{itemize}
    \item \textbf{BOOT I}: standard bootstrap;
    \item \textbf{BOOT II}: post-weighting bootstrap, i.e., Algorithm \ref{algo:bootII}; 
    \item \textbf{SAND}: sandwich variance estimator;
    \item  \textbf{WBExp I}: wild bootstrap (Algorithm \ref{algo:wildboot}) with $\xi\sim\text{EXP}(1)$ and the estimated IF $\widehat\phi_g$ computed using \eqref{eq:eif-I};
    \item  \textbf{WBExp II}: wild bootstrap (Algorithm \ref{algo:wildboot}) with $\xi\sim\text{EXP}(1)$ and the estimated IF $\widehat\phi_g$ computed using \eqref{eq:eif-II};
    \item  \textbf{WBRad I}: wild bootstrap (Algorithm \ref{algo:wildboot}) with $\xi$ following Rademacher distribution and the estimated IF $\widehat\phi_g$ computed using \eqref{eq:eif-I}; and
    \item  \textbf{WBRad II}: wild bootstrap (Algorithm \ref{algo:wildboot}) with $\xi$ following Rademacher distribution and the estimated IF $\widehat\phi_g$ computed using \eqref{eq:eif-II}. 
\end{itemize}
}

To assess the performance of the estimators we considered and investigate the sensitivity of their variance estimations when one or both models are misspecified, we included the following model specifications for PS and outcome regression (OR): (1) both PS and OR models are correctly specified; (2) only PS model is correctly specified; (3) only OR model is correctly specified; (4) both PS and OR model are misspecified. 

{For the misspecified PS and OR models, we intentionally excluded the higher-order terms $X_5 = X_1^2$, $X_6 = X_1X_2$, and $X_7 = X_2^2$ from the original (true) models. This omission preserves all confounders ($X_1$--$X_4$) in the PS and OR models, thereby maintaining Assumption \ref{assp:unconfound} for valid causal inference. However, it partially misspecifies the functional relationship between treatment (or outcomes) and covariates. This approach is consistent with prior work \citep{kang2007demystifying, li2021propensity, zhou2020propensity}.  }


We evaluate the simulation results and compare the performance of the estimators based on the following measures: \begin{enumerate}
    \item The relative bias, defined as $(\widehat\tau_g-\tau_g)/\tau_g$, where $\widehat\tau_g$ is the estimate of $\tau_g$ in a simulation replicate (subscript omitted). 
    The boxplots of the relative biases under Model 2 are reported in the main text. We also report the absolute relative
percent bias (ARBias\%) $ =  100\% *\left| \text{mean of }\left(\widehat\tau_g-\tau_g\right)/\tau_g\right|$, for each scenario, in Appendix \ref{apx:additional-sim}; 
\item The root mean square error (RMSE), defined as the square root of the average of $(\widehat\tau_g-\tau_g)^2$ over the 2000 data replicates;
\item The empirical standard deviation (ESD), $SD(\widehat\tau_g)$, is the standard deviation of estimates over the 2000 data replicates; 
\item The median standard error (SE), i.e, the median of the square root of estimated variance, for each competing method;
\item The relative efficiency (RE), which is the ratio of the empirical variance to the variance by a competing method. We report REs from all data replicates as boxplots and the median of REs out of 2000 data replicates;
\item The coverage probability (CP), i.e., the proportion of the number of times (out of the 2000 data replicates),  the true $\tau_g$ fell inside the 95\% Wald confidence intervals constructed with the corresponding variance estimation method. A CP would be considered significantly different from the 0.95 nominal level if it is outside the range of $[0.940,0.960] = 0.95\pm z_{0.975}\sqrt{{0.95(1-0.95)}/{2000}}$, where $z_{0.975}=1.96$ is the $0.025$-quantile of the standard normal distribution. 
\end{enumerate}

{
Based on these metrics, we further provide a step-by-step guide and clarification below to interpret results under possible model misspecifications, as designed in our simulations, when assessing the performance of a variance estimator in practice:
\begin{enumerate}
    \item Examine the estimation bias: Relevant metrics include ARBias\% and RMSE. These help determine whether the augmented estimator remains consistent under the various model specifications considered.
    \item Evaluate RE: If the ARBias\% and RMSE suggest that the estimator is consistent, RE can be used to assess how well the variance estimate captures the estimator’s sampling variability. An RE close to 1 indicates good alignment. However, if the estimator is biased, an RE near 1 does not guarantee the variance estimate is valid for the true WATE, and should be interpreted with caution.
    \item Check CP: Finally, examine the CP to determine whether the variance estimator supports valid inference. Nominal coverage (e.g., $\approx 95\%$ for a 95\% confidence interval) is expected if the variance is well estimated. However, a good CP alone does not confirm validity, as it can also result from a biased estimator paired with an overestimated variance. Ideally, when the bias is negligible, RE is close to 1, and CP is near the 95\% nominal level, one can be more confident that the variance is correctly estimated. 
\end{enumerate}
As a remark, the metrics ESD and SE can also serve as useful alternatives to RE. While RE is generally sufficient for comparing estimated variance with sampling variability, ESD and SE offer more direct while less summarized measures of the numerical values of the estimated variance and the empirical variability of the estimator. Furthermore, we emphasize that assessing variance estimators based solely on RE or CP is insufficient to establish their validity, especially under potential model misspecification. A comprehensive evaluation following our above step-by-step guide is necessary. The technical rationale behind this point is detailed in Appendix \ref{subapx:miss}. 
}

\subsection{Findings and results}\label{subsec:sim-results} 

To be succinct and concise, we only report the simulation results for Model 2 in the following Figures \ref{fig:sim-bias-RMSE}--\ref{fig:sim-CP}, under heterogeneous treatment effect with $N=1000$. This showcases we major contrasts across the different estimators; we defer the complete set of simulation results in Appendix \ref{apx:additional-sim}. 

\begin{figure}[H]
    \centering
    \includegraphics[trim=5 5 5 5, clip, width=\textwidth]{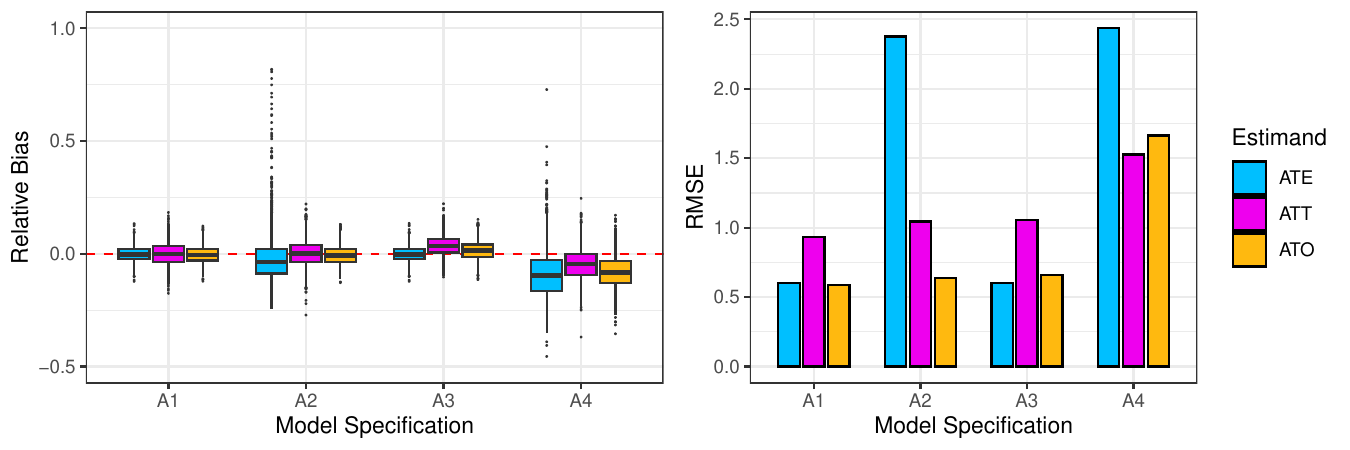}
    
    {\footnotesize A1: Both PS and OR models are correctly specified; A2: Only PS model is correctly specified; A3: Only OR model is correctly specified; A4: Both PS and OR models are misspecified.}\caption{{Point estimation results by the augmented estimator \eqref{eq:aug-wate} for ATE, ATT and ATO, } using relative biases and RMSEs, under Model 2 (medium $p$), with $N=1000$ and heterogeneous treatment effect. }\label{fig:sim-bias-RMSE}
\end{figure}

Because the ATO, ATM, and ATEN are similar estimands \citep{matsouaka2024causal} and they indeed yielded similar results (see Table \ref{tab:mediump-n1000-hets-results} and other results in Appendix \ref{apx:additional-sim}), we only report the results for ATE, ATT and ATO here. Figure \ref{fig:sim-bias-RMSE} reveals that, overall {the augmented estimator of} ATO has the smallest relative biases and RMSEs across the four model specifications, compared to {the augmented estimators of} ATE and ATT. The augmented estimator for all 3 estimands performs better when the OR model is correctly specified (cases A1 and A3) than when the PS model is correctly specified. These findings are consistent with most empirical studies about WATEs and their augmented estimators \citep{matsouaka2024causal,li2018balancing, zhou2020propensity}.

\begin{figure}[H]
	\centering
	\includegraphics[trim=5 5 6 5, clip, width=\textwidth]{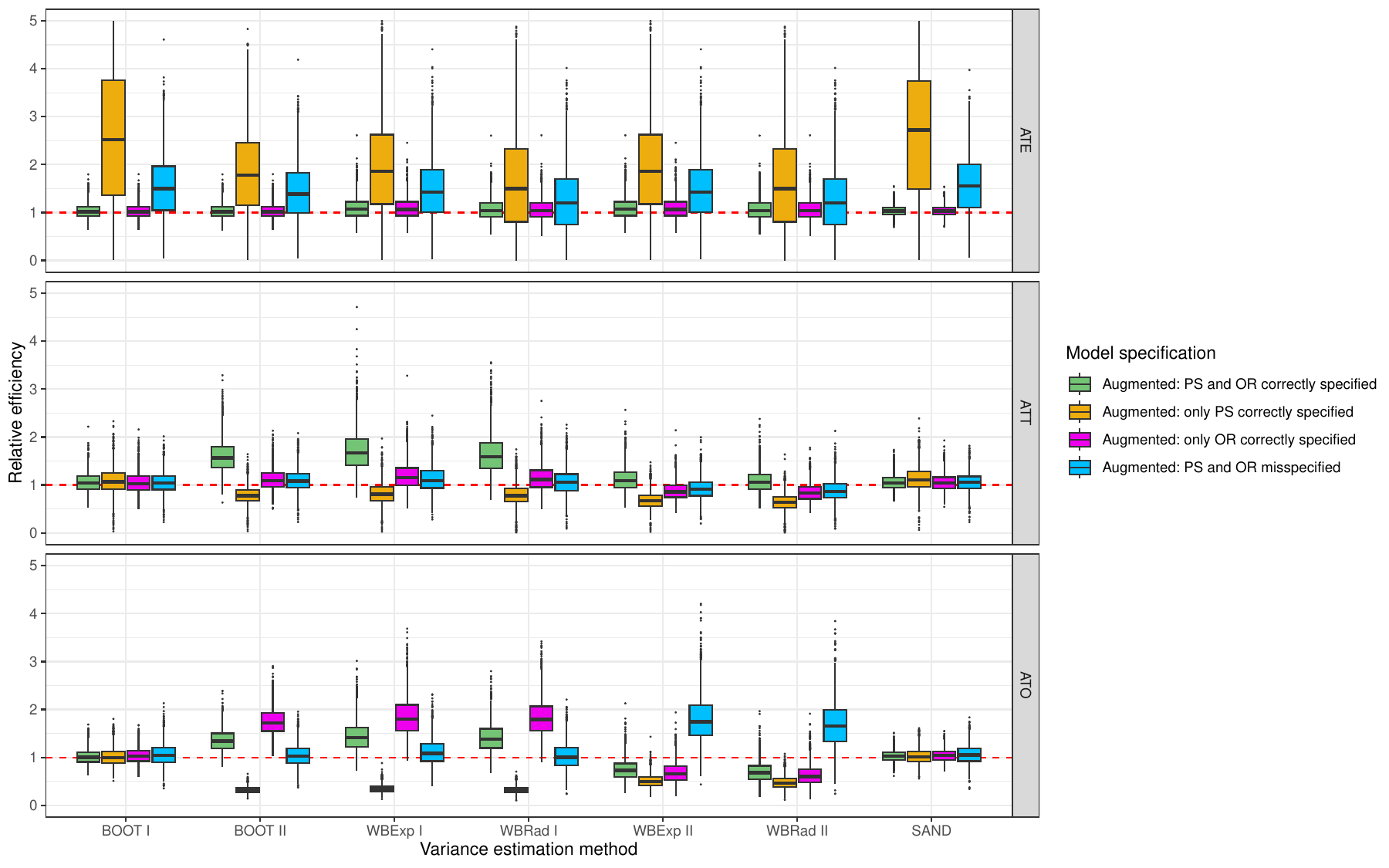}
 
	{\footnotesize BOOT I: standard bootstrap; BOOT II: post-weighting bootstrap; WBExp (resp. Rad): wild bootstrap via exponential (resp. Rademacher) distribution; SAND: sandwich variance estimation. The red dashed lines indicate the relative efficiency $=1$. }
	\caption{Relative efficiencies (REs) by different variance estimation methods for the model 2 with $N=1000$ and heterogeneous treatment effect. }\label{fig:sim-RE}
\end{figure}

Figure \ref{fig:sim-RE} shows the distributions of the REs by different estimands, model specifications and variance estimation methods. Interestingly, the performance of each variance estimation method depends on the estimand. For ATE, all methods are sensitive to OR model misspecifications as in these cases the boxplots are wider and the means of REs essentially deviate from 1. In addition, BOOT I and SAND methods perform slightly worse than other methods when only PS model is correctly specified. All WB I and WB II methods (WBExp I, WBRad I, WBExp II and WBRad II) perform similarly and are similar to BOOT II. We should note that WB I and II are the same for ATE because the term ${\partial g(\mb X)}/{\partial e(\mb X)}=0$ in the IF for WB II. Whenever the OR model is correctly specified, all methods result in good REs for ATE. Next, for ATT and ATO, we observed that BOOT I and SAND outperform in all model specifications. For BOOT II and WB methods on ATT and ATO, we found they generally fail to provide valid estimates of the variance, and only WBExp II and WBRad II for ATT perform well. 

From Tables \ref{tab:smallp-n1000-cons-results}, \ref{tab:mediump-n1000-cons-results}, \ref{tab:largep-n1000-cons-results} and \ref{tab:extremew-n1000-cons-results} in Appendix, we also note that when the treatment effect is homogeneous, all methods perform similarly and are valid for all ATE, ATT and ATO estimands. {This is because under a homogeneous treatment effect, all these estimands target the same quantity  (here the same treatment effect $\tau_g=4$ in our simulations).} Therefore, the level of heterogeneity of the treatment effect is also a factor that contributes to the uncertainty quantification. Intuitively, more heterogeneity means the individual treatment effects are more associated with covariates, which contributes to more uncertainty in estimating the estimand itself. 

Figure \ref{fig:sim-CP} depicts radar plots to compare the CPs; each corner in a radar plot represents a variance estimation method. We can see that for ATE, all WB methods and BOOT II result in approximately $95\%$ nominal coverage level, in most scenarios. The only exception are for when both PS and OR models are misspecified. BOOT I and SAND result in lower coverage when only PS model is correctly specified. In comparison, for both ATT and ATO, WB methods and BOOT II commonly result in coverage levels that are different from $95\%$ nominal level, either smaller or larger (i.e., anti-conservative or conservative confidence intervals, respectively). BOOT I and SAND for ATT and ATO result in valid coverage levels whenever the OR model is correctly specified. More insights and discussions for different variance estimation methods follow in Section \ref{sec:conclusion}. 
\begin{figure}[H]
	\centering
	\includegraphics[trim=10 10 10 2, clip, width=1\textwidth]{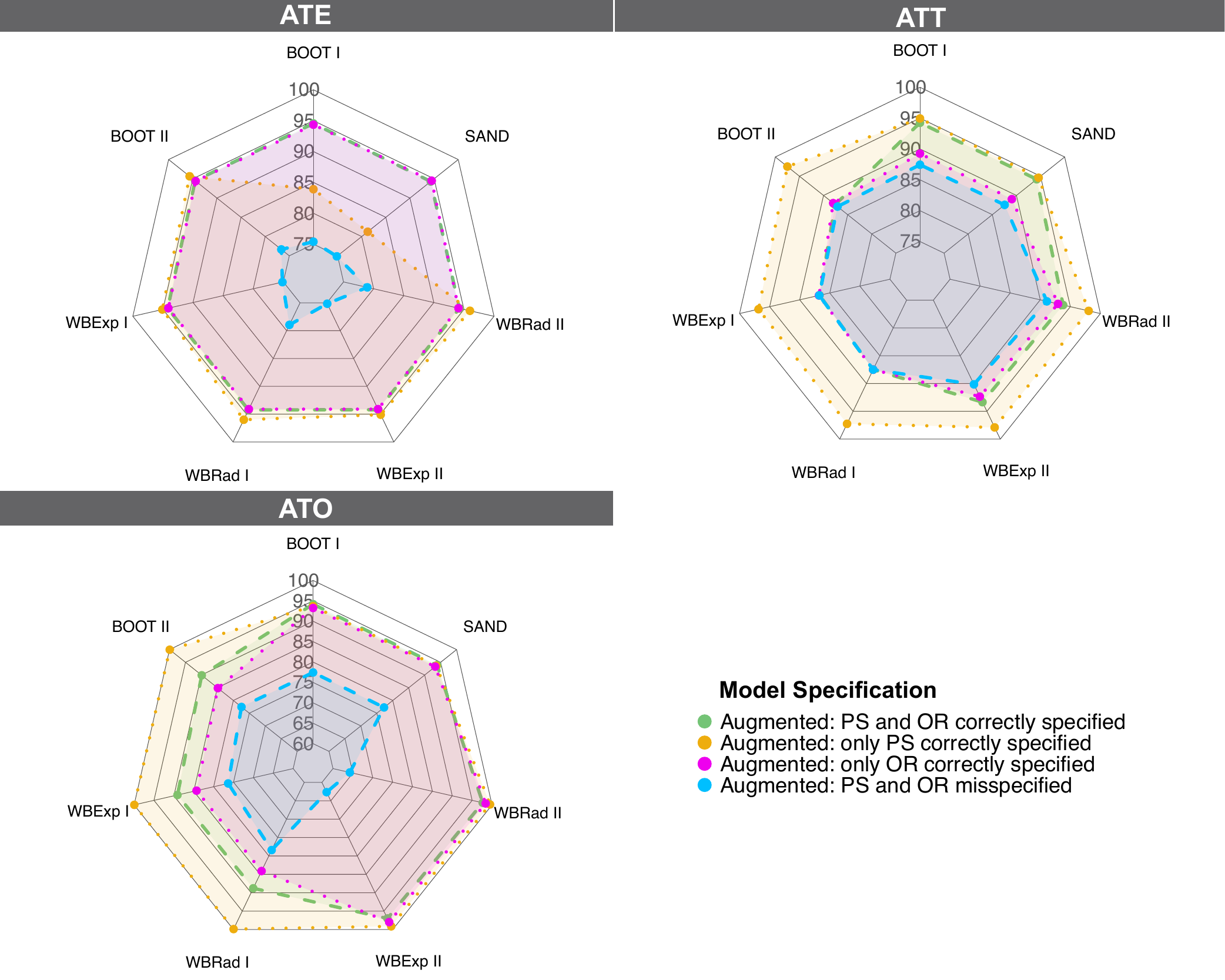}
	{\footnotesize BOOT I: standard bootstrap; BOOT II: post-weighting bootstrap; WBExp (resp. Rad): wild bootstrap via exponential (resp. Rademacher) distribution; SAND: sandwich variance estimation. }
	\caption{Coverage probabilities (\%) by different variance estimation methods by Model 2, with $N=1000$ and heterogeneous treatment effect. }\label{fig:sim-CP}
\end{figure} 

Finally, we briefly comment on additional results we obtained in Appendix \ref{apx:additional-sim} for all models. {When the treatment effect is homogeneous and across Models 1--3 (with varying $p$ from small to large), all competing methods yield similar and comparable variance estimates. This suggests that the variance estimators perform similarly in the absence of effect heterogeneity. Also, when the treatment effect is heterogeneous, Models 1 and 3 produce results consistent with those presented in this section, indicating that varying $p$ does not substantially impact the relative performance of the variance estimation methods. In the scenario with extreme PS weights and poorer overlap (Model 4), the results are broadly similar to those from Models 1–3. However, the point estimates for ATE and ATT exhibit greater bias across all model specifications. This result is expected given that these estimands rely on extreme PS weights, whereas other estimands employ bounded weights for all participants. Finally, for Model 5 (where SAND does not work), }Table \ref{tab:missingSAND} shows the frequency of unobtainable sandwich variance estimates over $2000$ simulations using Model 5; we can see that the frequency varies between 322 and 626 for all WATEs. This means that when the sample size is small and there is limited overlap, the sandwich variance estimator can be expected to not work properly. All the other methods also perform poorly under Model 5 (Tables \ref{tab:sandnot-n50-cons-results} and \ref{tab:sandnot-n50-hets-results}), except for BOOT I. Therefore, when the sample size is small, our finding suggests that standard bootstrap is still the optimal choice. 

{
\subsection{Results discussion}\label{subsec:simudiscuss}

When both the PS and OR models are correctly specified, we find that BOOT II and WB I yield RE values greater than 1 on average, whereas WB II tends to produce REs less than 1. For BOOT II, its strong performance for ATE contrasts with its behavior for ATT and ATO: BOOT II is valid for ATE but does not directly extend to other WATEs. In addition, WB I uses an IF that does not fully account for the role of the PS in the joint likelihood. Since the PS plays a non-ancillary role in defining ATT and ATO, the anti-conservative results observed for WB I are consistent with our theoretical understanding. For WB II, although the employed IF accounts for uncertainty in the PS model, we observe slightly conservative results for ATT and ATO. This suggests that while the IF used in WB II may not be the most efficient choice for the augmented estimator in Equation \eqref{eq:aug-wate}, it offers a computationally efficient, model-free, and slightly conservative alternative to BOOT I and BOOT II for WATE estimation. 

Furthermore, following the guideline provided in Section \ref{subsec:sim-measure}, we interpret the variance estimates through a comprehensive review of Figures \ref{fig:sim-bias-RMSE}--\ref{fig:sim-CP}. Interestingly, we observe that for some methods---such as BOOT II, WBExp I, and WBRad I for ATT---the RE values under both misspecified models (blue boxes) are closer to 1 than those under correctly specified models (green boxes). However, Figure \ref{fig:sim-CP} shows that the corresponding CPs are substantially below the nominal 95\% level. When combined with the estimation bias results in Figure \ref{fig:sim-bias-RMSE}, these findings suggest the following interpretation: the variance estimators yield RE values close to 1 because they approximate the sampling variability of $\tau_g^{\text{aug}}$ well, but $\tau_g^{\text{aug}}$ itself is impacted by misspecified PS and OR models. As a result, the poor CPs are driven by bias in estimating the true WATE, and the good REs merely indicate that the variance estimators are valid for a shifted estimand induced by the limits of the misspecified models. Similarly, as shown in Figure \ref{fig:sim-CP}, some CPs under model misspecification outperform those under correct model specification. For example, in the cases of ATT and ATO where only the PS model is correctly specified (orange radars), the CPs for BOOT II and WB I are closer to the nominal 95\% level than those under fully correctly specified models. However, Figure \ref{fig:sim-RE} reveals that the corresponding RE values for ATT and ATO in these cases are substantially greater than 1, indicating that the estimated variances are much larger than the empirical variances---thus reflecting conservative estimation of the true sampling variability. }

\section{Case Study}\label{sec:data}

We use the data from the National Health and Nutrition Examination Survey (NHANES) 2013–2014 to illustrate the proposed variance estimation methods \citep{zhao2019sensitivity}. 
The data can be found on the NHANES website: \url{https://wwwn.cdc.gov/Nchs/Nhanes/2013-2014/PBCD_H.htm}. 
{The dataset included a total of $N = 1,107$ participants, with 234 (21.14\%) in the treatment (or exposed) group and 873 (78.86\%) in the control group. Each participant completed a questionnaire on seafood consumption. High fish consumption (treatment group, $Z = 1$) was defined as consuming more than 12 servings of fish or shellfish in the past month, while low fish consumption (control group, $Z = 0$) was defined as 0 or 1 serving.} 

{To comprehensively evaluate the proposed variance estimation methods across different estimands, we conducted two separate analyses: one with the full data (in Section \ref{subsec:primary}) and the other using the subset of participants who are 40 years or older (see Section \ref{subsec:second}). In both analyses, we included 8 covariates: race, gender, age, income, missingness indicator for income, education, smoking history, and number of cigarettes smoked in the past month. The outcome of interest was the base-2 logarithm ($\log_2$) of total blood mercury level ($\mu$g/L).} 



\begin{figure}[H]
     \centering
     \begin{subfigure}[b]{0.45\textwidth}
         \centering
         \includegraphics[trim=5 5 5 5, clip, width=\textwidth]{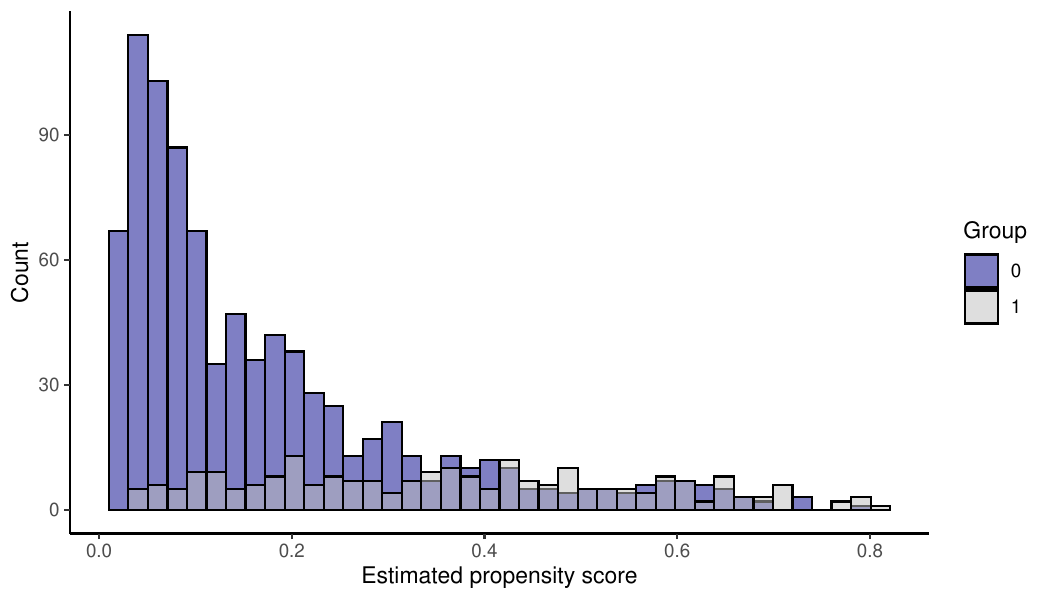}
         \caption{Estimated PSs by treatment group, where 0 = low fish consumption level, 1 = high fish consumption level}
         \label{fig:dat-ps}
     \end{subfigure}
     \hspace{-.1cm}
     \begin{subfigure}[b]{0.50\textwidth}
         \centering
         \includegraphics[trim=5 5 5 5, clip, width=\textwidth]{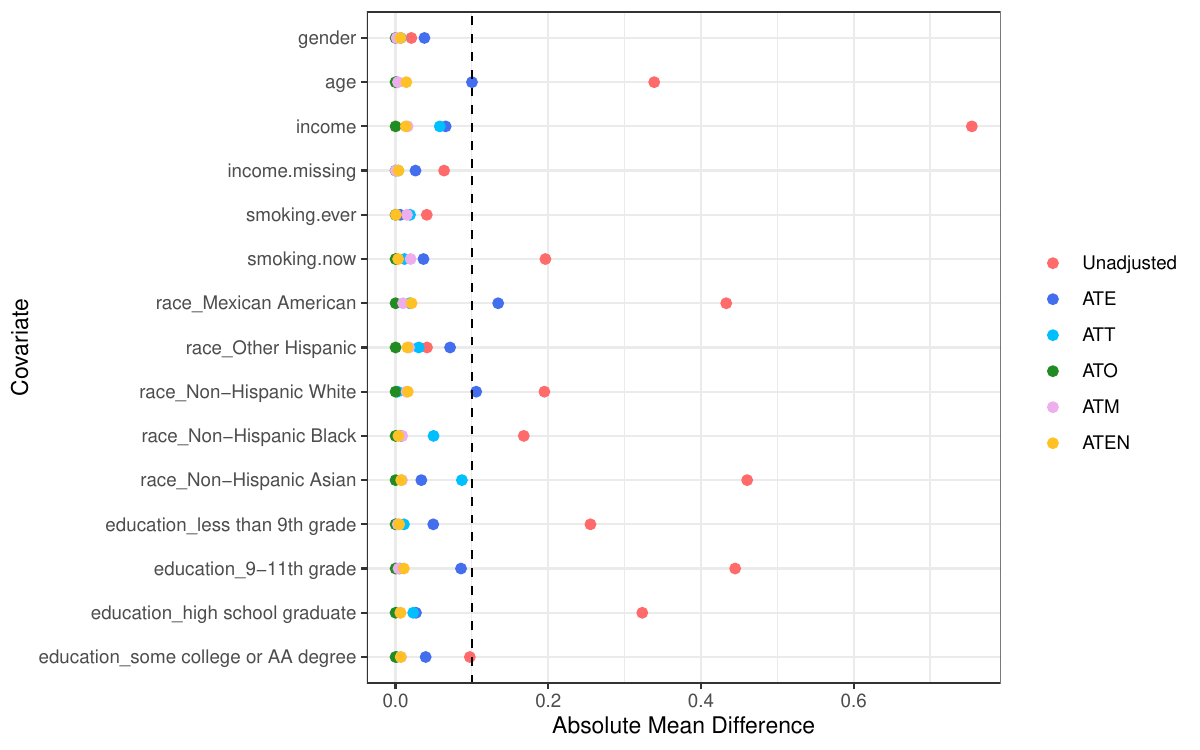}
         \caption{Covariates balance via different weights}
         \label{fig:dat-covbala}
     \end{subfigure}
        \caption{Estimated PS distributions by treatment group and covariates balance of the NHANES 2013-2014 data }
        \label{fig:Data_EX}
\end{figure}

{\subsection{Primary analysis}\label{subsec:primary}

Our primary analysis uses the full dataset and applies all variance estimation methods discussed in this paper to the augmented estimators for five WATEs (ATE, ATT, ATO, ATM, and ATEN). }
Figure \ref{fig:dat-ps} presents the estimated {PS} between the two treatment groups, estimated via logistic regression, {with a linear predictor for all 8 covariates}. The distribution of PSs for the control group (low fish consumption) is right-skewed, while that for the treated group (high fish consumption) is more uniformly distributed. {The distributions of the estimated PS of two groups are highly overlapped, except for a few observations whose estimated PSs are greater than 0.7. } Figure \ref{fig:dat-covbala} shows the covariates balance of each covariate, measured by {the absolute standardized mean difference (AMSD),  between the two treatment groups. The figure shows that all covariates are well balanced for treated, overlap, matching, and entropy weights, i.e., all weighted ASMDs are less than the rule-of-thumb threshold 0.1 \citep{austin2009balance}, compared to the unadjusted AMSD.} With the ATE weights, 
the ASMDs for age, Mexican American (race) and Non-Hispanic White (race) is at or {little over} the 0.1 threshold. 


\begin{table}[H]\small
	\begin{threeparttable}
		\centering
		\caption{Difference in ($\log_2$) blood mercury level by fish consumption*.}
		\label{tab:dat-varest}
		\begin{tabular}{llllllllllll}
			\toprule
			Method & Estimate & SE &  & Estimate & SE &  & Estimate & SE\\
            \midrule
            \addlinespace
            & \multicolumn{2}{c}{ATE (ESS:  403.40)}&&\multicolumn{2}{c}{ATT (ESS:  508.01)}&&\multicolumn{2}{c}{ATO (ESS: 590.28)} \\
			\cmidrule(lr){1-1}\cmidrule(lr){2-3}\cmidrule(lr){5-6}\cmidrule(lr){8-9}
            BOOT I & 1.74 & 0.128 && 2.12 & 0.122 && 1.98 & 0.105 &\\  
			BOOT II & 1.74 & 0.127  && 2.12 & 0.117 &&  1.98 & 0.102 &\\ 
			WBExp I & 1.74 & 0.091 & & 2.12 & 0.113 && 1.98 & 0.094  &\\ 
			WBRad I & 1.74 & 0.109 & & 2.12 & 0.103 && 1.98 & 0.093 &\\  
            WBExp II & 1.74 & 0.091 && 2.12 & 0.101 && 1.98 & 0.086 &\\ 
            WBRad II & 1.74 & 0.109 && 2.12 & 0.116 && 1.98 & 0.100 &\\ 
			SAND & 1.74 & 0.114 && 2.12 & 0.115 &&  1.98 & 0.097 &\\ 
			\addlinespace 
            &\multicolumn{2}{c}{ATM (ESS:  568.43)}&& \multicolumn{2}{c}{ATEN (ESS:  583.30)} & \\
			\cmidrule(lr){1-1}\cmidrule(lr){2-3}\cmidrule(lr){5-6}
			BOOT I &  2.03 & 0.108  & & 1.93 & 0.104 &\\ 
			BOOT II &  2.03 & 0.104 && 1.93 &  0.103 &\\ 
			WBExp I &  2.03 & 0.102 && 1.93 & 0.086 & &\multicolumn{2}{c}{*all p-values $<0.001$}\\  
			WBRad I &  2.03 & 0.096 && 1.93 & 0.094 &\\ 
            WBExp II &  2.03 & 0.102 && 1.93 & 0.092 &\\ 
            WBRad II  & 2.03 & 0.115 && 1.93 & 0.096 &\\ 
            SAND  & 2.03 & 0.102 && 1.93 & 0.096 &\\ 
			\bottomrule
		\end{tabular}
		\begin{tablenotes}
			\scriptsize
			\item ESS: effective sample size \citep{matsouaka2024overlap}; SE: standard error; BOOT I: standard bootstrap; BOOT II: post-weighting bootstrap; WBRad I (resp. Exp): wild bootstrap I via Rademacher (resp. exponential) distribution; WBRad II (resp. Exp): wild bootstrap II via Rademacher (resp. exponential) distribution; SAND: sandwich variance estimation.
		\end{tablenotes}
	\end{threeparttable}
\end{table}	


Table \ref{tab:dat-varest} provides the results of the difference in ($\log_2$) blood mercury level by fish consumption level. Among all the weighting methods, ATO has the largest effective sample size (ESS) of 590.28, while ATE has the smallest ESS of 403.40. The performance of ATO, ATM, and ATEN are very similar; overall, they are slightly more efficient (from the estimated SEs) than the other two estimands. The augmented estimates of ATO, ATM, and ATEN (1.98, 2.03, and 1.93, respectively) are overall closer to each other and are different from the ATE and ATT (1.74 and 2.12, respectively). The p-values are all smaller than 0.001 and agree with the conclusion that fish consumption statistically significantly impacts the ($\log_2$) blood mercury level. 

Regarding the variance estimation results, BOOT I and BOOT II generally induce {slightly }larger{, yet comparable,} SEs than other methods given the same estimand. 
For all estimands, BOOT I and BOOT II also have very close variance estimates, especially for ATE, aligning with the findings from our simulations. All the five estimands are significant based on multiple variance estimation methods, which further suggest that there is a treatment effect on different target populations (overall, treated, and equipoise), and thus we have evidence that the fish consumption is a strong factor in affecting the blood mercury level among different subpopulations. 

{Furthermore, we provide an explanation, supported by insights from our simulation study, for why the estimated variances are overall comparable across methods for each estimand. As shown by the point estimates for the five WATEs, all are close to 2.00, suggesting that the average causal effects across the populations defined by these estimands are similar. This implies that the treatment effect may be nearly homogeneous across individuals. In our simulation study, we similarly observed that when the treatment effect is homogeneous, all variance estimation methods tend to yield comparable estimates. This observation motivates our secondary analysis in Section \ref{subsec:second}, where we aim to identify a case in which the variance estimators yield more divergent results. }

{\subsection{Secondary subgroup analysis on a perturbed outcome}\label{subsec:second}

In this secondary analysis, we restrict our attention to a subgroup of older individuals (age $> 40$, $N = 591$), while maintaining the same covariates and treatment definition as in the primary analysis. Motivated by our earlier hypothesis that the treatment effect may be nearly homogeneous, we artificially perturb the outcome—$\log_2$ of blood mercury level—to introduce heterogeneity. Let $Y$ denote the original outcome and $Y_{\text{new}}$ the perturbed outcome, and we define
$$
Y_{\text{new}} = Y - 0.168\cdot Z \cdot(\text{age} + \text{gender}) + 8.56, 
$$
where the interaction term $-0.168 \cdot Z \cdot (\text{age} + \text{gender})$ introduces heterogeneity in the treatment effect across individuals by incorporating variation from age and gender. This design allows us to assess whether variance estimation methods yield more distinguishable results when treatment effects vary across the population. We totally acknowledge that this modification does not reflect the original data-generating process or underlying truth. However, our purpose is to further assess and validate the simulation finding that variance estimation methods differ in their sensitivity to treatment effect heterogeneity, enabling us to evaluate the advantages and potential limitations of our proposed approaches. For instance, applying different variance estimators may lead to different conclusions regarding the statistical significance of the treatment effect based on a prespecified  p-value threshold (e.g., 0.05), which is a critical practical concern as it may distort the interpretation of the treatment's impact. Below, we present and discuss the key findings from this analysis.

First, we refer readers to Appendix \ref{apx:data} for the estimated PS distributions and the covariate balance assessment (Figure \ref{fig:Data_EX2}), which yield results similar to those observed in the primary analysis in Figure \ref{fig:Data_EX}. In addition, Table \ref{tab:dat-varest2} below presents the results of the variance estimation methods applied in this secondary analysis.

\begin{table}[H]\small
	\begin{threeparttable}
		\centering
		\caption{Difference in $(\log_2)$ the perturbed blood mercury level by fish consumption among participants with age $> 40$. p-values that are $\leq 0.05$ are highlighted in bold. }
		\label{tab:dat-varest2}
        
		\begin{tabular}{llllllllllll}
			\toprule
			Method & Estimate & SE & p-value & Estimate & SE & p-value & Estimate & SE & p-value\\
            \midrule
            \addlinespace
            & \multicolumn{3}{c}{ATE (ESS: 251.27)}&\multicolumn{3}{c}{ATT (ESS: 262.77)}&\multicolumn{3}{c}{ATO (ESS: 359.35)} \\
			\cmidrule(lr){1-1}\cmidrule(lr){2-4}\cmidrule(lr){5-7}\cmidrule(lr){8-10}
            BOOT I & 0.21 & 0.145 & 0.15 & 0.54 & 0.228 & \textbf{0.02} & 0.28 & 0.163 & 0.09 \\ 
              BOOT II & 0.21 & 0.147 & 0.16 & 0.54 & 0.180 & \textbf{$<$0.01} & 0.28 & 0.144 & \textbf{0.05} \\ 
              WBExp I & 0.21 & 0.135 & 0.12 & 0.54 & 0.195 & \textbf{0.01} & 0.28 & 0.156 & 0.08 \\ 
              WBRad I & 0.21 & 0.134 & 0.12 & 0.54 & 0.200 & \textbf{0.01} & 0.28 & 0.143 & \textbf{0.05} \\ 
              WBExp II & 0.21 & 0.135 & 0.12 & 0.54 & 0.222 & \textbf{0.02} & 0.28 & 0.177 & 0.12 \\ 
              WBRad II & 0.21 & 0.134 & 0.12 & 0.54 & 0.253 & \textbf{0.03} & 0.28 & 0.182 & 0.13 \\ 
              SAND & 0.21 & 0.134 & 0.12 & 0.54 & 0.222 & \textbf{0.02} & 0.28 & 0.166 & 0.09 \\ 
			\addlinespace 
            &\multicolumn{3}{c}{ATM (ESS: 348.64)}&\multicolumn{3}{c}{ATEN (ESS: 354.53)} \\
			\cmidrule(lr){1-1}\cmidrule(lr){2-4}\cmidrule(lr){5-7}
			BOOT I & 0.32 & 0.174 & 0.07 & 0.25 & 0.155 & 0.10 \\ 
          BOOT II & 0.32 & 0.149 & \textbf{0.03} & 0.25 & 0.142 & 0.07 \\ 
          WBExp I & 0.32 & 0.160 & \textbf{0.05} & 0.25 & 0.153 & 0.10 \\ 
          WBRad I & 0.32 & 0.150 & \textbf{0.03} & 0.25 & 0.139 & 0.07 \\ 
          WBExp II & 0.32 & 0.197 & 0.11 & 0.25 & 0.160 & 0.11 \\ 
          WBRad II & 0.32 & 0.232 & 0.17 & 0.25 & 0.166 & 0.13 \\ 
          SAND & 0.32 & 0.179 & 0.07 & 0.25 & 0.155 & 0.10 \\ 
			\bottomrule
		\end{tabular}
		\begin{tablenotes}
			\scriptsize
			\item ESS: effective sample size; SE: standard error; BOOT I: standard bootstrap; BOOT II: post-weighting bootstrap; WBRad I (resp. Exp): wild bootstrap I via Rademacher (resp. exponential) distribution; WBRad II (resp. Exp): wild bootstrap II via Rademacher (resp. exponential) distribution; SAND: sandwich variance estimation.
		\end{tablenotes}
	\end{threeparttable}
\end{table}	
From Table \ref{tab:dat-varest2}, a key finding is that different variance estimators for ATO and ATM can lead to different conclusions regarding the significance of the treatment effect at the 0.05 level. We also observe that for all estimands except ATE, BOOT II, WBExp I, and WBRad I tend to produce smaller variance estimates than other methods, resulting in smaller corresponding p-values. This aligns with our simulation study, which showed that these methods yield similar variance estimates for ATE regardless of treatment effect heterogeneity. However, when the treatment effect is heterogeneous, BOOT II, WBExp I, and WBRad I may become anti-conservative.

In summary, based on our analysis of an artificially perturbed outcome with added heterogeneity, we find that certain variance estimation methods---particularly BOOT II, WBExp I, and WBRad I---may lead to anti-conservative inference for ATT, ATO, ATM and ATEN when treatment effects are heterogeneous. We therefore caution applied researchers against using these approaches for WATE estimands other than ATE, particularly when treatment effects are believed to be heterogeneous in the target population, as this may lead to biased conclusions regarding statistical significance and overconfidence in the estimated treatment effect.  
}

\section{Concluding Remarks}\label{sec:conclusion}

\subsection{Summary}\label{subsec:summ}

In this paper, we 
compared four variance estimation methods for a number of WATE estimands (using their augmented estimators): the standard bootstrap (BOOT I), the post-weighting bootstrap (BOOT II), the wild bootstrap (WB; using 2 different IFs), and the sandwich variance estimator (SAND). We assessed the performance of these variance estimation methods through extensive simulation studies and illustrated their application in a real-data analysis of the fish consumption effect on blood levels. 

Our findings indicate that the performance of a variance estimation method may have a different performance for the different WATEs; each method can also have its pros and cons dependent on the scenarios. For example, we found that BOOT II and WB methods are generally better for ATE estimators under different model specifications. However, for other estimands, BOOT I provides more accurate estimates. In addition, our study suggests that SAND has a number of practical limitations, including (but not be limited to)  small sample size or model-dependency, which often requires an acceptable model specification. It can also be problematic when there are violations of its regularity assumptions---while other methods under the same setting can perform well. 

In summary, based on the results we have obtained, we make the following suggestions. First, for ATE using the AIPW estimator, BOOT II and WB should be primarily considered in practice as it is model-free and more computing time-saving than the more conventional BOOT I. {This echos the finding of \cite{hahn1998role} that the role of PS in estimating ATE is ancillary, which means its knowledge does not affect achieving (asymptotically) the semiparametric efficiency bound using an augmented estimator. Thus, we can ignore its model re-fitting and predictions in bootstrap replicates,} as illustrated in Algorithm \ref{algo:bootII}. Second, for ATT and other estimands, some further investigations on its BOOT II and WB algorithms are needed in future research (see next Section \ref{subsec:disc}). Thus, BOOT I still has its non-negligible advantages. 

\subsection{Discussion}\label{subsec:disc}

Based on our extensive empirical studies, we provide alternative explanations on why BOOT II and WB methods fail for all the estimands but the ATE. {For BOOT II, the primary reason for these failures is directly connected to the definitions of the WATEs. Except for the ATE, the WATEs depend intrinsically on the true PS $e(\mb x)$; the true PS is included in their definitions through their tilting functions $g(\mb x)$, which are 
not trivial. Thus, the PS plays a key role in defining the target of inference. Despite the negative results for other WATEs, we believe our findings are still valuable and important. Our results highlight the distinct role of PS-induced variability across different estimands---a novel and practically important insight. For ATE, BOOT II offers a more efficient and robust alternative to BOOT I, avoiding possible random violations of the positivity assumption and reducing computational burden. For the other WATEs, we caution that PS-related variability remains a critical component to account for in variance estimation. }

{The above insight on the role of PS also results in two options for the IF to consider in the WB methods \citep{hirano2003efficient, crump2006moving} as discussed in Section \ref{subsec:IF-WB}}---treating the PS as known and unknown in the joint likelihood for the population model, respectively. Using the second IF derived by \cite{crump2006moving} often leads in a conservative variance estimate (i.e., using WB II) for the augmented estimator \eqref{eq:aug-wate}. Although this IF does not correspond to the form of the augmented estimator \eqref{eq:aug-wate} investigated in our paper, our purpose is to examine whether it can be a tool for uncertainty quantification for estimator \eqref{eq:aug-wate}. We found that for estimands other than ATE, the IF corresponding to the form of estimator \eqref{eq:aug-wate} (the first IF) provides anti-conservative variance estimates. Hence, we need to introduce additional components in the IF for perturbation in the WB algorithm. We did not investigate the estimator that corresponds to the second IF in this paper, as, to the best of our knowledge, it offers no clear advantage compared to the estimator \eqref{eq:aug-wate} for the following reasons. First, regarding  consistency, it has the same requirement for correct specifications on nuisance functions as estimator \eqref{eq:aug-wate}. Second, it is expected as less efficient than \eqref{eq:aug-wate}, as its variance estimation (based on investigation in using its IF in WB methods in our simulation) is larger than the expected variance for estimator \eqref{eq:aug-wate}. Third, it has a more complicated mathematical form, which is less favorable in practice for implementation and interpretation. 

In this paper, we have not studied the case where the variance is heterogeneous across participants (heteroscedasticity). In this case, the wild bootstrap is expected to fail  while the standard bootstrap does not \citep{wu1986jackknife, flachaire2005bootstrapping}. This implies a trade-off in choosing between higher robustness and efficiency in computing time. In future endeavors, we need to develop robust and efficient algorithms for variance estimation on WATEs to achieve the following objectives: (i) computing time-saving; (ii) robust to model misspecifications; (iii) flexible and perhaps model-free, even if these goals might not be able to attain simultaneously. As a result, our paper can serve as a motivating point for many future studies. Finally, 
{present work can potentially be extended to other types of outcome data, such as time-to-event outcomes \citep{cao2024using}, alternative causal estimands such as the weighted ATT \citep{liu2024average, liu2025assessing}, data fusion methods for multi-source studies \citep{liu2024multi, liu2025targeted, zhuang2025assessment, wang2025integrating}, randomized trials \citep{gao2024does, liu2025coadvise}, and integration with double machine learning approaches \citep{chernozhukov2018double} for estimating nuisance functions. }

\bibliography{_refs}

\newpage
\appendix
\newcounter{Appendix}[section]
\numberwithin{equation}{subsection}
\renewcommand\theequation{\Alph{section}.\arabic{subsection}.\arabic{equation}}
\numberwithin{table}{section}
\numberwithin{figure}{section}

\section{Appendix: Technical Details}\label{apx:details}

\subsection{Regularity conditions for sandwich variance estimation}\label{subapx:sandwich}

Denote the observed i.i.d. data $\mathcal{U} = \{\mb U_i=(Z_i,\mb X_i,Y_i),~i=1,\dots, N\}$. 
Consider the following estimating equation for $\bd\theta$ by an i.i.d. summation,
\begin{align*}
    \Psi_N(\bd\theta) = \frac1{\sqrt{N}}\sum_{i=1}^{N} \psi_{\bd\theta}(\mb U_i) = \bd 0,
\end{align*}
and $\widehat{\bd\theta} = \widehat{\bd\theta}(\mathcal{U})$ solves the above equation. 

Denote the true value of ${\bd\theta}$ by ${\bd\theta}^*$, and introduce an operator $\mathbb P$ such that $\mathbb P\{\psi_{\bd\theta}(\mb U)\} = \displaystyle\int\psi_{\bd\theta}(\mb u)dF(\mb u)$, where $F$ is a cdf. This notation is similar to but different from the common definition of expectation. For example, for $\widehat{\bd\theta} = \widehat{\bd\theta}(\mb U_1,\dots,\mb U_N)$, $\mathbb P\{\psi_{\widehat{\bd\theta}}(\mb U)\} = \displaystyle\int\psi_{\widehat{\bd\theta}}(\mb u)F(\mb u) \not= \Ex\{\psi_{\widehat{\bd\theta}}(\mb U)\}$, which does not involve $\widehat{\bd\theta}$ into randomness. 

To obtain the sandwich variance estimator of $\widehat{\bd\theta}$ from the estimating equation, the following ``regularity conditions'' are often made (or admitted by default) in literature, e.g., in \cite{van2000asymptotic}.
\begin{itemize}
    \item The parameter space $\Theta$ is an open subset of a Euclidian space $\mathbb R^d$, $d\geq 1$. 
    \item $\widehat{\bd\theta}$ is consistent to ${\bd\theta}^*$, i.e.,  $\widehat{\bd\theta}\to_p{\bd\theta}^*$. 
    \item The derivative $\dfrac{\partial\psi_{\bd\theta}(\mb u)}{\partial{\bd\theta}'}$ exists and $\mathbb P\left\{\dfrac{\partial\psi_{\bd\theta}(\mb U)}{\partial{\bd\theta}'}\right\}$ is continuous in ${\bd\theta}$.
    \item $\displaystyle\sup_{{\bd\theta}\in\Theta}\bigg|\dfrac1N\displaystyle\sum_{i=1}^N\dfrac{\partial\psi_{\bd\theta}(\mb U_i)}{\partial{\bd\theta}'} - \mathbb P\left\{\dfrac{\partial\psi_{\bd\theta}(\mb U)}{\partial{\bd\theta}'}\right\}\bigg|\to 0$ as $N\to\infty$. 
    \item The matrices $\mathbb P\left\{\dfrac{\partial\psi_{\bd\theta}(\mb U)}{\partial{\bd\theta}'}\right\}$ and $\dfrac1N\displaystyle\sum_{i=1}^{N}\frac{\partial\psi_{\widehat{\bd\theta}}(\mb U_i)}{\partial{{\bd\theta}'}}$ are nonsingular.
\end{itemize}
Then, a close-form consistent variance estimator for $\widehat{\bd\theta}$ can be given as follows. Denote $\mb A_N(\widehat{{\bd\theta}})=-\dfrac1N\displaystyle\sum_{i=1}^{N} \frac{\partial\psi_{\widehat{\bd\theta}}(\mb U_i)}{\partial{{\bd\theta}'}}$ and $\mb B_N(\widehat{{\bd\theta}})=\dfrac1N\displaystyle\sum_{i=1}^{N}\psi_{\widehat{\bd\theta}}(\mb U_i)\psi_{\bd\theta}(\mb U_i)'$, $\mb A({\bd\theta}) =\mathbb P\left\{\dfrac{\partial\psi_{\bd\theta}(\mb U)}{\partial{\bd\theta}'}\right\}$ and $\mb B({\bd\theta}) = \mathbb P\left\{\psi_{\bd\theta}(\mb U)\psi_{\bd\theta}(\mb U)'\right\}$. Then, under the above regularity conditions, 
\begin{align*}
    \widehat{\bd\Sigma}(\widehat{\bd\theta}) = \mb A_N(\widehat{\bd\theta})^{-1}\mb B_N(\widehat{\bd\theta})\{\mb A_N(\widehat{\bd\theta})^{-1}\}',
\end{align*}
is consistent to 
\begin{align*}
    {\bd\Sigma}({\bd\theta}^*) = \mb A({\bd\theta}^*)^{-1}\mb B({\bd\theta}^*)\{\mb A({\bd\theta}^*)^{-1}\}'.
\end{align*}
In addition, we have $\sqrt{N}(\widehat{\bd\theta}-{\bd\theta}^*)\to_d\mathcal N(\bd 0, {\bd\Sigma}({\bd\theta}^*))$. 

\subsection{Formula of the sandwich variance estimator}

We provide the estimating equation for $\widehat\tau_g^{\text{aug}}$ as follows. Denote $s_\gamma(\cdot)$ the score function of a parameter $\gamma$, then

 \begin{align*}
    \displaystyle\sum_{i=1}^{N} \psi_{{\bd\theta}_{\text{aug}}}(Z_i, \mb X_i, Y_i) & =
    \displaystyle\sum_{i=1}^{N} 
    \begin{bmatrix}
	s_{\bd\beta}(\mb X_i, Z_i)\\
	Z_is_{\bd\alpha_1}(\mb X_i, Y_i)\\
	(1-Z_i)s_{\bd\alpha_0}(\mb X_i, Y_i)\\
	g(\mb X_i)\{m_1(\mb X_i)-\tau_{1g}^m\}\\
	g(\mb X_i)\{m_0(\mb X_i)-\tau_{0g}^m\}\\
	Z_i\omega_1(\mb X_i)(Y_i-m_1(\mb X_i)-\mu_{1g}^m)\\
	(1-Z_i)\omega_0(\mb X_i)(Y_i-m_0(\mb X_i)-\mu_{0g}^m)\\
    \end{bmatrix} = 0
\end{align*}
with respect to ${\bd\theta}_{\text{aug}}=(\bd\beta',\bd\alpha_1', \bd\alpha_0', \tau_{1g}^{m}, \tau_{0g}^{m}, \mu_{1g}^m, \mu_{0g}^m)'$. For the estimator $\widehat\tau_{g}^{\text{aug}}$, we consider $\mb c=({\bf{0}},{\bf{0}},{\bf{0}},1,-1,1,-1)'$ such that $\widehat\tau_{g}^{\text{aug}}=\mb c'\widehat{\bd\theta}_{\text{aug}}=\widehat\tau_{1g}^m-\widehat\tau_{0g}^m+\widehat\mu_{1g}^m-\widehat\mu_{0g}^m$, where $\widehat{\bd\theta}_{g}^{\text{aug}}$ is the solution to this equation. Assuming that the same covariates appear as predictors in the regression models $m_z(\mb x)$, the non-zero components $\widehat A_{ij}$ of the matrix $ \mb A_N(\widehat{\bd\theta}_{\text{aug}})$ are given by 
\begin{align*}
    \widehat A_{11}&=\frac1N\sum_{i=1}^{N} \widehat e(\mb v_i)(1-\widehat e(\mb v_i))\mb v_i\mb v_i';~~
    \widehat A_{22}=\frac1N\sum_{i=1}^{N} z_i\mb w_i\mb w_i';~~ 
    \widehat A_{33}=\frac1N\sum_{i=1}^{N}(1-z_i)\mb w_i\mb w_i';\\ 
    \widehat A_{41}&=-N^{-1}\sum_{i=1}^{N} \frac{\partial g(\mb v_i)}{\partial\bd\beta'}\Big|_{\bd\beta=\widehat{\bd\beta}}\{\widehat m_1(\mb w_i)- \widehat \tau_{1g}^m\};~~
    \widehat A_{42}=\widehat A_{53}=-\frac1N\sum_{i=1}^{N} \widehat g(\mb v_i)\mb w_i'; \\
    \widehat A_{44}&= \widehat A_{55}=\frac1N\sum_{i=1}^{N} \widehat g(\mb v_i);~~
    \widehat A_{51}=-\frac1N\sum_{i=1}^{N} \frac{\partial g(\mb v_i)}{\partial\bd\beta'}\Big|_{\bd\beta=\widehat{\bd\beta}}\{\widehat m_0(\mb w_i)- \widehat \tau_{0g}^m\}; \\
    \widehat A_{61}&=-\frac1N\sum_{i=1}^{N} z_i\left[\frac{\partial g(\mb v_i)}{\partial\bd\beta'}\Big|_{\bd\beta=\widehat{\bd\beta}} - (1-\widehat e(\mb v_i))\widehat{g}(\mb v_i)\mb v_i'\right]\widehat e(\mb v_i)^{-1} \left( \widehat y_i-\widehat m_1(\mb w_i)-\widehat\mu_{1g}\right);\\
    \widehat A_{62}&=\frac1N\sum_{i=1}^{N} z_i\widehat \omega_1(\mb v_i)\mb w_i'; ~~
    \widehat A_{66}=\frac1N\sum_{i=1}^{N} z_i\widehat \omega_1(\mb v_i);\\
    \widehat A_{71}&=-\frac1N\sum_{i=1}^{N} (1-z_i)\left[\frac{\partial g(\mb v_i)}{\partial\bd\beta'}\Big|_{\bd\beta=\widehat{\bd\beta}} + \widehat e(\mb v_i)\widehat{g}(\mb v_i)\mb v_i'\right](1-\widehat e(\mb v_i))^{-1} ( \widehat y_i-\widehat m_0(\mb w_i)-\widehat\mu_{0g});\\
    \widehat A_{73}&=\frac1N\sum_{i=1}^{N} (1-z_i)\widehat \omega_0(\mb v_i)\mb w_i'; ~~ \widehat A_{77}=\frac1N\sum_{i=1}^{N}(1-z_i)\widehat \omega_0(\mb v_i).
\end{align*}  

An estimator of ${\bd\Sigma}({{\bd\theta}}_{\text{aug}})$ is then $ {\bd\Sigma}({\widehat{\bd\theta}}_{\text{aug}})=\mb A_N(\widehat{\bd\theta}_{\text{aug}})^{-1}\mb B_N(\widehat{\bd\theta}_{\text{aug}})\{\mb A_N(\widehat{\bd\theta}_{\text{aug}})'\}^{-1},$ from which we can derive the variance of $\widehat\tau_{g}^{\text{aug}}=\mb c'{\bd\theta}_{\text{aug}}$ as $\dfrac1N \mb c'{\bd\Sigma}({\widehat{{\bd\theta}}_{\text{aug}}})\mb c.$ Similarly as the case of H\'ajek-type estimator, we consider the same technique of the inverse of a $2\times 2$ block matrix, and now we write $\mb A_N(\widehat{\bd\theta}_{\text{aug}}) = \begin{bmatrix}\mb C_{11} & \mb C_{12} \\ \mb C_{21} & \mb C_{22}\end{bmatrix}$ with $\mb C_{11} = \left[\widehat A_{ij}\right]_{i,j\in\{1,2,3\}}$, the ``estimated information matrix'' of $(\bd\beta',\bd\alpha_0',\bd\alpha_1')'$ in presence of $\bd\eta = (\tau_{1g}^{m}, \tau_{0g}^{m}, \mu_{1g}^m, \mu_{0g}^m)'$, and $\mb C_{22} = \left[\widehat A_{ij}\right]_{i,j\in\{4,5,6,7\}}$, the ``estimated information matrix'' of $\bd\eta$ in presence of $\bd\beta$. Thus, similar to the case of H\'ajek-type estimator, we have 
\begin{align*}
    c'\mb A_N(\widehat{\bd\theta})^{-1} = (1,-1,1,-1)\begin{bmatrix}-\mb C_{22}^{-1}\mb C_{21}\mb C_{11}^{-1} & \mb C_{22}^{-1}\end{bmatrix}
\end{align*}
so we need the inverse of $\mb C_{11}=\left[\widehat A_{ij}\right]_{i,j\in\{1,2,3\}}$ and $\mb C_{22}=\left[\widehat A_{ij}\right]_{i,j\in\{4,5,6,7\}}$ and note that the latter is just a $4\times 4$ diagonal matrix. This technique simplifies the computation of $\mb c'\mb A_N(\widehat{\bd\theta})^{-1}$ for both H\'ajek-type and augmented estimators, compared to getting the inverse of $\mb A_N(\widehat{\bd\theta})$ first and then calculating the variance estimation. 

{
\subsection{Illustration of impact by model misspecifications}\label{subapx:miss}

Suppose that the estimated PS $\widehat{e}(\mb X)$ converges in probability to a general limit $\widetilde{e}(\mb X)$, and the estimated conditional outcome regression (OR) models $\widehat{m}_z(\mb X)$ converge in probability to general limits $\widetilde{m}_z(\mb X)$ for $z = 0, 1$. That is,
$$
\widehat{e}(\mb X) \xrightarrow{p} \widetilde{e}(\mb X), \quad \widehat{m}_z(\mb X) \xrightarrow{p} \widetilde{m}_z(\mb X).
$$
These limits $\widetilde{e}(\mb X)$ and $\widetilde{m}_z(\mb X)$ may not correspond to the true PS and outcome models, as the estimators may rely on misspecified models. Whenever the PS model is correctly specified, the augmented estimator \eqref{eq:aug-wate} in Section \ref{sec:setup} converges in probability to the true WATE $\tau_g$ in equation \eqref{eq:wate} \citep{zhou2020propensity}. However, if the PS model or all PS and the two outcome models are misspecified, the estimator is no longer consistent, even though it still converges to a limit $\widetilde{\tau}_g \neq \tau_g$.
Under such misspecifications, we can show that the augmented estimator \eqref{eq:aug-wate} converges in probability to 
$$
\widetilde{\tau}_g = \frac{\mathbb{E}[\widetilde{g}(\mb X)\{\widetilde{m}_1(\mb X) - \widetilde{m}_0(\mb X)\}]}{\mathbb{E}[\widetilde{g}(\mb X)]},
$$
where $\widetilde{g}(\mb X)$ is the tilting function obtained by plugging $\widetilde{e}(\mb X)$ into the definition of $g(\mb X)$. 

As a result, even if a variance estimator $\widehat{\Sigma}$ consistently estimates the sampling variability of the estimator, it corresponds to the variability around $\widetilde{\tau}_g$, not $\tau_g$. In this case, the usual central limit theorem yields
$$
\sqrt{n/\widehat{\Sigma}}\, (\widehat{\tau}_g - \widetilde{\tau}_g) \to_d \mathcal{N}(0, 1),
$$
instead of the more desirable
$$
\sqrt{n/\widehat{\Sigma}}\, (\widehat{\tau}_g - \tau_g) \to_d \mathcal{N}(0, 1),
$$
since the bias $\tau_g - \widetilde{\tau}_g$ is generally non-negligible.

Therefore, with some abuse of terminology, we note that the estimated variance may be consistent for the variance of a ``shifted estimand''---that is, the estimand $\widetilde\tau_g$ implicitly defined by the misspecified models---rather than the true WATE we aim to estimate. This illustration highlights that relying solely on RE, as defined in Section \ref{subsec:sim-measure}, is insufficient to assess the validity of a variance estimator. An RE close to 1 under model misspecification merely indicates that the variance is well estimated for the shifted estimand $\widetilde\tau_g$. Similarly, relying solely on CP is also not considerable enough, as the bias $\tau_g - \widetilde{\tau}_g$ plus a conservative variance estimate may also yield a good CP by chance. 
}

\section{Appendix: Additional Simulation Results}\label{apx:additional-sim}

\begin{figure}[H]
    \centering
    \includegraphics[width=\textwidth]{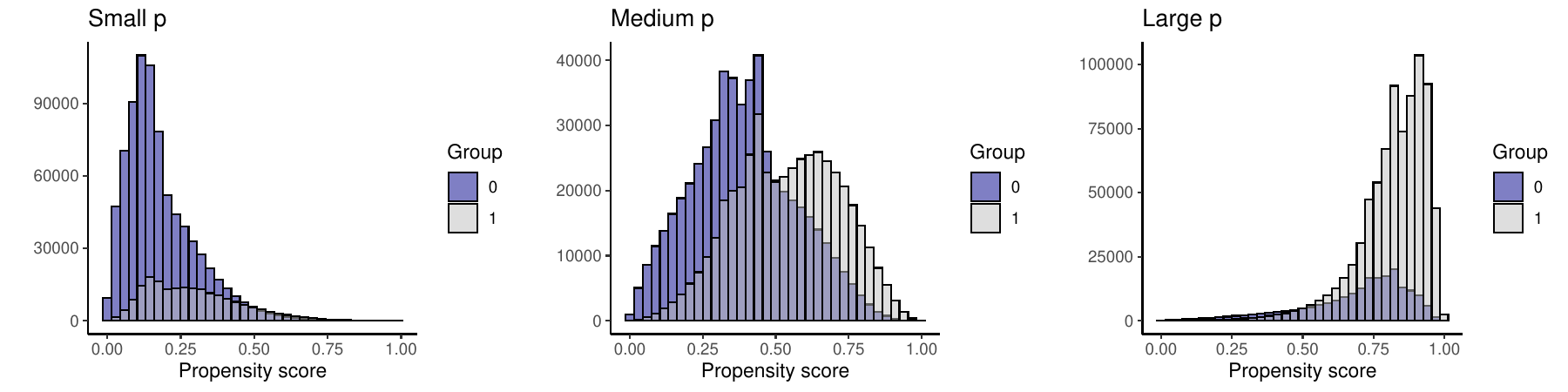}
    \caption{PS distributions under different proportions of treated participants $p$ (Models 1--3, corresponding to small $p=0.20$, medium $p=0.45$ and large $p=0.80$, respectively) using a random sample of $N=10^6$}
    \label{fig:sim-ps-dist1}
\end{figure}

\begin{figure}[H]
    \centering
\includegraphics[width=0.4\textwidth]{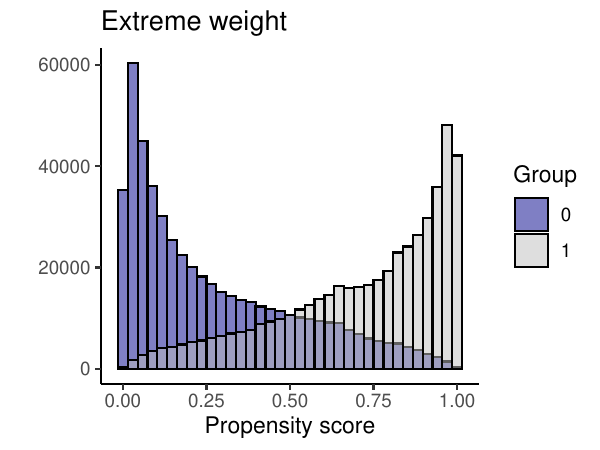}
    \caption{PS distributions under poor overlap and extreme weights (Model 4, $p=0.49$) using a random sample of $N=10^6$}
    \label{fig:sim-ps-dist2}
\end{figure}

\begin{table}[H]\footnotesize
        \caption{True WATE under heterogeneous treatment effects by different PS models in the simulation}
        \label{tab:truth}
        \centering
        \begin{tabular}{rrccccccc}
            \toprule
       Model & Case & ATE  &  ATT & ATO & ATM & ATEN \\ 
            \midrule
  1 & small $p$ (0.20) & 17.22  & 20.92 & 18.09  & 19.61 & 17.58 \\ 
 2 & medium $p$ (0.45) & 17.22  & 18.35 & 15.07 & 14.26 & 15.47 \\ 
 3 & large $p$ (0.80) & 17.22  & 16.85 & 15.42 & 15.84 & 15.58 \\ 
  4 & poor overlap and extreme weights exist (with a medium $p=0.49$) & 17.22 & 18.69 & 17.84  & 17.95 & 17.68 \\ 
  5 & sandwich variance does not work ($N=50$; with a large $p=0.80$)& 17.22  & 16.85 & 15.42  & 15.84 & 15.58 \\ 
            \bottomrule
        \end{tabular}
    \end{table}

\begin{table}[H]\footnotesize
    \centering
    \begin{tabular}{rrllllllllll}
    \toprule
    Model specifications & ATE & ATT & ATO & ATM & ATEN \\
    \midrule
    PS and OR models are correctly specified  & 626 & 626 & 626 & 626 & 626\\
    Only PS model is correctly specified & 322 & 322 & 322 & 322 & 322 \\
    Only OR model  is correctly specified & 626 & 622 & 622 & 622 & 622 \\
    PS and OR models are misspecified & 322 & 322 & 322 & 322 & 322 \\
 \bottomrule
    \end{tabular}
    \caption{Number of Monte Carlo replications when the sandwich variance is not obtainable from Model 5, under $N=50$ and $M=2000$ replicates}
    \label{tab:missingSAND}
\end{table}

\begin{table}[H]\tiny
	\begin{threeparttable}
		\centering
		\caption{Results of Model 1 (small $p=0.20$), $N=1000$, under heterogeneous treatment effect}
		\label{tab:smallp-n1000-hets-results}
		\begin{tabular}{rrcccccccccccc}
			\toprule
   Estimand & Method & ARBias\% & RMSE & CP\%  & SE & ESD & RE &ARBias\%& RMSE  & CP\% & SE & ESD & RE \\
			\cmidrule(lr){3-8}\cmidrule(lr){9-14}
			\addlinespace 
			& &\multicolumn{6}{c}{Augmented: PS and OR correctly specified} & \multicolumn{6}{c}{Augmented: Only PS correctly specified} \\\cmidrule(lr){3-8}\cmidrule(lr){9-14}
			\cmidrule(lr){1-2}\cmidrule(lr){3-8}\cmidrule(lr){9-14}
			\multirow{7}{*}{ATE} 
 & BOOT I & 0.14 & 0.61 & 94.45 & 0.603 & 0.610 & 1.022 & 5.72 & 4.08 & 77.65 & 2.154 & 3.962 & 3.382 \\ 
   & BOOT II & 0.14 & 0.61 & 94.55 & 0.603 & 0.610 & 1.022 & 5.72 & 4.08 & 88.75 & 2.875 & 3.962 & 1.899 \\ 
   & WBExp I & 0.14 & 0.61 & 93.60 & 0.585 & 0.610 & 1.087 & 5.72 & 4.08 & 86.40 & 2.680 & 3.962 & 2.185 \\ 
   & WBRad I & 0.14 & 0.61 & 93.80 & 0.595 & 0.610 & 1.052 & 5.72 & 4.08 & 89.15 & 3.216 & 3.962 & 1.517 \\ 
   & WBExp II & 0.14 & 0.61 & 93.60 & 0.585 & 0.610 & 1.087 & 5.72 & 4.08 & 86.40 & 2.680 & 3.962 & 2.185 \\ 
   & WBRad II & 0.14 & 0.61 & 93.80 & 0.595 & 0.610 & 1.052 & 5.72 & 4.08 & 89.15 & 3.216 & 3.962 & 1.517 \\ 
   & SAND & 0.14 & 0.61 & 94.45 & 0.599 & 0.610 & 1.039 & 5.72 & 4.08 & 72.25 & 1.762 & 3.962 & 5.057 \\ 
    \addlinespace
   			\multirow{7}{*}{ATT} 
& BOOT I & 0.17 & 1.52 & 94.45 & 1.498 & 1.517 & 1.026 & 0.21 & 1.58 & 94.45 & 1.564 & 1.575 & 1.014 \\ 
   & BOOT II & 0.17 & 1.52 & 80.20 & 0.978 & 1.517 & 2.407 & 0.21 & 1.58 & 95.25 & 1.621 & 1.575 & 0.944 \\ 
   & WBExp I & 0.17 & 1.52 & 78.45 & 0.947 & 1.517 & 2.565 & 0.21 & 1.58 & 94.50 & 1.557 & 1.575 & 1.023 \\ 
   & WBRad I & 0.17 & 1.52 & 79.80 & 0.981 & 1.517 & 2.393 & 0.21 & 1.58 & 94.95 & 1.613 & 1.575 & 0.953 \\ 
   & WBExp II & 0.17 & 1.52 & 93.40 & 1.450 & 1.517 & 1.095 & 0.21 & 1.58 & 97.90 & 1.920 & 1.575 & 0.673 \\ 
   & WBRad II & 0.17 & 1.52 & 94.05 & 1.488 & 1.517 & 1.039 & 0.21 & 1.58 & 98.25 & 1.974 & 1.575 & 0.637 \\ 
   & SAND & 0.17 & 1.52 & 94.10 & 1.491 & 1.517 & 1.035 & 0.21 & 1.58 & 94.15 & 1.542 & 1.575 & 1.043 \\ 
       \addlinespace
   			\multirow{7}{*}{ATO} 
& BOOT I & 0.30 & 0.88 & 94.65 & 0.880 & 0.883 & 1.007 & 0.55 & 0.95 & 94.55 & 0.951 & 0.948 & 0.994 \\ 
    & BOOT II & 0.30 & 0.88 & 87.75 & 0.709 & 0.883 & 1.550 & 0.55 & 0.95 & 99.45 & 1.489 & 0.948 & 0.405 \\ 
   & WBExp I & 0.30 & 0.88 & 86.80 & 0.692 & 0.883 & 1.626 & 0.55 & 0.95 & 99.20 & 1.415 & 0.948 & 0.449 \\ 
   & WBRad I & 0.30 & 0.88 & 87.55 & 0.701 & 0.883 & 1.585 & 0.55 & 0.95 & 99.35 & 1.466 & 0.948 & 0.418 \\ 
   & WBExp II & 0.30 & 0.88 & 96.30 & 0.983 & 0.883 & 0.807 & 0.55 & 0.95 & 98.70 & 1.295 & 0.948 & 0.536 \\ 
   & WBRad II & 0.30 & 0.88 & 96.95 & 0.999 & 0.883 & 0.780 & 0.55 & 0.95 & 99.10 & 1.343 & 0.948 & 0.499 \\ 
   & SAND & 0.30 & 0.88 & 94.65 & 0.868 & 0.883 & 1.034 & 0.55 & 0.95 & 93.60 & 0.930 & 0.948 & 1.039 \\ 
\addlinespace   
			\multirow{7}{*}{ATM}
& BOOT I & 0.66 & 1.09 & 94.25 & 1.079 & 1.083 & 1.007 & 0.77 & 1.13 & 94.25 & 1.129 & 1.121 & 0.985 \\ 
    & BOOT II & 0.66 & 1.09 & 85.05 & 0.803 & 1.083 & 1.818 & 0.77 & 1.13 & 97.70 & 1.447 & 1.121 & 0.600 \\ 
   & WBExp I & 0.66 & 1.09 & 83.80 & 0.783 & 1.083 & 1.916 & 0.77 & 1.13 & 97.10 & 1.381 & 1.121 & 0.659 \\ 
   & WBRad I & 0.66 & 1.09 & 84.35 & 0.797 & 1.083 & 1.847 & 0.77 & 1.13 & 97.70 & 1.420 & 1.121 & 0.623 \\ 
   & WBExp II & 0.66 & 1.09 & 97.50 & 1.355 & 1.083 & 0.639 & 0.77 & 1.13 & 98.50 & 1.558 & 1.121 & 0.517 \\ 
   & WBRad II & 0.66 & 1.09 & 97.80 & 1.399 & 1.083 & 0.600 & 0.77 & 1.13 & 98.90 & 1.609 & 1.121 & 0.485 \\ 
   & SAND & 0.66 & 1.09 & 93.85 & 1.069 & 1.083 & 1.027 & 0.77 & 1.13 & 94.30 & 1.110 & 1.121 & 1.020 \\ 
			\addlinespace 
			\multirow{7}{*}{ATEN} 
   & BOOT I & 0.28 & 0.79 & 94.70 & 0.785 & 0.789 & 1.009 & 0.87 & 0.96 & 93.50 & 0.921 & 0.945 & 1.053 \\ 
   & BOOT II & 0.28 & 0.79 & 89.65 & 0.667 & 0.789 & 1.398 & 0.87 & 0.96 & 99.80 & 1.640 & 0.945 & 0.332 \\ 
   & WBExp I & 0.28 & 0.79 & 88.50 & 0.650 & 0.789 & 1.472 & 0.87 & 0.96 & 99.65 & 1.555 & 0.945 & 0.369 \\ 
   & WBRad I & 0.28 & 0.79 & 88.60 & 0.660 & 0.789 & 1.431 & 0.87 & 0.96 & 99.75 & 1.638 & 0.945 & 0.333 \\ 
   & WBExp II & 0.28 & 0.79 & 95.70 & 0.854 & 0.789 & 0.854 & 0.87 & 0.96 & 98.65 & 1.283 & 0.945 & 0.543 \\ 
   & WBRad II & 0.28 & 0.79 & 96.45 & 0.877 & 0.789 & 0.809 & 0.87 & 0.96 & 98.95 & 1.340 & 0.945 & 0.497 \\ 
   & SAND & 0.28 & 0.79 & 94.55 & 0.775 & 0.789 & 1.037 & 0.87 & 0.96 & 93.20 & 0.896 & 0.945 & 1.113 \\ 
			\addlinespace 
			& &\multicolumn{6}{c}{Augmented: only OR correctly specified} & \multicolumn{6}{c}{Augmented: PS and OR misspecified}\\ 
			\cmidrule(lr){1-2}\cmidrule(lr){3-8}\cmidrule(lr){9-14}
	\multirow{7}{*}{ATE} 
& BOOT I & 0.14 & 0.61 & 94.40 & 0.602 & 0.610 & 1.023 & 3.31 & 4.40 & 83.70 & 2.914 & 4.364 & 2.243 \\ 
   & BOOT II & 0.14 & 0.61 & 94.40 & 0.602 & 0.610 & 1.024 & 3.31 & 4.40 & 84.55 & 2.949 & 4.364 & 2.190 \\ 
   & WBExp I & 0.14 & 0.61 & 93.60 & 0.588 & 0.610 & 1.075 & 3.31 & 4.40 & 84.50 & 3.072 & 4.364 & 2.018 \\ 
   & WBRad I & 0.14 & 0.61 & 93.95 & 0.594 & 0.610 & 1.052 & 3.31 & 4.40 & 87.80 & 3.713 & 4.364 & 1.382 \\ 
   & WBExp II & 0.14 & 0.61 & 93.60 & 0.588 & 0.610 & 1.075 & 3.31 & 4.40 & 84.50 & 3.072 & 4.364 & 2.018 \\ 
   & WBRad II & 0.14 & 0.61 & 93.95 & 0.594 & 0.610 & 1.052 & 3.31 & 4.40 & 87.80 & 3.713 & 4.364 & 1.382 \\ 
   & SAND & 0.14 & 0.61 & 94.45 & 0.599 & 0.610 & 1.036 & 3.31 & 4.40 & 82.45 & 2.771 & 4.364 & 2.481 \\ 
			\addlinespace 
   \multirow{7}{*}{ATT} 
& BOOT I & 3.86 & 1.50 & 92.95 & 1.264 & 1.263 & 0.998 & 3.02 & 1.92 & 91.05 & 1.796 & 1.816 & 1.022 \\ 
   & BOOT II & 3.86 & 1.50 & 85.90 & 1.053 & 1.263 & 1.438 & 3.02 & 1.92 & 88.35 & 1.633 & 1.816 & 1.237 \\ 
   & WBExp I & 3.86 & 1.50 & 84.05 & 1.014 & 1.263 & 1.551 & 3.02 & 1.92 & 86.60 & 1.578 & 1.816 & 1.324 \\ 
   & WBRad I & 3.86 & 1.50 & 86.05 & 1.049 & 1.263 & 1.449 & 3.02 & 1.92 & 88.50 & 1.637 & 1.816 & 1.231 \\ 
   & WBExp II & 3.86 & 1.50 & 95.60 & 1.448 & 1.263 & 0.760 & 3.02 & 1.92 & 93.20 & 1.938 & 1.816 & 0.878 \\ 
   & WBRad II & 3.86 & 1.50 & 96.30 & 1.495 & 1.263 & 0.714 & 3.02 & 1.92 & 94.70 & 1.994 & 1.816 & 0.830 \\ 
   & SAND & 3.86 & 1.50 & 93.35 & 1.258 & 1.263 & 1.009 & 3.02 & 1.92 & 91.25 & 1.776 & 1.816 & 1.046 \\ 
			\addlinespace  
   \multirow{7}{*}{ATO} 
& BOOT I & 3.84 & 1.07 & 88.50 & 0.807 & 0.809 & 1.006 & 3.91 & 1.67 & 89.30 & 1.473 & 1.517 & 1.060 \\ 
 & BOOT II & 3.84 & 1.07 & 84.75 & 0.750 & 0.809 & 1.164 & 3.91 & 1.67 & 89.30 & 1.483 & 1.517 & 1.047 \\ 
   & WBExp I & 3.84 & 1.07 & 83.85 & 0.733 & 0.809 & 1.220 & 3.91 & 1.67 & 87.85 & 1.425 & 1.517 & 1.132 \\ 
   & WBRad I & 3.84 & 1.07 & 84.00 & 0.743 & 0.809 & 1.186 & 3.91 & 1.67 & 89.90 & 1.467 & 1.517 & 1.068 \\ 
   & WBExp II & 3.84 & 1.07 & 94.35 & 0.961 & 0.809 & 0.709 & 3.91 & 1.67 & 81.95 & 1.189 & 1.517 & 1.628 \\ 
   & WBRad II & 3.84 & 1.07 & 95.25 & 0.988 & 0.809 & 0.671 & 3.91 & 1.67 & 83.15 & 1.219 & 1.517 & 1.548 \\ 
   & SAND & 3.84 & 1.07 & 88.45 & 0.803 & 0.809 & 1.016 & 3.91 & 1.67 & 89.70 & 1.465 & 1.517 & 1.071 \\ 
	\addlinespace 
	\multirow{7}{*}{ATM} 
& BOOT I & 5.81 & 1.51 & 82.85 & 1.003 & 0.998 & 0.991 & 1.22 & 1.58 & 93.20 & 1.534 & 1.564 & 1.038 \\ 
    & BOOT II & 5.81 & 1.51 & 75.20 & 0.882 & 0.998 & 1.280 & 1.22 & 1.58 & 92.15 & 1.465 & 1.564 & 1.140 \\ 
   & WBExp I & 5.81 & 1.51 & 73.30 & 0.860 & 0.998 & 1.347 & 1.22 & 1.58 & 90.90 & 1.403 & 1.564 & 1.242 \\ 
   & WBRad I & 5.81 & 1.51 & 74.15 & 0.876 & 0.998 & 1.299 & 1.22 & 1.58 & 91.80 & 1.445 & 1.564 & 1.171 \\ 
   & WBExp II & 5.81 & 1.51 & 96.60 & 1.421 & 0.998 & 0.493 & 1.22 & 1.58 & 92.40 & 1.491 & 1.564 & 1.099 \\ 
   & WBRad II & 5.81 & 1.51 & 97.05 & 1.474 & 0.998 & 0.458 & 1.22 & 1.58 & 92.85 & 1.531 & 1.564 & 1.043 \\ 
   & SAND & 5.81 & 1.51 & 82.55 & 1.001 & 0.998 & 0.995 & 1.22 & 1.58 & 93.20 & 1.530 & 1.564 & 1.045 \\ 
		\addlinespace 
		\multirow{7}{*}{ATEN} 
& BOOT I & 3.21 & 0.92 & 89.75 & 0.726 & 0.729 & 1.008 & 4.51 & 1.86 & 88.35 & 1.577 & 1.683 & 1.138 \\  
    & BOOT II & 3.21 & 0.92 & 87.75 & 0.698 & 0.729 & 1.090 & 4.51 & 1.86 & 89.60 & 1.648 & 1.683 & 1.043 \\ 
   & WBExp I & 3.21 & 0.92 & 86.35 & 0.679 & 0.729 & 1.152 & 4.51 & 1.86 & 88.30 & 1.587 & 1.683 & 1.125 \\ 
   & WBRad I & 3.21 & 0.92 & 87.40 & 0.690 & 0.729 & 1.115 & 4.51 & 1.86 & 89.95 & 1.678 & 1.683 & 1.007 \\ 
   & WBExp II & 3.21 & 0.92 & 95.10 & 0.857 & 0.729 & 0.722 & 4.51 & 1.86 & 79.35 & 1.225 & 1.683 & 1.887 \\ 
   & WBRad II & 3.21 & 0.92 & 95.85 & 0.883 & 0.729 & 0.680 & 4.51 & 1.86 & 80.45 & 1.274 & 1.683 & 1.746 \\ 
   & SAND & 3.21 & 0.92 & 88.95 & 0.721 & 0.729 & 1.022 & 4.51 & 1.86 & 88.00 & 1.566 & 1.683 & 1.156 \\
	\bottomrule
    \end{tabular}
    \begin{tablenotes}
        \scriptsize
        \item ARBias\%: absolute percent relative bias; RMSE: root mean square error; CP\%: coverage probability (\%); SE: median of standard errors by proposed method; ESD: empirical standard deviation; RE: median relative efficiency; BOOT I: standard bootstrap; BOOT II: post-weighting bootstrap;  WBExp (resp. ExpRad): wild bootstrap via exponential (resp. Rademacher) distribution; SAND: sandwich variance estimator;  PS: propensity score; OR: outcome regression.
    \end{tablenotes}
\end{threeparttable}
\end{table}

\begin{table}[H]\tiny
	\begin{threeparttable}
		\centering
		\caption{Results of Model 1 (small $p=0.20$), $N=1000$, under homogeneous treatment effect}
		\label{tab:smallp-n1000-cons-results}
		\begin{tabular}{rrcccccccccccc}
			\toprule
   Estimand & Method & ARBias\%& RMSE & CP\%  & SE & ESD & RE &ARBias\%& RMSE  & CP\% & SE & ESD & RE \\
			\cmidrule(lr){3-8}\cmidrule(lr){9-14}
			\addlinespace 
			& &\multicolumn{6}{c}{Augmented: PS and OR correctly specified} & \multicolumn{6}{c}{Augmented: Only PS correctly specified} \\\cmidrule(lr){3-8}\cmidrule(lr){9-14}
			\cmidrule(lr){1-2}\cmidrule(lr){3-8}\cmidrule(lr){9-14}
			\multirow{7}{*}{ATE} 
 & BOOT I & 0.01 & 0.10 & 94.55 & 0.101 & 0.105 & 1.062 & 5.11 & 1.13 &  79.05 & 0.576 & 1.114 & 3.746 \\ 
   & BOOT II & 0.01 & 0.10 & 94.55 & 0.101 & 0.105 & 1.073 & 5.11 & 1.13 &  92.10 & 0.804 & 1.114 & 1.919 \\ 
   & WBExp I & 0.01 & 0.10 & 91.45 & 0.089 & 0.105 & 1.364 & 5.11 & 1.13 &  89.00 & 0.764 & 1.114 & 2.128 \\ 
   & WBRad I & 0.01 & 0.10 & 92.75 & 0.093 & 0.105 & 1.271 & 5.11 & 1.13 &  91.40 & 0.906 & 1.114 & 1.510 \\ 
   & WBExp II & 0.01 & 0.10 & 91.45 & 0.089 & 0.105 & 1.364 & 5.11 & 1.13 &  89.00 & 0.764 & 1.114 & 2.128 \\ 
   & WBRad II & 0.01 & 0.10 & 92.75 & 0.093 & 0.105 & 1.271 & 5.11 & 1.13 &  91.40 & 0.906 & 1.114 & 1.510 \\ 
   & SAND & 0.01 & 0.10 & 92.70 & 0.095 & 0.105 & 1.220 & 5.11 & 1.13 &  74.65 & 0.475 & 1.114 & 5.507 \\ 
    \addlinespace
   			\multirow{7}{*}{ATT} 
& BOOT I & 0.00 & 0.09 & 95.25 & 0.088 & 0.087 & 0.990 & 0.54 & 0.33 &  96.85 & 0.263 & 0.334 & 1.607 \\  
   & BOOT II & 0.00 & 0.09 & 94.90 & 0.087 & 0.087 & 1.009 & 0.54 & 0.33 & 100.00 & 0.539 & 0.334 & 0.384 \\
   & WBExp I & 0.00 & 0.09 & 93.75 & 0.084 & 0.087 & 1.079 & 0.54 & 0.33 &  99.90 & 0.523 & 0.334 & 0.407 \\ 
   & WBRad I & 0.00 & 0.09 & 94.75 & 0.085 & 0.087 & 1.049 & 0.54 & 0.33 & 100.00 & 0.555 & 0.334 & 0.362 \\ 
   & WBExp II & 0.00 & 0.09 & 94.25 & 0.085 & 0.087 & 1.061 & 0.54 & 0.33 & 100.00 & 0.642 & 0.334 & 0.270 \\ 
   & WBRad II & 0.00 & 0.09 & 94.85 & 0.086 & 0.087 & 1.028 & 0.54 & 0.33 & 100.00 & 0.673 & 0.334 & 0.246 \\ 
   & SAND & 0.00 & 0.09 & 94.90 & 0.086 & 0.087 & 1.023 & 0.54 & 0.33 &  95.95 & 0.241 & 0.334 & 1.918 \\ 
       \addlinespace
   			\multirow{7}{*}{ATO} 
 & BOOT I & 0.02 & 0.08 & 94.90 & 0.084 & 0.083 & 0.982 & 0.09 & 0.09 &  96.00 & 0.094 & 0.091 & 0.935 \\ 
   & BOOT II & 0.02 & 0.08 & 95.20 & 0.084 & 0.083 & 0.977 & 0.09 & 0.09 & 100.00 & 0.442 & 0.091 & 0.043 \\ 
   & WBExp I & 0.02 & 0.08 & 94.00 & 0.081 & 0.083 & 1.048 & 0.09 & 0.09 & 100.00 & 0.423 & 0.091 & 0.047 \\ 
   & WBRad I & 0.02 & 0.08 & 94.30 & 0.082 & 0.083 & 1.024 & 0.09 & 0.09 & 100.00 & 0.436 & 0.091 & 0.044 \\ 
   & WBExp II & 0.02 & 0.08 & 94.20 & 0.081 & 0.083 & 1.037 & 0.09 & 0.09 & 100.00 & 0.392 & 0.091 & 0.054 \\ 
   & WBRad II & 0.02 & 0.08 & 94.80 & 0.083 & 0.083 & 1.003 & 0.09 & 0.09 & 100.00 & 0.401 & 0.091 & 0.052 \\ 
   & SAND & 0.02 & 0.08 & 95.15 & 0.083 & 0.083 & 0.997 & 0.09 & 0.09 &  95.75 & 0.092 & 0.091 & 0.978 \\ 
			\addlinespace 
			\multirow{7}{*}{ATM}
& BOOT I & 0.02 & 0.08 & 95.20 & 0.085 & 0.085 & 0.980 & 0.22 & 0.12 &  95.75 & 0.128 & 0.123 & 0.921 \\ 
    & BOOT II & 0.02 & 0.08 & 95.15 & 0.085 & 0.085 & 0.989 & 0.22 & 0.12 & 100.00 & 0.447 & 0.123 & 0.076 \\
   & WBExp I & 0.02 & 0.08 & 94.25 & 0.083 & 0.085 & 1.050 & 0.22 & 0.12 & 100.00 & 0.432 & 0.123 & 0.081 \\ 
   & WBRad I & 0.02 & 0.08 & 94.30 & 0.083 & 0.085 & 1.031 & 0.22 & 0.12 & 100.00 & 0.446 & 0.123 & 0.076 \\ 
   & WBExp II & 0.02 & 0.08 & 94.75 & 0.083 & 0.085 & 1.030 & 0.22 & 0.12 & 100.00 & 0.458 & 0.123 & 0.072 \\ 
   & WBRad II & 0.02 & 0.08 & 94.70 & 0.084 & 0.085 & 1.005 & 0.22 & 0.12 & 100.00 & 0.467 & 0.123 & 0.069 \\ 
   & SAND & 0.02 & 0.08 & 95.05 & 0.085 & 0.085 & 0.996 & 0.22 & 0.12 &  96.35 & 0.132 & 0.123 & 0.868 \\ 
			\addlinespace 
			\multirow{7}{*}{ATEN} 
   & BOOT I & 0.01 & 0.08 & 95.15 & 0.084 & 0.084 & 0.990 & 0.21 & 0.12 &  94.15 & 0.118 & 0.124 & 1.090 \\ 
   & BOOT II & 0.01 & 0.08 & 95.30 & 0.085 & 0.084 & 0.977 & 0.21 & 0.12 & 100.00 & 0.476 & 0.124 & 0.067 \\
   & WBExp I & 0.01 & 0.08 & 93.85 & 0.081 & 0.084 & 1.061 & 0.21 & 0.12 & 100.00 & 0.450 & 0.124 & 0.075 \\ 
   & WBRad I & 0.01 & 0.08 & 94.40 & 0.083 & 0.084 & 1.024 & 0.21 & 0.12 & 100.00 & 0.473 & 0.124 & 0.068 \\ 
   & WBExp II & 0.01 & 0.08 & 94.00 & 0.082 & 0.084 & 1.043 & 0.21 & 0.12 & 100.00 & 0.390 & 0.124 & 0.101 \\ 
   & WBRad II & 0.01 & 0.08 & 94.50 & 0.083 & 0.084 & 1.007 & 0.21 & 0.12 & 100.00 & 0.402 & 0.124 & 0.095 \\ 
   & SAND & 0.01 & 0.08 & 94.95 & 0.083 & 0.084 & 1.010 & 0.21 & 0.12 &  94.20 & 0.115 & 0.124 & 1.155 \\  
			\addlinespace 
			& &\multicolumn{6}{c}{Augmented: only OR correctly specified} & \multicolumn{6}{c}{Augmented: PS and OR misspecified}\\ 
			\cmidrule(lr){1-2}\cmidrule(lr){3-8}\cmidrule(lr){9-14}
	\multirow{7}{*}{ATE} 
& BOOT I & 0.01 & 0.10 & 95.30 & 0.100 & 0.100 & 1.002 &  4.53 & 1.23 & 84.70 & 0.837 & 1.221 & 2.124 \\ 
   & BOOT II & 0.01 & 0.10 & 95.35 & 0.100 & 0.100 & 1.008 &  4.53 & 1.23 & 84.55 & 0.834 & 1.221 & 2.142 \\ 
   & WBExp I & 0.01 & 0.10 & 92.30 & 0.091 & 0.100 & 1.225 &  4.53 & 1.23 & 84.85 & 0.875 & 1.221 & 1.945 \\ 
   & WBRad I & 0.01 & 0.10 & 93.05 & 0.095 & 0.100 & 1.128 &  4.53 & 1.23 & 87.90 & 1.035 & 1.221 & 1.392 \\ 
   & WBExp II & 0.01 & 0.10 & 92.30 & 0.091 & 0.100 & 1.225 &  4.53 & 1.23 & 84.85 & 0.875 & 1.221 & 1.945 \\ 
   & WBRad II & 0.01 & 0.10 & 93.05 & 0.095 & 0.100 & 1.128 &  4.53 & 1.23 & 87.90 & 1.035 & 1.221 & 1.392 \\ 
   & SAND & 0.01 & 0.10 & 94.25 & 0.095 & 0.100 & 1.117 &  4.53 & 1.23 & 83.20 & 0.806 & 1.221 & 2.292 \\ 
			\addlinespace 
   \multirow{7}{*}{ATT} 
& BOOT I & 0.00 & 0.09 & 95.30 & 0.088 & 0.087 & 0.984 & 11.79 & 0.76 & 87.20 & 0.560 & 0.599 & 1.142 \\ 
   & BOOT II & 0.00 & 0.09 & 95.20 & 0.088 & 0.087 & 0.990 & 11.79 & 0.76 & 88.10 & 0.568 & 0.599 & 1.111 \\
   & WBExp I & 0.00 & 0.09 & 93.00 & 0.082 & 0.087 & 1.144 & 11.79 & 0.76 & 87.55 & 0.556 & 0.599 & 1.160 \\ 
   & WBRad I & 0.00 & 0.09 & 92.90 & 0.082 & 0.087 & 1.120 & 11.79 & 0.76 & 90.55 & 0.589 & 0.599 & 1.033 \\ 
   & WBExp II & 0.00 & 0.09 & 93.35 & 0.083 & 0.087 & 1.116 & 11.79 & 0.76 & 94.65 & 0.690 & 0.599 & 0.752 \\ 
   & WBRad II & 0.00 & 0.09 & 92.90 & 0.084 & 0.087 & 1.089 & 11.79 & 0.76 & 96.35 & 0.721 & 0.599 & 0.690 \\ 
   & SAND & 0.00 & 0.09 & 94.65 & 0.086 & 0.087 & 1.021 & 11.79 & 0.76 & 86.80 & 0.555 & 0.599 & 1.165 \\ 
			\addlinespace  
   \multirow{7}{*}{ATO} 
& BOOT I & 0.00 & 0.09 & 95.00 & 0.086 & 0.086 & 0.982 & 10.39 & 0.63 & 82.15 & 0.454 & 0.471 & 1.075 \\ 
    & BOOT II & 0.00 & 0.09 & 95.20 & 0.087 & 0.086 & 0.978 & 10.39 & 0.63 & 82.30 & 0.457 & 0.471 & 1.061 \\ 
   & WBExp I & 0.00 & 0.09 & 92.65 & 0.079 & 0.086 & 1.163 & 10.39 & 0.63 & 80.75 & 0.441 & 0.471 & 1.141 \\ 
   & WBRad I & 0.00 & 0.09 & 93.10 & 0.081 & 0.086 & 1.127 & 10.39 & 0.63 & 83.00 & 0.459 & 0.471 & 1.052 \\ 
   & WBExp II & 0.00 & 0.09 & 93.45 & 0.080 & 0.086 & 1.137 & 10.39 & 0.63 & 74.35 & 0.386 & 0.471 & 1.488 \\ 
   & WBRad II & 0.00 & 0.09 & 93.65 & 0.081 & 0.086 & 1.106 & 10.39 & 0.63 & 76.10 & 0.397 & 0.471 & 1.407 \\ 
   & SAND & 0.00 & 0.09 & 94.65 & 0.085 & 0.086 & 1.020 & 10.39 & 0.63 & 81.95 & 0.453 & 0.471 & 1.080 \\ 
			\addlinespace 
            \multirow{7}{*}{ATM} 
	 & BOOT I & 0.01 & 0.09 & 95.40 & 0.087 & 0.086 & 0.980 & 10.76 & 0.65 & 82.55 & 0.468 & 0.481 & 1.058 \\ 
   & BOOT II & 0.01 & 0.09 & 95.50 & 0.087 & 0.086 & 0.986 & 10.76 & 0.65 & 84.00 & 0.478 & 0.481 & 1.015 \\ 
   & WBExp I & 0.01 & 0.09 & 93.05 & 0.081 & 0.086 & 1.138 & 10.76 & 0.65 & 83.25 & 0.467 & 0.481 & 1.063 \\ 
   & WBRad I & 0.01 & 0.09 & 93.25 & 0.082 & 0.086 & 1.107 & 10.76 & 0.65 & 85.25 & 0.483 & 0.481 & 0.994 \\ 
   & WBExp II & 0.01 & 0.09 & 93.50 & 0.082 & 0.086 & 1.111 & 10.76 & 0.65 & 84.05 & 0.471 & 0.481 & 1.046 \\ 
   & WBRad II & 0.01 & 0.09 & 93.65 & 0.083 & 0.086 & 1.078 & 10.76 & 0.65 & 84.95 & 0.485 & 0.481 & 0.986 \\ 
   & SAND & 0.01 & 0.09 & 94.95 & 0.086 & 0.086 & 1.012 & 10.76 & 0.65 & 82.85 & 0.469 & 0.481 & 1.054 \\ 
			\addlinespace 
			\multirow{7}{*}{ATEN} 
& BOOT I & 0.01 & 0.09 & 95.10 & 0.087 & 0.087 & 0.985 &  9.98 & 0.65 & 83.20 & 0.485 & 0.518 & 1.141 \\ 
   & BOOT II & 0.01 & 0.09 & 95.25 & 0.088 & 0.087 & 0.975 &  9.98 & 0.65 & 83.35 & 0.489 & 0.518 & 1.120 \\
   & WBExp I & 0.01 & 0.09 & 92.70 & 0.080 & 0.087 & 1.171 &  9.98 & 0.65 & 81.85 & 0.474 & 0.518 & 1.194 \\ 
   & WBRad I & 0.01 & 0.09 & 93.25 & 0.081 & 0.087 & 1.138 &  9.98 & 0.65 & 84.55 & 0.503 & 0.518 & 1.060 \\ 
   & WBExp II & 0.01 & 0.09 & 92.90 & 0.081 & 0.087 & 1.152 &  9.98 & 0.65 & 72.90 & 0.391 & 0.518 & 1.752 \\ 
   & WBRad II & 0.01 & 0.09 & 93.40 & 0.082 & 0.087 & 1.120 &  9.98 & 0.65 & 75.50 & 0.409 & 0.518 & 1.607 \\ 
   & SAND & 0.01 & 0.09 & 94.90 & 0.085 & 0.087 & 1.031 &  9.98 & 0.65 & 82.55 & 0.484 & 0.518 & 1.144 \\ 
			\bottomrule
		\end{tabular}
		\begin{tablenotes}
			\scriptsize
\item ARBias\%: absolute percent relative bias; RMSE: root mean square error; CP\%: coverage probability (\%); SE: median of standard errors by proposed method; ESD: empirical standard deviation; RE: median relative efficiency; BOOT I: standard bootstrap; BOOT II: post-weighting bootstrap;  WBExp (resp. ExpRad): wild bootstrap via exponential (resp. Rademacher) distribution; SAND: sandwich variance estimator;  PS: propensity score; OR: outcome regression.
		\end{tablenotes}
	\end{threeparttable}
\end{table}

\begin{table}[H]\tiny
	\begin{threeparttable}
		\centering
		\caption{Results of Model 2 (medium $p=0.45$), $N=1000$, under heterogeneous treatment effect}
		\label{tab:mediump-n1000-hets-results}
		\begin{tabular}{rrcccccccccccc}
			\toprule
			Estimand & Method & ARBias\%& RMSE & CP\%  & SE & ESD & RE & ARBias\%& RMSE  & CP\% & SE & ESD & RE \\
			\cmidrule(lr){3-8}\cmidrule(lr){9-14}
			\addlinespace 
			& &\multicolumn{6}{c}{Augmented: PS and OR correctly specified} & \multicolumn{6}{c}{Augmented: Only PS correctly specified} \\\cmidrule(lr){3-8}\cmidrule(lr){9-14}
			\cmidrule(lr){1-2}\cmidrule(lr){3-8}\cmidrule(lr){9-14}
			\multirow{7}{*}{ATE} 
  & BOOT I & 0.12 & 0.60 & 94.55 & 0.599 & 0.604 & 1.017 & 1.70 & 2.38 & 83.90 & 1.160 & 2.359 & 4.135 \\ 
   & BOOT II & 0.12 & 0.60 & 94.50 & 0.599 & 0.604 & 1.017 & 1.70 & 2.38 & 95.65 & 1.769 & 2.359 & 1.780 \\
   & WBExp I & 0.12 & 0.60 & 94.20 & 0.585 & 0.604 & 1.068 & 1.70 & 2.38 & 95.10 & 1.727 & 2.359 & 1.866 \\ 
   & WBRad I & 0.12 & 0.60 & 94.25 & 0.591 & 0.604 & 1.044 & 1.70 & 2.38 & 96.00 & 1.925 & 2.359 & 1.502 \\ 
   & WBExp II & 0.12 & 0.60 & 94.20 & 0.585 & 0.604 & 1.068 & 1.70 & 2.38 & 95.10 & 1.727 & 2.359 & 1.866 \\ 
   & WBRad II & 0.12 & 0.60 & 94.25 & 0.591 & 0.604 & 1.044 & 1.70 & 2.38 & 96.00 & 1.925 & 2.359 & 1.502 \\ 
   & SAND & 0.12 & 0.60 & 94.30 & 0.594 & 0.604 & 1.033 & 1.70 & 2.38 & 81.25 & 1.049 & 2.359 & 5.063 \\
   \addlinespace
   			\multirow{7}{*}{ATT} 
  & BOOT I & 0.13 & 0.93 & 94.25 & 0.914 & 0.932 & 1.038 & 0.14 & 1.05 & 94.95 & 1.014 & 1.045 & 1.063 \\ 
    & BOOT II & 0.13 & 0.93 & 87.60 & 0.745 & 0.932 & 1.565 & 0.14 & 1.05 & 97.50 & 1.189 & 1.045 & 0.773 \\ 
   & WBExp I & 0.13 & 0.93 & 86.70 & 0.722 & 0.932 & 1.665 & 0.14 & 1.05 & 96.85 & 1.164 & 1.045 & 0.806 \\ 
   & WBRad I & 0.13 & 0.93 & 87.50 & 0.738 & 0.932 & 1.593 & 0.14 & 1.05 & 97.25 & 1.188 & 1.045 & 0.774 \\ 
   & WBExp II & 0.13 & 0.93 & 93.35 & 0.892 & 0.932 & 1.092 & 0.14 & 1.05 & 97.90 & 1.277 & 1.045 & 0.670 \\ 
   & WBRad II & 0.13 & 0.93 & 93.80 & 0.904 & 0.932 & 1.061 & 0.14 & 1.05 & 98.05 & 1.306 & 1.045 & 0.640 \\ 
   & SAND & 0.13 & 0.93 & 94.20 & 0.912 & 0.932 & 1.044 & 0.14 & 1.05 & 94.60 & 0.994 & 1.045 & 1.106 \\ 
       \addlinespace
   			\multirow{7}{*}{ATO} 
     & BOOT I & 0.33 & 0.59 & 94.20 & 0.584 & 0.586 & 1.004 & 0.57 & 0.64 & 93.70 & 0.636 & 0.634 & 0.996 \\ 
    & BOOT II & 0.33 & 0.59 & 89.90 & 0.505 & 0.586 & 1.342 & 0.57 & 0.64 & 99.95 & 1.128 & 0.634 & 0.316 \\
   & WBExp I & 0.33 & 0.59 & 89.10 & 0.493 & 0.586 & 1.409 & 0.57 & 0.64 & 99.95 & 1.094 & 0.634 & 0.337 \\ 
   & WBRad I & 0.33 & 0.59 & 88.80 & 0.498 & 0.586 & 1.384 & 0.57 & 0.64 & 99.90 & 1.132 & 0.634 & 0.314 \\ 
   & WBExp II & 0.33 & 0.59 & 97.00 & 0.687 & 0.586 & 0.726 & 0.57 & 0.64 & 99.15 & 0.899 & 0.634 & 0.498 \\ 
   & WBRad II & 0.33 & 0.59 & 97.65 & 0.710 & 0.586 & 0.681 & 0.57 & 0.64 & 99.50 & 0.931 & 0.634 & 0.465 \\ 
   & SAND & 0.33 & 0.59 & 93.90 & 0.579 & 0.586 & 1.023 & 0.57 & 0.64 & 93.65 & 0.630 & 0.634 & 1.015 \\ 
\addlinespace 
			\multirow{7}{*}{ATM}
  & BOOT I & 0.07 & 0.60 & 94.65 & 0.602 & 0.600 & 0.995 & 0.18 & 0.64 & 94.35 & 0.646 & 0.638 & 0.976 \\ 
   & BOOT II & 0.07 & 0.60 & 88.10 & 0.492 & 0.600 & 1.488 & 0.18 & 0.64 & 99.90 & 1.005 & 0.638 & 0.403 \\ 
   & WBExp I & 0.07 & 0.60 & 87.95 & 0.479 & 0.600 & 1.572 & 0.18 & 0.64 & 99.80 & 0.978 & 0.638 & 0.426 \\ 
   & WBRad I & 0.07 & 0.60 & 88.05 & 0.484 & 0.600 & 1.542 & 0.18 & 0.64 & 99.80 & 0.998 & 0.638 & 0.409 \\ 
   & WBExp II & 0.07 & 0.60 & 99.15 & 0.827 & 0.600 & 0.527 & 0.18 & 0.64 & 99.65 & 0.953 & 0.638 & 0.448 \\ 
   & WBRad II & 0.07 & 0.60 & 99.20 & 0.852 & 0.600 & 0.497 & 0.18 & 0.64 & 99.90 & 0.976 & 0.638 & 0.427 \\ 
   & SAND & 0.07 & 0.60 & 94.70 & 0.595 & 0.600 & 1.019 & 0.18 & 0.64 & 94.50 & 0.642 & 0.638 & 0.989 \\ 
			\addlinespace 
			\multirow{7}{*}{ATEN}
    & BOOT I & 0.36 & 0.58 & 94.60 & 0.574 & 0.575 & 1.004 & 0.73 & 0.69 & 93.05 & 0.663 & 0.677 & 1.043 \\  
   & BOOT II & 0.36 & 0.58 & 90.80 & 0.516 & 0.575 & 1.242 & 0.73 & 0.69 & 99.95 & 1.228 & 0.677 & 0.304 \\
   & WBExp I & 0.36 & 0.58 & 90.00 & 0.505 & 0.575 & 1.296 & 0.73 & 0.69 & 99.85 & 1.191 & 0.677 & 0.323 \\ 
   & WBRad I & 0.36 & 0.58 & 90.50 & 0.509 & 0.575 & 1.276 & 0.73 & 0.69 & 99.85 & 1.244 & 0.677 & 0.296 \\ 
   & WBExp II & 0.36 & 0.58 & 96.60 & 0.647 & 0.575 & 0.788 & 0.73 & 0.69 & 99.10 & 0.976 & 0.677 & 0.480 \\ 
   & WBRad II & 0.36 & 0.58 & 97.25 & 0.665 & 0.575 & 0.748 & 0.73 & 0.69 & 99.30 & 1.017 & 0.677 & 0.443 \\ 
   & SAND & 0.36 & 0.58 & 94.15 & 0.568 & 0.575 & 1.023 & 0.73 & 0.69 & 92.80 & 0.651 & 0.677 & 1.080 \\ 
			\addlinespace 
			& &\multicolumn{6}{c}{Augmented: only OR correctly specified} & \multicolumn{6}{c}{Augmented: PS and OR misspecified}\\ 
			\cmidrule(lr){1-2}\cmidrule(lr){3-8}\cmidrule(lr){9-14}
	\multirow{7}{*}{ATE} 
 & BOOT I & 0.12 & 0.60 & 94.35 & 0.599 & 0.604 & 1.017 & 8.86 & 2.44 & 75.40 & 1.551 & 1.899 & 1.499 \\ 
      & BOOT II & 0.12 & 0.60 & 94.40 & 0.599 & 0.604 & 1.016 & 8.86 & 2.44 & 76.65 & 1.612 & 1.899 & 1.387 \\ 
   & WBExp I & 0.12 & 0.60 & 94.10 & 0.585 & 0.604 & 1.065 & 8.86 & 2.44 & 75.15 & 1.591 & 1.899 & 1.425 \\ 
   & WBRad I & 0.12 & 0.60 & 94.10 & 0.591 & 0.604 & 1.042 & 8.86 & 2.44 & 78.95 & 1.732 & 1.899 & 1.201 \\ 
   & WBExp II & 0.12 & 0.60 & 94.10 & 0.585 & 0.604 & 1.065 & 8.86 & 2.44 & 75.15 & 1.591 & 1.899 & 1.425 \\ 
   & WBRad II & 0.12 & 0.60 & 94.10 & 0.591 & 0.604 & 1.042 & 8.86 & 2.44 & 78.95 & 1.732 & 1.899 & 1.201 \\ 
   & SAND & 0.12 & 0.60 & 94.50 & 0.594 & 0.604 & 1.033 & 8.86 & 2.44 & 74.85 & 1.524 & 1.899 & 1.552 \\ 
			\addlinespace 
   \multirow{7}{*}{ATT} 
   & BOOT I & 3.58 & 1.05 & 89.25 & 0.813 & 0.824 & 1.027 & 4.60 & 1.53 & 87.45 & 1.247 & 1.273 & 1.043 \\ 
    & BOOT II & 3.58 & 1.05 & 88.00 & 0.788 & 0.824 & 1.093 & 4.60 & 1.53 & 87.10 & 1.224 & 1.273 & 1.082 \\ 
   & WBExp I & 3.58 & 1.05 & 86.80 & 0.767 & 0.824 & 1.155 & 4.60 & 1.53 & 86.75 & 1.218 & 1.273 & 1.093 \\ 
   & WBRad I & 3.58 & 1.05 & 87.50 & 0.782 & 0.824 & 1.112 & 4.60 & 1.53 & 87.50 & 1.240 & 1.273 & 1.054 \\ 
   & WBExp II & 3.58 & 1.05 & 92.35 & 0.891 & 0.824 & 0.856 & 4.60 & 1.53 & 90.15 & 1.330 & 1.273 & 0.917 \\ 
   & WBRad II & 3.58 & 1.05 & 92.95 & 0.906 & 0.824 & 0.828 & 4.60 & 1.53 & 91.10 & 1.367 & 1.273 & 0.867 \\ 
   & SAND & 3.58 & 1.05 & 89.05 & 0.806 & 0.824 & 1.045 & 4.60 & 1.53 & 87.55 & 1.240 & 1.273 & 1.055 \\ 
	\addlinespace 
   \multirow{7}{*}{ATO} 
   & BOOT I & 1.45 & 0.66 & 93.25 & 0.617 & 0.625 & 1.026 & 8.19 & 1.66 & 77.45 & 1.090 & 1.115 & 1.046 \\  
   & BOOT II & 1.45 & 0.66 & 84.85 & 0.476 & 0.625 & 1.723 & 8.19 & 1.66 & 77.45 & 1.101 & 1.115 & 1.027 \\
   & WBExp I & 1.45 & 0.66 & 84.30 & 0.465 & 0.625 & 1.803 & 8.19 & 1.66 & 76.35 & 1.073 & 1.115 & 1.081 \\ 
   & WBRad I & 1.45 & 0.66 & 84.10 & 0.467 & 0.625 & 1.789 & 8.19 & 1.66 & 78.40 & 1.112 & 1.115 & 1.005 \\ 
   & WBExp II & 1.45 & 0.66 & 98.00 & 0.769 & 0.625 & 0.659 & 8.19 & 1.66 & 62.65 & 0.844 & 1.115 & 1.746 \\ 
   & WBRad II & 1.45 & 0.66 & 98.40 & 0.805 & 0.625 & 0.602 & 8.19 & 1.66 & 64.25 & 0.867 & 1.115 & 1.654 \\ 
   & SAND & 1.45 & 0.66 & 93.30 & 0.612 & 0.625 & 1.043 & 8.19 & 1.66 & 77.25 & 1.087 & 1.115 & 1.052 \\ 
			\addlinespace 
			\multirow{7}{*}{ATM} 
   & BOOT I & 0.31 & 0.68 & 94.05 & 0.662 & 0.675 & 1.040 & 9.07 & 1.63 & 72.95 & 0.990 & 0.988 & 0.995 \\  
   & BOOT II & 0.31 & 0.68 & 79.75 & 0.434 & 0.675 & 2.416 & 9.07 & 1.63 & 69.60 & 0.938 & 0.988 & 1.110 \\ 
   & WBExp I & 0.31 & 0.68 & 78.35 & 0.423 & 0.675 & 2.540 & 9.07 & 1.63 & 68.00 & 0.909 & 0.988 & 1.182 \\ 
   & WBRad I & 0.31 & 0.68 & 79.30 & 0.426 & 0.675 & 2.512 & 9.07 & 1.63 & 69.05 & 0.932 & 0.988 & 1.123 \\ 
   & WBExp II & 0.31 & 0.68 & 98.70 & 0.853 & 0.675 & 0.625 & 9.07 & 1.63 & 61.25 & 0.819 & 0.988 & 1.456 \\ 
   & WBRad II & 0.31 & 0.68 & 98.65 & 0.881 & 0.675 & 0.587 & 9.07 & 1.63 & 62.25 & 0.830 & 0.988 & 1.417 \\ 
   & SAND & 0.31 & 0.68 & 94.00 & 0.660 & 0.675 & 1.046 & 9.07 & 1.63 & 72.80 & 0.983 & 0.988 & 1.010 \\ 
			\addlinespace 
			\multirow{7}{*}{ATEN} 
    & BOOT I & 1.54 & 0.65 & 93.60 & 0.594 & 0.600 & 1.020 & 8.00 & 1.73 & 78.50 & 1.160 & 1.216 & 1.098 \\ 
    & BOOT II & 1.54 & 0.65 & 88.35 & 0.497 & 0.600 & 1.457 & 8.00 & 1.73 & 79.65 & 1.198 & 1.216 & 1.030 \\
   & WBExp I & 1.54 & 0.65 & 87.15 & 0.486 & 0.600 & 1.525 & 8.00 & 1.73 & 78.10 & 1.166 & 1.216 & 1.088 \\ 
   & WBRad I & 1.54 & 0.65 & 87.60 & 0.489 & 0.600 & 1.505 & 8.00 & 1.73 & 80.25 & 1.226 & 1.216 & 0.984 \\ 
   & WBExp II & 1.54 & 0.65 & 97.20 & 0.709 & 0.600 & 0.717 & 8.00 & 1.73 & 67.15 & 0.946 & 1.216 & 1.651 \\ 
   & WBRad II & 1.54 & 0.65 & 97.10 & 0.740 & 0.600 & 0.659 & 8.00 & 1.73 & 69.05 & 0.985 & 1.216 & 1.524 \\ 
   & SAND & 1.54 & 0.65 & 93.50 & 0.589 & 0.600 & 1.038 & 8.00 & 1.73 & 78.20 & 1.154 & 1.216 & 1.111 \\ 
			\bottomrule
		\end{tabular}
		\begin{tablenotes}
			\scriptsize
			\item ARBias\%: absolute percent relative bias; RMSE: root mean square error; CP\%: coverage probability (\%); SE: median of standard errors by proposed method; ESD: empirical standard deviation; RE: median relative efficiency; BOOT I: standard bootstrap; BOOT II: post-weighting bootstrap;  WBExp (resp. ExpRad): wild bootstrap via exponential (resp. Rademacher) distribution; SAND: sandwich variance estimator;  PS: propensity score; OR: outcome regression.
		\end{tablenotes}
	\end{threeparttable}
\end{table}	

\begin{table}[H]\tiny
	\begin{threeparttable}
		\centering
		\caption{Results of Model 2 (medium $p=0.45$), $N=1000$, under homogeneous treatment effect}
		\label{tab:mediump-n1000-cons-results}
		\begin{tabular}{rrcccccccccccc}
			\toprule
			Estimand & Method & ARBias\%& RMSE & CP\%  & SE & ESD & RE &ARBias\%& RMSE  & CP\% & SE & ESD & RE \\
			\cmidrule(lr){3-8}\cmidrule(lr){9-14}
			\addlinespace 
			& &\multicolumn{6}{c}{Augmented: PS and OR correctly specified} & \multicolumn{6}{c}{Augmented: Only PS correctly specified} \\\cmidrule(lr){3-8}\cmidrule(lr){9-14}
			\cmidrule(lr){1-2}\cmidrule(lr){3-8}\cmidrule(lr){9-14}
			\multirow{7}{*}{ATE} 
   & BOOT I & 0.03 & 0.07 & 94.55 & 0.072 & 0.073 & 1.031 & 1.06 & 0.66 &  90.25 & 0.326 & 0.659 & 4.074 \\ 
    & BOOT II & 0.03 & 0.07 & 94.40 & 0.072 & 0.073 & 1.036 & 1.06 & 0.66 &  99.80 & 0.583 & 0.659 & 1.277 \\ 
   & WBExp I & 0.03 & 0.07 & 93.65 & 0.070 & 0.073 & 1.111 & 1.06 & 0.66 &  99.60 & 0.568 & 0.659 & 1.344 \\ 
   & WBRad I & 0.03 & 0.07 & 94.65 & 0.071 & 0.073 & 1.076 & 1.06 & 0.66 &  99.70 & 0.616 & 0.659 & 1.143 \\ 
   & WBExp II & 0.03 & 0.07 & 93.65 & 0.070 & 0.073 & 1.111 & 1.06 & 0.66 &  99.60 & 0.568 & 0.659 & 1.344 \\ 
   & WBRad II & 0.03 & 0.07 & 94.65 & 0.071 & 0.073 & 1.076 & 1.06 & 0.66 &  99.70 & 0.616 & 0.659 & 1.143 \\ 
   & SAND & 0.03 & 0.07 & 94.30 & 0.071 & 0.073 & 1.069 & 1.06 & 0.66 &  87.00 & 0.290 & 0.659 & 5.178 \\ 
    \addlinespace
   			\multirow{7}{*}{ATT} 
& BOOT I & 0.06 & 0.08 & 94.55 & 0.077 & 0.077 & 1.006 & 1.32 & 0.45 &  92.85 & 0.320 & 0.442 & 1.908 \\ 
   & BOOT II & 0.06 & 0.08 & 94.50 & 0.076 & 0.077 & 1.025 & 1.32 & 0.45 &  99.45 & 0.538 & 0.442 & 0.675 \\ 
   & WBExp I & 0.06 & 0.08 & 94.10 & 0.074 & 0.077 & 1.097 & 1.32 & 0.45 &  99.15 & 0.529 & 0.442 & 0.698 \\ 
   & WBRad I & 0.06 & 0.08 & 94.25 & 0.076 & 0.077 & 1.047 & 1.32 & 0.45 &  99.25 & 0.566 & 0.442 & 0.610 \\ 
   & WBExp II & 0.06 & 0.08 & 94.25 & 0.074 & 0.077 & 1.094 & 1.32 & 0.45 &  99.50 & 0.584 & 0.442 & 0.572 \\ 
   & WBRad II & 0.06 & 0.08 & 94.15 & 0.076 & 0.077 & 1.044 & 1.32 & 0.45 &  99.45 & 0.623 & 0.442 & 0.504 \\ 
   & SAND & 0.06 & 0.08 & 94.25 & 0.075 & 0.077 & 1.052 & 1.32 & 0.45 &  91.60 & 0.296 & 0.442 & 2.237 \\ 
       \addlinespace
   			\multirow{7}{*}{ATO} 
& BOOT I & 0.03 & 0.07 & 95.05 & 0.069 & 0.068 & 0.977 & 0.00 & 0.08 &  95.35 & 0.080 & 0.079 & 0.968 \\ 
     & BOOT II & 0.03 & 0.07 & 95.10 & 0.069 & 0.068 & 0.975 & 0.00 & 0.08 & 100.00 & 0.381 & 0.079 & 0.043 \\ 
   & WBExp I & 0.03 & 0.07 & 94.55 & 0.067 & 0.068 & 1.026 & 0.00 & 0.08 & 100.00 & 0.369 & 0.079 & 0.046 \\ 
   & WBRad I & 0.03 & 0.07 & 94.95 & 0.068 & 0.068 & 0.999 & 0.00 & 0.08 & 100.00 & 0.378 & 0.079 & 0.044 \\ 
   & WBExp II & 0.03 & 0.07 & 94.35 & 0.068 & 0.068 & 1.018 & 0.00 & 0.08 & 100.00 & 0.295 & 0.079 & 0.072 \\ 
   & WBRad II & 0.03 & 0.07 & 95.05 & 0.069 & 0.068 & 0.988 & 0.00 & 0.08 & 100.00 & 0.303 & 0.079 & 0.068 \\ 
   & SAND & 0.03 & 0.07 & 95.15 & 0.069 & 0.068 & 0.986 & 0.00 & 0.08 &  95.15 & 0.079 & 0.079 & 0.999 \\ 
\addlinespace 
			\multirow{7}{*}{ATM}
& BOOT I & 0.03 & 0.07 & 95.70 & 0.070 & 0.069 & 0.971 & 0.04 & 0.08 &  96.00 & 0.088 & 0.084 & 0.908 \\ 
    & BOOT II & 0.03 & 0.07 & 95.70 & 0.070 & 0.069 & 0.977 & 0.04 & 0.08 & 100.00 & 0.341 & 0.084 & 0.061 \\ 
   & WBExp I & 0.03 & 0.07 & 94.35 & 0.068 & 0.069 & 1.030 & 0.04 & 0.08 & 100.00 & 0.333 & 0.084 & 0.064 \\ 
   & WBRad I & 0.03 & 0.07 & 95.15 & 0.069 & 0.069 & 0.995 & 0.04 & 0.08 & 100.00 & 0.339 & 0.084 & 0.062 \\ 
   & WBExp II & 0.03 & 0.07 & 94.50 & 0.068 & 0.069 & 1.017 & 0.04 & 0.08 & 100.00 & 0.294 & 0.084 & 0.082 \\ 
   & WBRad II & 0.03 & 0.07 & 95.30 & 0.070 & 0.069 & 0.982 & 0.04 & 0.08 & 100.00 & 0.301 & 0.084 & 0.078 \\ 
   & SAND & 0.03 & 0.07 & 95.50 & 0.070 & 0.069 & 0.979 & 0.04 & 0.08 &  96.20 & 0.089 & 0.084 & 0.885 \\ 
			\addlinespace 
			\multirow{7}{*}{ATEN} 
    & BOOT I & 0.03 & 0.07 & 94.90 & 0.069 & 0.068 & 0.984 & 0.07 & 0.11 &  95.20 & 0.101 & 0.105 & 1.088 \\ 
    & BOOT II & 0.03 & 0.07 & 95.00 & 0.069 & 0.068 & 0.978 & 0.07 & 0.11 & 100.00 & 0.411 & 0.105 & 0.066 \\ 
   & WBExp I & 0.03 & 0.07 & 94.35 & 0.067 & 0.068 & 1.036 & 0.07 & 0.11 & 100.00 & 0.400 & 0.105 & 0.069 \\ 
   & WBRad I & 0.03 & 0.07 & 94.95 & 0.068 & 0.068 & 1.003 & 0.07 & 0.11 & 100.00 & 0.410 & 0.105 & 0.066 \\ 
   & WBExp II & 0.03 & 0.07 & 94.40 & 0.067 & 0.068 & 1.029 & 0.07 & 0.11 & 100.00 & 0.326 & 0.105 & 0.104 \\ 
   & WBRad II & 0.03 & 0.07 & 95.05 & 0.069 & 0.068 & 0.994 & 0.07 & 0.11 & 100.00 & 0.333 & 0.105 & 0.100 \\ 
   & SAND & 0.03 & 0.07 & 95.15 & 0.069 & 0.068 & 0.993 & 0.07 & 0.11 &  94.55 & 0.099 & 0.105 & 1.133 \\ 
   \addlinespace 
   & &\multicolumn{6}{c}{Augmented: only OR correctly specified} & \multicolumn{6}{c}{Augmented: PS and OR misspecified}\\ 
   \cmidrule(lr){1-2}\cmidrule(lr){3-8}\cmidrule(lr){9-14}
	\multirow{7}{*}{ATE} 
& BOOT I & 0.03 & 0.07 & 94.95 & 0.071 & 0.071 & 0.990 & 14.88 & 0.89 & 78.85 & 0.582 & 0.661 & 1.292 \\  
    & BOOT II & 0.03 & 0.07 & 94.85 & 0.071 & 0.071 & 0.990 & 14.88 & 0.89 & 78.50 & 0.575 & 0.661 & 1.321 \\
   & WBExp I & 0.03 & 0.07 & 94.00 & 0.068 & 0.071 & 1.098 & 14.88 & 0.89 & 78.10 & 0.571 & 0.661 & 1.340 \\ 
   & WBRad I & 0.03 & 0.07 & 93.95 & 0.069 & 0.071 & 1.068 & 14.88 & 0.89 & 81.30 & 0.616 & 0.661 & 1.154 \\ 
   & WBExp II & 0.03 & 0.07 & 94.00 & 0.068 & 0.071 & 1.098 & 14.88 & 0.89 & 78.10 & 0.571 & 0.661 & 1.340 \\ 
   & WBRad II & 0.03 & 0.07 & 93.95 & 0.069 & 0.071 & 1.068 & 14.88 & 0.89 & 81.30 & 0.616 & 0.661 & 1.154 \\ 
   & SAND & 0.03 & 0.07 & 94.85 & 0.070 & 0.071 & 1.012 & 14.88 & 0.89 & 78.55 & 0.570 & 0.661 & 1.348 \\
  \addlinespace 
   \multirow{7}{*}{ATT} 
   & BOOT I & 0.05 & 0.07 & 95.50 & 0.074 & 0.073 & 0.960 & 17.29 & 0.96 & 84.20 & 0.594 & 0.672 & 1.279 \\  
    & BOOT II & 0.05 & 0.07 & 95.35 & 0.074 & 0.073 & 0.968 & 17.29 & 0.96 & 85.95 & 0.614 & 0.672 & 1.198 \\ 
   & WBExp I & 0.05 & 0.07 & 94.10 & 0.071 & 0.073 & 1.060 & 17.29 & 0.96 & 87.50 & 0.625 & 0.672 & 1.156 \\ 
   & WBRad I & 0.05 & 0.07 & 94.55 & 0.072 & 0.073 & 1.023 & 17.29 & 0.96 & 91.75 & 0.680 & 0.672 & 0.978 \\ 
   & WBExp II & 0.05 & 0.07 & 94.15 & 0.071 & 0.073 & 1.057 & 17.29 & 0.96 & 91.65 & 0.691 & 0.672 & 0.946 \\ 
   & WBRad II & 0.05 & 0.07 & 94.75 & 0.072 & 0.073 & 1.016 & 17.29 & 0.96 & 94.50 & 0.746 & 0.672 & 0.812 \\ 
   & SAND & 0.05 & 0.07 & 95.20 & 0.073 & 0.073 & 0.988 & 17.29 & 0.96 & 83.15 & 0.582 & 0.672 & 1.333 \\ 
	\addlinespace   
   \multirow{7}{*}{ATO} 
   & BOOT I & 0.03 & 0.07 & 95.20 & 0.070 & 0.069 & 0.973 & 13.30 & 0.67 & 73.25 & 0.397 & 0.409 & 1.060 \\  
    & BOOT II & 0.03 & 0.07 & 95.30 & 0.070 & 0.069 & 0.973 & 13.30 & 0.67 & 73.80 & 0.399 & 0.409 & 1.048 \\ 
   & WBExp I & 0.03 & 0.07 & 93.40 & 0.066 & 0.069 & 1.075 & 13.30 & 0.67 & 72.45 & 0.392 & 0.409 & 1.088 \\ 
   & WBRad I & 0.03 & 0.07 & 94.10 & 0.067 & 0.069 & 1.046 & 13.30 & 0.67 & 74.55 & 0.402 & 0.409 & 1.030 \\ 
   & WBExp II & 0.03 & 0.07 & 93.05 & 0.066 & 0.069 & 1.067 & 13.30 & 0.67 & 52.95 & 0.288 & 0.409 & 2.012 \\ 
   & WBRad II & 0.03 & 0.07 & 94.30 & 0.067 & 0.069 & 1.042 & 13.30 & 0.67 & 54.90 & 0.295 & 0.409 & 1.918 \\ 
   & SAND & 0.03 & 0.07 & 95.35 & 0.069 & 0.069 & 0.987 & 13.30 & 0.67 & 73.35 & 0.397 & 0.409 & 1.057 \\ 
			\addlinespace 
			\multirow{7}{*}{ATM} 
& BOOT I & 0.03 & 0.07 & 95.50 & 0.070 & 0.069 & 0.970 & 11.87 & 0.59 & 72.00 & 0.341 & 0.347 & 1.034 \\  
    & BOOT II & 0.03 & 0.07 & 95.45 & 0.070 & 0.069 & 0.977 & 11.87 & 0.59 & 71.60 & 0.340 & 0.347 & 1.041 \\ 
   & WBExp I & 0.03 & 0.07 & 93.75 & 0.067 & 0.069 & 1.069 & 11.87 & 0.59 & 70.60 & 0.333 & 0.347 & 1.082 \\ 
   & WBRad I & 0.03 & 0.07 & 94.60 & 0.068 & 0.069 & 1.045 & 11.87 & 0.59 & 72.00 & 0.339 & 0.347 & 1.048 \\ 
   & WBExp II & 0.03 & 0.07 & 93.80 & 0.067 & 0.069 & 1.062 & 11.87 & 0.59 & 53.00 & 0.253 & 0.347 & 1.879 \\ 
   & WBRad II & 0.03 & 0.07 & 94.65 & 0.068 & 0.069 & 1.033 & 11.87 & 0.59 & 53.70 & 0.258 & 0.347 & 1.812 \\ 
   & SAND & 0.03 & 0.07 & 95.50 & 0.070 & 0.069 & 0.982 & 11.87 & 0.59 & 71.85 & 0.339 & 0.347 & 1.043 \\ 
	\addlinespace 
	\multirow{7}{*}{ATEN} 
   & BOOT I & 0.03 & 0.07 & 95.00 & 0.070 & 0.069 & 0.977 & 13.72 & 0.71 & 74.80 & 0.430 & 0.449 & 1.088 \\ 
    & BOOT II & 0.03 & 0.07 & 95.05 & 0.070 & 0.069 & 0.975 & 13.72 & 0.71 & 75.70 & 0.433 & 0.449 & 1.075 \\ 
   & WBExp I & 0.03 & 0.07 & 93.30 & 0.066 & 0.069 & 1.082 & 13.72 & 0.71 & 74.20 & 0.426 & 0.449 & 1.111 \\ 
   & WBRad I & 0.03 & 0.07 & 93.85 & 0.067 & 0.069 & 1.050 & 13.72 & 0.71 & 76.50 & 0.444 & 0.449 & 1.023 \\ 
   & WBExp II & 0.03 & 0.07 & 93.15 & 0.066 & 0.069 & 1.074 & 13.72 & 0.71 & 58.50 & 0.335 & 0.449 & 1.793 \\ 
   & WBRad II & 0.03 & 0.07 & 94.15 & 0.067 & 0.069 & 1.044 & 13.72 & 0.71 & 60.30 & 0.346 & 0.449 & 1.683 \\ 
   & SAND & 0.03 & 0.07 & 95.10 & 0.069 & 0.069 & 0.991 & 13.72 & 0.71 & 75.20 & 0.430 & 0.449 & 1.088 \\ 
			\bottomrule
		\end{tabular}
		\begin{tablenotes}
			\scriptsize
   \item ARBias\%: absolute percent relative bias; RMSE: root mean square error; CP\%: coverage probability (\%); SE: median of standard errors by proposed method; ESD: empirical standard deviation; RE: median relative efficiency; BOOT I: standard bootstrap; BOOT II: post-weighting bootstrap;  WBExp (resp. ExpRad): wild bootstrap via exponential (resp. Rademacher) distribution; SAND: sandwich variance estimator;  PS: propensity score; OR: outcome regression.
		\end{tablenotes}
	\end{threeparttable}
\end{table}

\begin{table}[H]\tiny
	\begin{threeparttable}
		\centering
		\caption{Results of Model 3 (large $p=0.80$), $N=1000$, under heterogeneous treatment effect}
		\label{tab:largep-n1000-hets-results}
				\begin{tabular}{rrcccccccccccc}
			\toprule
   Estimand & Method & ARBias\%& RMSE & CP\%  & SE & ESD & RE &ARBias\%& RMSE  & CP\% & SE & ESD & RE \\
			\cmidrule(lr){3-8}\cmidrule(lr){9-14}
			\addlinespace 
			& &\multicolumn{6}{c}{Augmented: PS and OR correctly specified} & \multicolumn{6}{c}{Augmented: Only PS correctly specified} \\\cmidrule(lr){3-8}\cmidrule(lr){9-14}
			\cmidrule(lr){1-2}\cmidrule(lr){3-8}\cmidrule(lr){9-14}
			\multirow{7}{*}{ATE} 
 & BOOT I & 0.13 & 0.61 & 94.65 & 0.602 & 0.609 & 1.024 & 0.06 & 1.19 & 95.60 & 0.879 & 1.187 & 1.822 \\ 
    & BOOT II & 0.13 & 0.61 & 94.65 & 0.602 & 0.609 & 1.025 & 0.06 & 1.19 & 99.45 & 1.294 & 1.187 & 0.840 \\ 
   & WBExp I & 0.13 & 0.61 & 93.60 & 0.587 & 0.609 & 1.078 & 0.06 & 1.19 & 99.40 & 1.261 & 1.187 & 0.886 \\ 
   & WBRad I & 0.13 & 0.61 & 94.25 & 0.594 & 0.609 & 1.054 & 0.06 & 1.19 & 99.40 & 1.340 & 1.187 & 0.784 \\ 
   & WBExp II & 0.13 & 0.61 & 93.60 & 0.587 & 0.609 & 1.078 & 0.06 & 1.19 & 99.40 & 1.261 & 1.187 & 0.886 \\ 
   & WBRad II & 0.13 & 0.61 & 94.25 & 0.594 & 0.609 & 1.054 & 0.06 & 1.19 & 99.40 & 1.340 & 1.187 & 0.784 \\ 
   & SAND & 0.13 & 0.61 & 94.70 & 0.598 & 0.609 & 1.040 & 0.06 & 1.19 & 93.95 & 0.811 & 1.187 & 2.142 \\ 
    \addlinespace
   			\multirow{7}{*}{ATT} 
& BOOT I & 0.15 & 0.67 & 94.75 & 0.659 & 0.666 & 1.020 & 0.63 & 1.10 & 95.00 & 0.938 & 1.093 & 1.359 \\ 
    & BOOT II & 0.15 & 0.67 & 93.20 & 0.609 & 0.666 & 1.195 & 0.63 & 1.10 & 97.75 & 1.157 & 1.093 & 0.893 \\ 
   & WBExp I & 0.15 & 0.67 & 91.70 & 0.593 & 0.666 & 1.258 & 0.63 & 1.10 & 97.20 & 1.122 & 1.093 & 0.950 \\ 
   & WBRad I & 0.15 & 0.67 & 92.40 & 0.600 & 0.666 & 1.231 & 0.63 & 1.10 & 97.50 & 1.168 & 1.093 & 0.876 \\ 
   & WBExp II & 0.15 & 0.67 & 94.00 & 0.642 & 0.666 & 1.076 & 0.63 & 1.10 & 97.80 & 1.162 & 1.093 & 0.884 \\ 
   & WBRad II & 0.15 & 0.67 & 94.25 & 0.652 & 0.666 & 1.042 & 0.63 & 1.10 & 98.00 & 1.215 & 1.093 & 0.809 \\ 
   & SAND & 0.15 & 0.67 & 94.85 & 0.654 & 0.666 & 1.037 & 0.63 & 1.10 & 93.50 & 0.875 & 1.093 & 1.561 \\ 
       \addlinespace
   			\multirow{7}{*}{ATO} 
& BOOT I & 0.46 & 0.84 & 92.95 & 0.807 & 0.837 & 1.077 & 0.82 & 0.93 & 91.95 & 0.868 & 0.917 & 1.115 \\ 
    & BOOT II & 0.46 & 0.84 & 80.55 & 0.562 & 0.837 & 2.217 & 0.82 & 0.93 & 98.75 & 1.273 & 0.917 & 0.519 \\ 
   & WBExp I & 0.46 & 0.84 & 79.25 & 0.545 & 0.837 & 2.361 & 0.82 & 0.93 & 98.40 & 1.261 & 0.917 & 0.528 \\ 
   & WBRad I & 0.46 & 0.84 & 80.20 & 0.555 & 0.837 & 2.279 & 0.82 & 0.93 & 98.65 & 1.320 & 0.917 & 0.482 \\ 
   & WBExp II & 0.46 & 0.84 & 95.35 & 0.919 & 0.837 & 0.830 & 0.82 & 0.93 & 98.30 & 1.213 & 0.917 & 0.571 \\ 
   & WBRad II & 0.46 & 0.84 & 96.20 & 0.966 & 0.837 & 0.751 & 0.82 & 0.93 & 98.55 & 1.251 & 0.917 & 0.537 \\ 
   & SAND & 0.46 & 0.84 & 92.95 & 0.800 & 0.837 & 1.095 & 0.82 & 0.93 & 91.80 & 0.864 & 0.917 & 1.125 \\ 
\addlinespace 
			\multirow{7}{*}{ATM}
& BOOT I & 0.59 & 0.94 & 93.55 & 0.903 & 0.932 & 1.066 & 1.03 & 1.04 & 92.60 & 0.988 & 1.026 & 1.079 \\ 
    & BOOT II & 0.59 & 0.94 & 78.85 & 0.604 & 0.932 & 2.384 & 1.03 & 1.04 & 98.40 & 1.360 & 1.026 & 0.569 \\ 
   & WBExp I & 0.59 & 0.94 & 77.40 & 0.582 & 0.932 & 2.560 & 1.03 & 1.04 & 98.15 & 1.374 & 1.026 & 0.558 \\ 
   & WBRad I & 0.59 & 0.94 & 78.45 & 0.598 & 0.932 & 2.429 & 1.03 & 1.04 & 98.30 & 1.430 & 1.026 & 0.515 \\ 
   & WBExp II & 0.59 & 0.94 & 96.75 & 1.107 & 0.932 & 0.708 & 1.03 & 1.04 & 98.80 & 1.420 & 1.026 & 0.522 \\ 
   & WBRad II & 0.59 & 0.94 & 97.15 & 1.155 & 0.932 & 0.651 & 1.03 & 1.04 & 98.80 & 1.451 & 1.026 & 0.500 \\ 
   & SAND & 0.59 & 0.94 & 93.35 & 0.903 & 0.932 & 1.066 & 1.03 & 1.04 & 92.80 & 0.990 & 1.026 & 1.076 \\ 
			\addlinespace 
			\multirow{7}{*}{ATEN} 
& BOOT I & 0.47 & 0.78 & 93.35 & 0.750 & 0.781 & 1.086 & 0.76 & 0.88 & 91.40 & 0.819 & 0.871 & 1.131 \\ 
     & BOOT II & 0.47 & 0.78 & 82.45 & 0.548 & 0.781 & 2.030 & 0.76 & 0.88 & 98.85 & 1.239 & 0.871 & 0.494 \\ 
   & WBExp I & 0.47 & 0.78 & 81.75 & 0.532 & 0.781 & 2.154 & 0.76 & 0.88 & 98.55 & 1.221 & 0.871 & 0.508 \\ 
   & WBRad I & 0.47 & 0.78 & 82.25 & 0.539 & 0.781 & 2.101 & 0.76 & 0.88 & 98.60 & 1.275 & 0.871 & 0.466 \\ 
   & WBExp II & 0.47 & 0.78 & 95.15 & 0.831 & 0.781 & 0.884 & 0.76 & 0.88 & 98.30 & 1.144 & 0.871 & 0.579 \\ 
   & WBRad II & 0.47 & 0.78 & 96.30 & 0.885 & 0.781 & 0.780 & 0.76 & 0.88 & 98.40 & 1.190 & 0.871 & 0.535 \\ 
   & SAND & 0.47 & 0.78 & 93.00 & 0.743 & 0.781 & 1.105 & 0.76 & 0.88 & 91.10 & 0.808 & 0.871 & 1.160 \\ 
			\addlinespace 
			& &\multicolumn{6}{c}{Augmented: only OR correctly specified} & \multicolumn{6}{c}{Augmented: PS and OR misspecified}\\ 
			\cmidrule(lr){1-2}\cmidrule(lr){3-8}\cmidrule(lr){9-14}
	\multirow{7}{*}{ATE} 
& BOOT I & 0.12 & 0.61 & 94.70 & 0.602 & 0.609 & 1.025 & 11.73 & 2.55 & 71.55 & 1.342 & 1.560 & 1.350 \\ 
    & BOOT II & 0.12 & 0.61 & 94.70 & 0.601 & 0.609 & 1.026 & 11.73 & 2.55 & 70.00 & 1.316 & 1.560 & 1.406 \\ 
   & WBExp I & 0.12 & 0.61 & 93.75 & 0.586 & 0.609 & 1.080 & 11.73 & 2.55 & 73.65 & 1.357 & 1.560 & 1.321 \\ 
   & WBRad I & 0.12 & 0.61 & 94.20 & 0.594 & 0.609 & 1.051 & 11.73 & 2.55 & 80.70 & 1.475 & 1.560 & 1.119 \\ 
   & WBExp II & 0.12 & 0.61 & 93.75 & 0.586 & 0.609 & 1.080 & 11.73 & 2.55 & 73.65 & 1.357 & 1.560 & 1.321 \\ 
   & WBRad II & 0.12 & 0.61 & 94.20 & 0.594 & 0.609 & 1.051 & 11.73 & 2.55 & 80.70 & 1.475 & 1.560 & 1.119 \\ 
   & SAND & 0.12 & 0.61 & 94.75 & 0.597 & 0.609 & 1.040 & 11.73 & 2.55 & 69.80 & 1.318 & 1.560 & 1.402 \\ 
			\addlinespace 
   \multirow{7}{*}{ATT} 
& BOOT I & 2.21 & 0.74 & 91.90 & 0.632 & 0.639 & 1.024 &  8.92 & 2.23 & 84.60 & 1.355 & 1.646 & 1.477 \\ 
    & BOOT II & 2.21 & 0.74 & 91.65 & 0.624 & 0.639 & 1.050 &  8.92 & 2.23 & 83.10 & 1.330 & 1.646 & 1.531 \\ 
   & WBExp I & 2.21 & 0.74 & 90.30 & 0.608 & 0.639 & 1.104 &  8.92 & 2.23 & 89.85 & 1.417 & 1.646 & 1.349 \\ 
   & WBRad I & 2.21 & 0.74 & 90.75 & 0.618 & 0.639 & 1.068 &  8.92 & 2.23 & 93.95 & 1.560 & 1.646 & 1.113 \\ 
   & WBExp II & 2.21 & 0.74 & 92.50 & 0.644 & 0.639 & 0.986 &  8.92 & 2.23 & 91.45 & 1.472 & 1.646 & 1.250 \\ 
   & WBRad II & 2.21 & 0.74 & 92.75 & 0.652 & 0.639 & 0.961 &  8.92 & 2.23 & 95.35 & 1.617 & 1.646 & 1.036 \\ 
   & SAND & 2.21 & 0.74 & 91.80 & 0.628 & 0.639 & 1.035 &  8.92 & 2.23 & 82.95 & 1.324 & 1.646 & 1.547 \\ 
			\addlinespace   
   \multirow{7}{*}{ATO} 
& BOOT I & 3.06 & 0.81 & 85.35 & 0.643 & 0.662 & 1.060 & 13.92 & 2.46 & 54.05 & 1.171 & 1.211 & 1.069 \\ 
    & BOOT II & 3.06 & 0.81 & 78.55 & 0.552 & 0.662 & 1.438 & 13.92 & 2.46 & 56.20 & 1.208 & 1.211 & 1.005 \\ 
   & WBExp I & 3.06 & 0.81 & 77.40 & 0.537 & 0.662 & 1.521 & 13.92 & 2.46 & 53.20 & 1.171 & 1.211 & 1.069 \\ 
   & WBRad I & 3.06 & 0.81 & 78.25 & 0.546 & 0.662 & 1.467 & 13.92 & 2.46 & 57.65 & 1.233 & 1.211 & 0.965 \\ 
   & WBExp II & 3.06 & 0.81 & 96.45 & 0.977 & 0.662 & 0.459 & 13.92 & 2.46 & 57.35 & 1.234 & 1.211 & 0.962 \\ 
   & WBRad II & 3.06 & 0.81 & 97.20 & 1.045 & 0.662 & 0.401 & 13.92 & 2.46 & 59.55 & 1.269 & 1.211 & 0.910 \\ 
   & SAND & 3.06 & 0.81 & 85.10 & 0.639 & 0.662 & 1.073 & 13.92 & 2.46 & 53.45 & 1.159 & 1.211 & 1.091 \\ 
			\addlinespace 
			\multirow{7}{*}{ATM} 
& BOOT I & 1.26 & 0.81 & 92.05 & 0.754 & 0.781 & 1.074 & 12.49 & 2.40 & 65.75 & 1.316 & 1.357 & 1.063 \\ 
     & BOOT II & 1.26 & 0.81 & 85.20 & 0.622 & 0.781 & 1.577 & 12.49 & 2.40 & 68.60 & 1.379 & 1.357 & 0.968 \\ 
   & WBExp I & 1.26 & 0.81 & 84.30 & 0.605 & 0.781 & 1.665 & 12.49 & 2.40 & 67.70 & 1.353 & 1.357 & 1.006 \\ 
   & WBRad I & 1.26 & 0.81 & 84.85 & 0.620 & 0.781 & 1.586 & 12.49 & 2.40 & 69.65 & 1.424 & 1.357 & 0.908 \\ 
   & WBExp II & 1.26 & 0.81 & 98.30 & 1.128 & 0.781 & 0.480 & 12.49 & 2.40 & 75.50 & 1.545 & 1.357 & 0.771 \\ 
   & WBRad II & 1.26 & 0.81 & 98.90 & 1.185 & 0.781 & 0.435 & 12.49 & 2.40 & 76.35 & 1.583 & 1.357 & 0.735 \\ 
   & SAND & 1.26 & 0.81 & 91.90 & 0.755 & 0.781 & 1.072 & 12.49 & 2.40 & 66.35 & 1.322 & 1.357 & 1.053 \\ 
			\addlinespace 
			\multirow{7}{*}{ATEN} 
& BOOT I & 2.78 & 0.77 & 86.15 & 0.613 & 0.630 & 1.056 & 13.65 & 2.44 & 53.85 & 1.148 & 1.193 & 1.080 \\ 
    & BOOT II & 2.78 & 0.77 & 79.70 & 0.531 & 0.630 & 1.410 & 13.65 & 2.44 & 54.35 & 1.160 & 1.193 & 1.057 \\ 
   & WBExp I & 2.78 & 0.77 & 78.75 & 0.517 & 0.630 & 1.486 & 13.65 & 2.44 & 51.70 & 1.129 & 1.193 & 1.117 \\ 
   & WBRad I & 2.78 & 0.77 & 79.20 & 0.525 & 0.630 & 1.439 & 13.65 & 2.44 & 55.90 & 1.187 & 1.193 & 1.010 \\ 
   & WBExp II & 2.78 & 0.77 & 96.75 & 0.955 & 0.630 & 0.436 & 13.65 & 2.44 & 51.70 & 1.120 & 1.193 & 1.134 \\ 
   & WBRad II & 2.78 & 0.77 & 97.70 & 1.035 & 0.630 & 0.371 & 13.65 & 2.44 & 53.80 & 1.160 & 1.193 & 1.057 \\ 
   & SAND & 2.78 & 0.77 & 85.65 & 0.609 & 0.630 & 1.071 & 13.65 & 2.44 & 52.85 & 1.139 & 1.193 & 1.097 \\ 
			\bottomrule
		\end{tabular}
		\begin{tablenotes}
			\scriptsize
			\item ARBias\%: absolute percent relative bias; RMSE: root mean square error; CP\%: coverage probability (\%); SE: median of standard errors by proposed method; ESD: empirical standard deviation; RE: median relative efficiency; BOOT I: standard bootstrap; BOOT II: post-weighting bootstrap;  WBExp (resp. ExpRad): wild bootstrap via exponential (resp. Rademacher) distribution; SAND: sandwich variance estimator;  PS: propensity score; OR: outcome regression.
		\end{tablenotes}
	\end{threeparttable}
\end{table}

\begin{table}[H]\tiny
	\begin{threeparttable}
		\centering
		\caption{Results of Model 3 (large $p=0.80$), $N=1000$, under homogeneous treatment effect}
		\label{tab:largep-n1000-cons-results}
				\begin{tabular}{rrcccccccccccc}
			\toprule
   Estimand & Method & ARBias\%& RMSE & CP\%  & SE & ESD & RE &ARBias\%& RMSE  & CP\% & SE & ESD & RE \\
			\cmidrule(lr){3-8}\cmidrule(lr){9-14}
			\addlinespace 
			& &\multicolumn{6}{c}{Augmented: PS and OR correctly specified} & \multicolumn{6}{c}{Augmented: Only PS correctly specified} \\\cmidrule(lr){3-8}\cmidrule(lr){9-14}
			\cmidrule(lr){1-2}\cmidrule(lr){3-8}\cmidrule(lr){9-14}
			\multirow{7}{*}{ATE} 
 & BOOT I & 0.02 & 0.10 & 94.60 & 0.098 & 0.099 & 1.019 & 1.35 & 0.74 &  92.45 & 0.494 & 0.733 & 2.201 \\ 
     & BOOT II & 0.02 & 0.10 & 94.55 & 0.097 & 0.099 & 1.032 & 1.35 & 0.74 &  98.50 & 0.776 & 0.733 & 0.893 \\ 
   & WBExp I & 0.02 & 0.10 & 92.50 & 0.090 & 0.099 & 1.221 & 1.35 & 0.74 &  97.65 & 0.739 & 0.733 & 0.985 \\ 
   & WBRad I & 0.02 & 0.10 & 93.15 & 0.093 & 0.099 & 1.129 & 1.35 & 0.74 &  98.00 & 0.829 & 0.733 & 0.783 \\ 
   & WBExp II & 0.02 & 0.10 & 92.50 & 0.090 & 0.099 & 1.221 & 1.35 & 0.74 &  97.65 & 0.739 & 0.733 & 0.985 \\ 
   & WBRad II & 0.02 & 0.10 & 93.15 & 0.093 & 0.099 & 1.129 & 1.35 & 0.74 &  98.00 & 0.829 & 0.733 & 0.783 \\ 
   & SAND & 0.02 & 0.10 & 93.55 & 0.093 & 0.099 & 1.128 & 1.35 & 0.74 &  88.80 & 0.429 & 0.733 & 2.918 \\ 
    \addlinespace
   			\multirow{7}{*}{ATT} 
& BOOT I & 0.01 & 0.11 & 94.85 & 0.108 & 0.109 & 1.027 & 2.46 & 0.89 &  89.95 & 0.602 & 0.882 & 2.148 \\ 
    & BOOT II & 0.01 & 0.11 & 94.60 & 0.106 & 0.109 & 1.050 & 2.46 & 0.89 &  95.90 & 0.842 & 0.882 & 1.095 \\ 
   & WBExp I & 0.01 & 0.11 & 91.55 & 0.096 & 0.109 & 1.290 & 2.46 & 0.89 &  94.75 & 0.805 & 0.882 & 1.201 \\ 
   & WBRad I & 0.01 & 0.11 & 92.80 & 0.101 & 0.109 & 1.162 & 2.46 & 0.89 &  95.90 & 0.914 & 0.882 & 0.931 \\ 
   & WBExp II & 0.01 & 0.11 & 91.40 & 0.096 & 0.109 & 1.290 & 2.46 & 0.89 &  95.25 & 0.832 & 0.882 & 1.124 \\ 
   & WBRad II & 0.01 & 0.11 & 92.65 & 0.102 & 0.109 & 1.153 & 2.46 & 0.89 &  96.30 & 0.942 & 0.882 & 0.877 \\ 
   & SAND & 0.01 & 0.11 & 93.55 & 0.101 & 0.109 & 1.161 & 2.46 & 0.89 &  86.65 & 0.528 & 0.882 & 2.784 \\ 
       \addlinespace
   			\multirow{7}{*}{ATO} 
& BOOT I & 0.05 & 0.08 & 94.90 & 0.084 & 0.084 & 0.996 & 0.20 & 0.10 &  95.60 & 0.101 & 0.099 & 0.964 \\ 
    & BOOT II & 0.05 & 0.08 & 94.80 & 0.084 & 0.084 & 0.992 & 0.20 & 0.10 & 100.00 & 0.518 & 0.099 & 0.037 \\ 
   & WBExp I & 0.05 & 0.08 & 93.60 & 0.081 & 0.084 & 1.078 & 0.20 & 0.10 & 100.00 & 0.497 & 0.099 & 0.040 \\ 
   & WBRad I & 0.05 & 0.08 & 94.40 & 0.082 & 0.084 & 1.048 & 0.20 & 0.10 & 100.00 & 0.516 & 0.099 & 0.037 \\ 
   & WBExp II & 0.05 & 0.08 & 94.10 & 0.081 & 0.084 & 1.066 & 0.20 & 0.10 & 100.00 & 0.393 & 0.099 & 0.064 \\ 
   & WBRad II & 0.05 & 0.08 & 94.25 & 0.083 & 0.084 & 1.028 & 0.20 & 0.10 & 100.00 & 0.405 & 0.099 & 0.060 \\ 
   & SAND & 0.05 & 0.08 & 95.25 & 0.083 & 0.084 & 1.015 & 0.20 & 0.10 &  95.40 & 0.099 & 0.099 & 1.009 \\ 
\addlinespace 
			\multirow{7}{*}{ATM}
& BOOT I & 0.05 & 0.09 & 95.55 & 0.086 & 0.085 & 0.985 & 0.36 & 0.12 &  95.80 & 0.127 & 0.121 & 0.903 \\ 
    & BOOT II & 0.05 & 0.09 & 95.25 & 0.085 & 0.085 & 0.991 & 0.36 & 0.12 & 100.00 & 0.511 & 0.121 & 0.056 \\ 
   & WBExp I & 0.05 & 0.09 & 93.90 & 0.082 & 0.085 & 1.065 & 0.36 & 0.12 & 100.00 & 0.497 & 0.121 & 0.059 \\ 
   & WBRad I & 0.05 & 0.09 & 94.60 & 0.084 & 0.085 & 1.034 & 0.36 & 0.12 & 100.00 & 0.513 & 0.121 & 0.056 \\ 
   & WBExp II & 0.05 & 0.09 & 94.90 & 0.083 & 0.085 & 1.042 & 0.36 & 0.12 & 100.00 & 0.432 & 0.121 & 0.079 \\ 
   & WBRad II & 0.05 & 0.09 & 94.90 & 0.085 & 0.085 & 1.009 & 0.36 & 0.12 & 100.00 & 0.445 & 0.121 & 0.074 \\ 
   & SAND & 0.05 & 0.09 & 95.55 & 0.085 & 0.085 & 0.998 & 0.36 & 0.12 &  96.65 & 0.131 & 0.121 & 0.859 \\ 
			\addlinespace 
			\multirow{7}{*}{ATEN} 
& BOOT I & 0.04 & 0.08 & 95.10 & 0.084 & 0.085 & 1.004 & 0.03 & 0.13 &  95.05 & 0.129 & 0.133 & 1.057 \\ 
    & BOOT II & 0.04 & 0.08 & 95.10 & 0.085 & 0.085 & 0.993 & 0.03 & 0.13 & 100.00 & 0.545 & 0.133 & 0.060 \\ 
   & WBExp I & 0.04 & 0.08 & 93.85 & 0.081 & 0.085 & 1.085 & 0.03 & 0.13 & 100.00 & 0.520 & 0.133 & 0.065 \\ 
   & WBRad I & 0.04 & 0.08 & 94.20 & 0.082 & 0.085 & 1.053 & 0.03 & 0.13 & 100.00 & 0.547 & 0.133 & 0.059 \\ 
   & WBExp II & 0.04 & 0.08 & 94.10 & 0.082 & 0.085 & 1.072 & 0.03 & 0.13 & 100.00 & 0.413 & 0.133 & 0.104 \\ 
   & WBRad II & 0.04 & 0.08 & 94.35 & 0.083 & 0.085 & 1.042 & 0.03 & 0.13 & 100.00 & 0.430 & 0.133 & 0.096 \\ 
   & SAND & 0.04 & 0.08 & 94.90 & 0.083 & 0.085 & 1.026 & 0.03 & 0.13 &  94.50 & 0.126 & 0.133 & 1.115 \\ 
			\addlinespace 
			& &\multicolumn{6}{c}{Augmented: only OR correctly specified} & \multicolumn{6}{c}{Augmented: PS and OR misspecified}\\ 
			\cmidrule(lr){1-2}\cmidrule(lr){3-8}\cmidrule(lr){9-14}
	\multirow{7}{*}{ATE} 
& BOOT I & 0.04 & 0.10 & 95.10 & 0.097 & 0.097 & 1.004 & 34.55 & 1.87 & 81.50 & 0.968 & 1.268 & 1.713 \\ 
   & BOOT II & 0.04 & 0.10 & 95.05 & 0.096 & 0.097 & 1.011 & 34.55 & 1.87 & 81.00 & 0.974 & 1.268 & 1.695 \\
   & WBExp I & 0.04 & 0.10 & 93.00 & 0.090 & 0.097 & 1.157 & 34.55 & 1.87 & 89.10 & 1.036 & 1.268 & 1.498 \\ 
   & WBRad I & 0.04 & 0.10 & 93.90 & 0.094 & 0.097 & 1.059 & 34.55 & 1.87 & 94.95 & 1.221 & 1.268 & 1.077 \\ 
   & WBExp II & 0.04 & 0.10 & 93.00 & 0.090 & 0.097 & 1.157 & 34.55 & 1.87 & 89.10 & 1.036 & 1.268 & 1.498 \\ 
   & WBRad II & 0.04 & 0.10 & 93.90 & 0.094 & 0.097 & 1.059 & 34.55 & 1.87 & 94.95 & 1.221 & 1.268 & 1.077 \\ 
   & SAND & 0.04 & 0.10 & 94.35 & 0.093 & 0.097 & 1.090 & 34.55 & 1.87 & 78.80 & 0.934 & 1.268 & 1.841 \\ 
			\addlinespace 
   \multirow{7}{*}{ATT} 
& BOOT I & 0.03 & 0.10 & 95.25 & 0.104 & 0.104 & 1.007 & 36.54 & 2.07 & 84.55 & 1.084 & 1.463 & 1.819 \\ 
    & BOOT II & 0.03 & 0.10 & 95.10 & 0.103 & 0.104 & 1.021 & 36.54 & 2.07 & 84.50 & 1.100 & 1.463 & 1.767 \\ 
   & WBExp I & 0.03 & 0.10 & 93.45 & 0.097 & 0.104 & 1.151 & 36.54 & 2.07 & 93.55 & 1.221 & 1.463 & 1.436 \\ 
   & WBRad I & 0.03 & 0.10 & 94.00 & 0.103 & 0.104 & 1.035 & 36.54 & 2.07 & 97.25 & 1.450 & 1.463 & 1.017 \\ 
   & WBExp II & 0.03 & 0.10 & 93.55 & 0.098 & 0.104 & 1.143 & 36.54 & 2.07 & 94.00 & 1.263 & 1.463 & 1.342 \\ 
   & WBRad II & 0.03 & 0.10 & 93.85 & 0.103 & 0.104 & 1.033 & 36.54 & 2.07 & 97.85 & 1.485 & 1.463 & 0.970 \\ 
   & SAND & 0.03 & 0.10 & 94.15 & 0.099 & 0.104 & 1.104 & 36.54 & 2.07 & 82.00 & 1.041 & 1.463 & 1.974 \\ 
			\addlinespace   
   \multirow{7}{*}{ATO} 
& BOOT I & 0.06 & 0.09 & 94.90 & 0.087 & 0.087 & 1.003 & 20.44 & 0.98 & 66.15 & 0.529 & 0.542 & 1.052 \\ 
   & BOOT II & 0.06 & 0.09 & 95.00 & 0.087 & 0.087 & 1.002 & 20.44 & 0.98 & 66.55 & 0.531 & 0.542 & 1.043 \\ 
   & WBExp I & 0.06 & 0.09 & 92.85 & 0.079 & 0.087 & 1.193 & 20.44 & 0.98 & 64.05 & 0.512 & 0.542 & 1.123 \\ 
   & WBRad I & 0.06 & 0.09 & 92.50 & 0.080 & 0.087 & 1.166 & 20.44 & 0.98 & 66.75 & 0.531 & 0.542 & 1.044 \\ 
   & WBExp II & 0.06 & 0.09 & 92.75 & 0.080 & 0.087 & 1.179 & 20.44 & 0.98 & 49.50 & 0.415 & 0.542 & 1.706 \\ 
   & WBRad II & 0.06 & 0.09 & 93.15 & 0.081 & 0.087 & 1.148 & 20.44 & 0.98 & 51.05 & 0.426 & 0.542 & 1.619 \\ 
   & SAND & 0.06 & 0.09 & 94.65 & 0.085 & 0.087 & 1.044 & 20.44 & 0.98 & 66.20 & 0.526 & 0.542 & 1.062 \\ 
			\addlinespace 
			\multirow{7}{*}{ATM} 
& BOOT I & 0.07 & 0.09 & 95.40 & 0.088 & 0.088 & 0.998 & 20.02 & 0.96 & 68.30 & 0.531 & 0.538 & 1.025 \\ 
    & BOOT II & 0.07 & 0.09 & 95.45 & 0.088 & 0.088 & 1.003 & 20.02 & 0.96 & 68.35 & 0.534 & 0.538 & 1.016 \\ 
   & WBExp I & 0.07 & 0.09 & 92.90 & 0.081 & 0.088 & 1.190 & 20.02 & 0.96 & 66.40 & 0.516 & 0.538 & 1.086 \\ 
   & WBRad I & 0.07 & 0.09 & 92.85 & 0.082 & 0.088 & 1.159 & 20.02 & 0.96 & 68.45 & 0.530 & 0.538 & 1.028 \\ 
   & WBExp II & 0.07 & 0.09 & 92.85 & 0.082 & 0.088 & 1.149 & 20.02 & 0.96 & 58.80 & 0.469 & 0.538 & 1.314 \\ 
   & WBRad II & 0.07 & 0.09 & 93.10 & 0.083 & 0.088 & 1.131 & 20.02 & 0.96 & 60.05 & 0.475 & 0.538 & 1.284 \\ 
   & SAND & 0.07 & 0.09 & 94.85 & 0.087 & 0.088 & 1.031 & 20.02 & 0.96 & 68.45 & 0.531 & 0.538 & 1.028 \\ 
			\addlinespace 
			\multirow{5}{*}{ATEN} 
& BOOT I & 0.05 & 0.09 & 94.95 & 0.087 & 0.087 & 1.005 & 22.33 & 1.08 & 67.00 & 0.572 & 0.600 & 1.102 \\ 
    & BOOT II & 0.05 & 0.09 & 94.95 & 0.087 & 0.087 & 1.000 & 22.33 & 1.08 & 68.45 & 0.583 & 0.600 & 1.062 \\ 
   & WBExp I & 0.05 & 0.09 & 93.10 & 0.080 & 0.087 & 1.193 & 22.33 & 1.08 & 65.65 & 0.565 & 0.600 & 1.129 \\ 
   & WBRad I & 0.05 & 0.09 & 92.60 & 0.081 & 0.087 & 1.149 & 22.33 & 1.08 & 71.25 & 0.602 & 0.600 & 0.996 \\ 
   & WBExp II & 0.05 & 0.09 & 93.10 & 0.081 & 0.087 & 1.170 & 22.33 & 1.08 & 47.95 & 0.443 & 0.600 & 1.834 \\ 
   & WBRad II & 0.05 & 0.09 & 92.70 & 0.082 & 0.087 & 1.136 & 22.33 & 1.08 & 52.30 & 0.460 & 0.600 & 1.705 \\ 
   & SAND & 0.05 & 0.09 & 94.45 & 0.085 & 0.087 & 1.050 & 22.33 & 1.08 & 66.70 & 0.569 & 0.600 & 1.114 \\ 
			\bottomrule
		\end{tabular}
		\begin{tablenotes}
			\scriptsize
			\item ARBias\%: absolute percent relative bias; RMSE: root mean square error; CP\%: coverage probability (\%); SE: median of standard errors by proposed method; ESD: empirical standard deviation; RE: median relative efficiency; BOOT I: standard bootstrap; BOOT II: post-weighting bootstrap;  WBExp (resp. ExpRad): wild bootstrap via exponential (resp. Rademacher) distribution; SAND: sandwich variance estimator;  PS: propensity score; OR: outcome regression.
		\end{tablenotes}
	\end{threeparttable}
\end{table}

\begin{table}[H]\tiny
	\begin{threeparttable}
		\centering
		\caption{Results of Model 4 (poor overlap with a medium $p=0.49$), $N=1000$, under heterogeneous treatment effect}
		\label{tab:extremew-n1000-hets-results}
		\begin{tabular}{rrcccccccccccc}
			\toprule
   Estimand & Method & ARBias\%& RMSE & CP\%  & SE & ESD & RE &ARBias\%& RMSE  & CP\% & SE & ESD & RE \\
			\cmidrule(lr){3-8}\cmidrule(lr){9-14}
			\addlinespace 
			& &\multicolumn{6}{c}{Augmented: PS and OR correctly specified} & \multicolumn{6}{c}{Augmented: Only PS correctly specified} \\\cmidrule(lr){3-8}\cmidrule(lr){9-14}
			\cmidrule(lr){1-2}\cmidrule(lr){3-8}\cmidrule(lr){9-14}
			\multirow{7}{*}{ATE} 
& BOOT I & 0.13 & 0.61 & 94.35 & 0.606 & 0.614 & 1.025 & 0.15 & 2.02 & 95.10 & 1.337 & 2.016 & 2.271 \\ 
    & BOOT II & 0.13 & 0.61 & 94.35 & 0.606 & 0.614 & 1.028 & 0.15 & 2.02 & 98.15 & 1.638 & 2.016 & 1.513 \\ 
   & WBExp I & 0.13 & 0.61 & 94.25 & 0.596 & 0.614 & 1.063 & 0.15 & 2.02 & 96.65 & 1.449 & 2.016 & 1.936 \\ 
   & WBRad I & 0.13 & 0.61 & 94.40 & 0.602 & 0.614 & 1.039 & 0.15 & 2.02 & 97.45 & 1.550 & 2.016 & 1.690 \\ 
   & WBExp II & 0.13 & 0.61 & 94.25 & 0.596 & 0.614 & 1.063 & 0.15 & 2.02 & 96.65 & 1.449 & 2.016 & 1.936 \\ 
   & WBRad II & 0.13 & 0.61 & 94.40 & 0.602 & 0.614 & 1.039 & 0.15 & 2.02 & 97.45 & 1.550 & 2.016 & 1.690 \\ 
   & SAND & 0.13 & 0.61 & 94.45 & 0.605 & 0.614 & 1.032 & 0.15 & 2.02 & 93.15 & 1.208 & 2.016 & 2.783 \\ 
    \addlinespace
   			\multirow{7}{*}{ATT} 
& BOOT I & 0.18 & 0.96 & 94.15 & 0.922 & 0.956 & 1.074 & 1.64 & 1.50 & 93.35 & 1.172 & 1.465 & 1.563 \\ 
    & BOOT II & 0.18 & 0.96 & 89.90 & 0.801 & 0.956 & 1.423 & 1.64 & 1.50 & 97.20 & 1.355 & 1.465 & 1.169 \\ 
   & WBExp I & 0.18 & 0.96 & 89.30 & 0.785 & 0.956 & 1.481 & 1.64 & 1.50 & 96.10 & 1.293 & 1.465 & 1.285 \\ 
   & WBRad I & 0.18 & 0.96 & 89.70 & 0.796 & 0.956 & 1.441 & 1.64 & 1.50 & 96.30 & 1.337 & 1.465 & 1.201 \\ 
   & WBExp II & 0.18 & 0.96 & 93.55 & 0.897 & 0.956 & 1.136 & 1.64 & 1.50 & 96.55 & 1.308 & 1.465 & 1.255 \\ 
   & WBRad II & 0.18 & 0.96 & 93.95 & 0.915 & 0.956 & 1.091 & 1.64 & 1.50 & 96.75 & 1.353 & 1.465 & 1.172 \\ 
   & SAND & 0.18 & 0.96 & 94.20 & 0.918 & 0.956 & 1.084 & 1.64 & 1.50 & 92.35 & 1.125 & 1.465 & 1.697 \\ 
       \addlinespace
   			\multirow{7}{*}{ATO} 
& BOOT I & 0.29 & 0.75 & 94.90 & 0.745 & 0.751 & 1.017 & 0.59 & 0.92 & 93.85 & 0.912 & 0.918 & 1.014 \\ 
    & BOOT II & 0.29 & 0.75 & 93.25 & 0.705 & 0.751 & 1.137 & 0.59 & 0.92 & 99.75 & 1.459 & 0.918 & 0.396 \\ 
   & WBExp I & 0.29 & 0.75 & 92.30 & 0.691 & 0.751 & 1.184 & 0.59 & 0.92 & 99.45 & 1.396 & 0.918 & 0.433 \\ 
   & WBRad I & 0.29 & 0.75 & 93.20 & 0.699 & 0.751 & 1.157 & 0.59 & 0.92 & 99.50 & 1.445 & 0.918 & 0.404 \\ 
   & WBExp II & 0.29 & 0.75 & 98.50 & 0.973 & 0.751 & 0.596 & 0.59 & 0.92 & 99.25 & 1.373 & 0.918 & 0.447 \\ 
   & WBRad II & 0.29 & 0.75 & 98.75 & 1.016 & 0.751 & 0.546 & 0.59 & 0.92 & 99.40 & 1.416 & 0.918 & 0.421 \\ 
   & SAND & 0.29 & 0.75 & 94.90 & 0.742 & 0.751 & 1.025 & 0.59 & 0.92 & 93.50 & 0.899 & 0.918 & 1.044 \\ 
\addlinespace 
			\multirow{7}{*}{ATM}
  & BOOT I & 0.23 & 0.83 & 95.05 & 0.817 & 0.825 & 1.019 & 0.50 & 0.98 & 93.70 & 0.967 & 0.975 & 1.016 \\ 
     & BOOT II & 0.23 & 0.83 & 92.85 & 0.749 & 0.825 & 1.212 & 0.50 & 0.98 & 99.45 & 1.487 & 0.975 & 0.430 \\ 
   & WBExp I & 0.23 & 0.83 & 91.75 & 0.732 & 0.825 & 1.272 & 0.50 & 0.98 & 99.30 & 1.437 & 0.975 & 0.460 \\ 
   & WBRad I & 0.23 & 0.83 & 92.75 & 0.746 & 0.825 & 1.224 & 0.50 & 0.98 & 99.35 & 1.480 & 0.975 & 0.434 \\ 
   & WBExp II & 0.23 & 0.83 & 99.25 & 1.258 & 0.825 & 0.430 & 0.50 & 0.98 & 99.35 & 1.473 & 0.975 & 0.438 \\ 
   & WBRad II & 0.23 & 0.83 & 99.60 & 1.311 & 0.825 & 0.396 & 0.50 & 0.98 & 99.45 & 1.516 & 0.975 & 0.414 \\ 
   & SAND & 0.23 & 0.83 & 95.20 & 0.821 & 0.825 & 1.009 & 0.50 & 0.98 & 94.25 & 0.972 & 0.975 & 1.007 \\ 
			\addlinespace 
			\multirow{7}{*}{ATEN} 
& BOOT I & 0.20 & 0.71 & 95.10 & 0.705 & 0.711 & 1.016 & 0.51 & 0.93 & 93.65 & 0.916 & 0.924 & 1.017 \\ 
    & BOOT II & 0.20 & 0.71 & 93.60 & 0.677 & 0.711 & 1.104 & 0.51 & 0.93 & 99.70 & 1.452 & 0.924 & 0.405 \\ 
   & WBExp I & 0.20 & 0.71 & 92.75 & 0.661 & 0.711 & 1.157 & 0.51 & 0.93 & 99.50 & 1.388 & 0.924 & 0.443 \\ 
   & WBRad I & 0.20 & 0.71 & 93.90 & 0.670 & 0.711 & 1.126 & 0.51 & 0.93 & 99.55 & 1.440 & 0.924 & 0.412 \\ 
   & WBExp II & 0.20 & 0.71 & 97.85 & 0.878 & 0.711 & 0.655 & 0.51 & 0.93 & 99.20 & 1.340 & 0.924 & 0.475 \\ 
   & WBRad II & 0.20 & 0.71 & 98.50 & 0.912 & 0.711 & 0.609 & 0.51 & 0.93 & 99.50 & 1.391 & 0.924 & 0.441 \\ 
   & SAND & 0.20 & 0.71 & 94.90 & 0.702 & 0.711 & 1.028 & 0.51 & 0.93 & 93.85 & 0.898 & 0.924 & 1.059 \\ 
			\addlinespace 
			& &\multicolumn{6}{c}{Augmented: only OR correctly specified} & \multicolumn{6}{c}{Augmented: PS and OR misspecified}\\ 
			\cmidrule(lr){1-2}\cmidrule(lr){3-8}\cmidrule(lr){9-14}
	\multirow{7}{*}{ATE} 
& BOOT I & 0.11 & 0.62 & 95.00 & 0.612 & 0.624 & 1.041 &  9.08 & 5.02 & 91.35 & 2.920 & 4.769 & 2.667 \\ 
    & BOOT II & 0.11 & 0.62 & 94.95 & 0.611 & 0.624 & 1.045 &  9.08 & 5.02 & 90.40 & 2.862 & 4.769 & 2.777 \\ 
   & WBExp I & 0.11 & 0.62 & 95.10 & 0.618 & 0.624 & 1.022 &  9.08 & 5.02 & 93.00 & 3.046 & 4.769 & 2.451 \\ 
   & WBRad I & 0.11 & 0.62 & 95.95 & 0.628 & 0.624 & 0.988 &  9.08 & 5.02 & 96.80 & 3.874 & 4.769 & 1.515 \\ 
   & WBExp II & 0.11 & 0.62 & 95.10 & 0.618 & 0.624 & 1.022 &  9.08 & 5.02 & 93.00 & 3.046 & 4.769 & 2.451 \\ 
   & WBRad II & 0.11 & 0.62 & 95.95 & 0.628 & 0.624 & 0.988 &  9.08 & 5.02 & 96.80 & 3.874 & 4.769 & 1.515 \\ 
   & SAND & 0.11 & 0.62 & 94.85 & 0.610 & 0.624 & 1.047 &  9.08 & 5.02 & 88.15 & 2.736 & 4.769 & 3.038 \\ 
			\addlinespace 
   \multirow{7}{*}{ATT} 
& BOOT I & 2.35 & 0.94 & 89.55 & 0.806 & 0.826 & 1.051 & 16.90 & 5.33 & 84.65 & 1.972 & 4.290 & 4.731 \\ 
    & BOOT II & 2.35 & 0.94 & 88.20 & 0.766 & 0.826 & 1.164 & 16.90 & 5.33 & 82.85 & 1.895 & 4.290 & 5.123 \\ 
   & WBExp I & 2.35 & 0.94 & 89.30 & 0.792 & 0.826 & 1.088 & 16.90 & 5.33 & 97.55 & 2.689 & 4.290 & 2.545 \\ 
   & WBRad I & 2.35 & 0.94 & 90.25 & 0.814 & 0.826 & 1.031 & 16.90 & 5.33 & 98.85 & 3.739 & 4.290 & 1.316 \\ 
   & WBExp II & 2.35 & 0.94 & 94.55 & 0.954 & 0.826 & 0.751 & 16.90 & 5.33 & 97.75 & 2.703 & 4.290 & 2.519 \\ 
   & WBRad II & 2.35 & 0.94 & 95.10 & 0.975 & 0.826 & 0.718 & 16.90 & 5.33 & 98.95 & 3.746 & 4.290 & 1.312 \\ 
   & SAND & 2.35 & 0.94 & 89.30 & 0.801 & 0.826 & 1.064 & 16.90 & 5.33 & 78.30 & 1.856 & 4.290 & 5.344 \\ 
			\addlinespace 
   
   \multirow{7}{*}{ATO} 
& BOOT I & 1.66 & 0.80 & 94.05 & 0.734 & 0.743 & 1.026 & 12.23 & 2.77 & 77.25 & 1.626 & 1.712 & 1.108 \\ 
    & BOOT II & 1.66 & 0.80 & 93.85 & 0.720 & 0.743 & 1.065 & 12.23 & 2.77 & 77.50 & 1.636 & 1.712 & 1.095 \\ 
   & WBExp I & 1.66 & 0.80 & 92.80 & 0.708 & 0.743 & 1.101 & 12.23 & 2.77 & 73.75 & 1.557 & 1.712 & 1.208 \\ 
   & WBRad I & 1.66 & 0.80 & 93.25 & 0.715 & 0.743 & 1.078 & 12.23 & 2.77 & 77.30 & 1.631 & 1.712 & 1.102 \\ 
   & WBExp II & 1.66 & 0.80 & 99.00 & 1.049 & 0.743 & 0.501 & 12.23 & 2.77 & 70.60 & 1.487 & 1.712 & 1.325 \\ 
   & WBRad II & 1.66 & 0.80 & 99.30 & 1.093 & 0.743 & 0.462 & 12.23 & 2.77 & 74.25 & 1.551 & 1.712 & 1.218 \\ 
   & SAND & 1.66 & 0.80 & 94.05 & 0.729 & 0.743 & 1.040 & 12.23 & 2.77 & 76.80 & 1.613 & 1.712 & 1.126 \\ 
			\addlinespace 
			\multirow{7}{*}{ATM} 
  & BOOT I & 2.19 & 0.89 & 93.80 & 0.791 & 0.799 & 1.020 & 11.75 & 2.70 & 77.75 & 1.611 & 1.680 & 1.087 \\ 
    & BOOT II & 2.19 & 0.89 & 93.05 & 0.772 & 0.799 & 1.073 & 11.75 & 2.70 & 77.10 & 1.598 & 1.680 & 1.106 \\ 
   & WBExp I & 2.19 & 0.89 & 91.40 & 0.755 & 0.799 & 1.120 & 11.75 & 2.70 & 73.40 & 1.527 & 1.680 & 1.210 \\ 
   & WBRad I & 2.19 & 0.89 & 92.60 & 0.767 & 0.799 & 1.086 & 11.75 & 2.70 & 77.25 & 1.589 & 1.680 & 1.118 \\ 
   & WBExp II & 2.19 & 0.89 & 99.80 & 1.264 & 0.799 & 0.400 & 11.75 & 2.70 & 74.25 & 1.543 & 1.680 & 1.185 \\ 
   & WBRad II & 2.19 & 0.89 & 99.70 & 1.306 & 0.799 & 0.375 & 11.75 & 2.70 & 76.90 & 1.588 & 1.680 & 1.120 \\ 
   & SAND & 2.19 & 0.89 & 93.35 & 0.784 & 0.799 & 1.039 & 11.75 & 2.70 & 77.45 & 1.602 & 1.680 & 1.099 \\ 
			\addlinespace 
			\multirow{7}{*}{ATEN} 
& BOOT I & 1.62 & 0.77 & 94.10 & 0.701 & 0.709 & 1.024 & 13.07 & 2.94 & 77.35 & 1.693 & 1.815 & 1.149 \\ 
   & BOOT II & 1.62 & 0.77 & 93.75 & 0.690 & 0.709 & 1.057 & 13.07 & 2.94 & 78.40 & 1.726 & 1.815 & 1.106 \\ 
   & WBExp I & 1.62 & 0.77 & 92.95 & 0.678 & 0.709 & 1.096 & 13.07 & 2.94 & 75.50 & 1.649 & 1.815 & 1.211 \\ 
   & WBRad I & 1.62 & 0.77 & 93.40 & 0.685 & 0.709 & 1.072 & 13.07 & 2.94 & 79.30 & 1.741 & 1.815 & 1.087 \\ 
   & WBExp II & 1.62 & 0.77 & 98.95 & 1.024 & 0.709 & 0.480 & 13.07 & 2.94 & 70.35 & 1.545 & 1.815 & 1.380 \\ 
   & WBRad II & 1.62 & 0.77 & 99.30 & 1.087 & 0.709 & 0.426 & 13.07 & 2.94 & 75.05 & 1.628 & 1.815 & 1.243 \\ 
   & SAND & 1.62 & 0.77 & 94.10 & 0.695 & 0.709 & 1.042 & 13.07 & 2.94 & 76.70 & 1.683 & 1.815 & 1.163 \\ 
			\bottomrule
		\end{tabular}
		\begin{tablenotes}
			\scriptsize
			\item SE: median of standard errors by proposed method; ESD: empirical standard deviation; RE: relative efficiency; CP: coverage probability; SAND: sandwich variance; WBRad (resp. Exp) : wild bootstrap via Rademacher (resp. exponential) distribution; BOOT I: standard bootstrap; BOOT II: post-weighting bootstrap; SAND: sandwich variance estimator;  PS: propensity score; OR: outcome regression.
		\end{tablenotes}
	\end{threeparttable}
\end{table}

\begin{table}[H]\tiny
	\begin{threeparttable}
		\centering
		\caption{Results of Model 4 (poor overlap with a medium $p=0.49$), $N=1000$, under homogeneous treatment effect}
		\label{tab:extremew-n1000-cons-results}
		\begin{tabular}{rrcccccccccccc}
			\toprule
   Estimand & Method & ARBias\%& RMSE & CP\%  & SE & ESD & RE &ARBias\%& RMSE  & CP\% & SE & ESD & RE \\
			\cmidrule(lr){3-8}\cmidrule(lr){9-14}
			\addlinespace 
			& &\multicolumn{6}{c}{Augmented: PS and OR correctly specified} & \multicolumn{6}{c}{Augmented: Only PS correctly specified} \\\cmidrule(lr){3-8}\cmidrule(lr){9-14}
			\cmidrule(lr){1-2}\cmidrule(lr){3-8}\cmidrule(lr){9-14}
			\multirow{7}{*}{ATE} 
& BOOT I & 0.08 & 0.14 & 94.05 & 0.118 & 0.143 & 1.461 & 1.08 & 0.79 &  95.75 & 0.403 & 0.785 & 3.783 \\ 
    & BOOT II & 0.08 & 0.14 & 93.65 & 0.117 & 0.143 & 1.492 & 1.08 & 0.79 &  99.65 & 0.606 & 0.785 & 1.677 \\ 
   & WBExp I & 0.08 & 0.14 & 89.15 & 0.099 & 0.143 & 2.111 & 1.08 & 0.79 &  99.75 & 0.568 & 0.785 & 1.907 \\ 
   & WBRad I & 0.08 & 0.14 & 91.65 & 0.108 & 0.143 & 1.751 & 1.08 & 0.79 &  99.80 & 0.619 & 0.785 & 1.605 \\ 
   & WBExp II & 0.08 & 0.14 & 89.15 & 0.099 & 0.143 & 2.111 & 1.08 & 0.79 &  99.75 & 0.568 & 0.785 & 1.907 \\ 
   & WBRad II & 0.08 & 0.14 & 91.65 & 0.108 & 0.143 & 1.751 & 1.08 & 0.79 &  99.80 & 0.619 & 0.785 & 1.605 \\ 
   & SAND & 0.08 & 0.14 & 91.70 & 0.111 & 0.143 & 1.663 & 1.08 & 0.79 &  93.00 & 0.354 & 0.785 & 4.915 \\ 
    \addlinespace
   			\multirow{7}{*}{ATT} 
& BOOT I & 0.26 & 0.19 & 93.15 & 0.156 & 0.187 & 1.439 & 2.61 & 1.09 &  93.20 & 0.614 & 1.081 & 3.093 \\ 
    & BOOT II & 0.26 & 0.19 & 92.95 & 0.154 & 0.187 & 1.484 & 2.61 & 1.09 &  98.15 & 0.789 & 1.081 & 1.876 \\ 
   & WBExp I & 0.26 & 0.19 & 82.90 & 0.111 & 0.187 & 2.838 & 2.61 & 1.09 &  97.50 & 0.710 & 1.081 & 2.317 \\ 
   & WBRad I & 0.26 & 0.19 & 86.95 & 0.126 & 0.187 & 2.220 & 2.61 & 1.09 &  98.55 & 0.798 & 1.081 & 1.836 \\ 
   & WBExp II & 0.26 & 0.19 & 83.40 & 0.112 & 0.187 & 2.804 & 2.61 & 1.09 &  97.65 & 0.712 & 1.081 & 2.303 \\ 
   & WBRad II & 0.26 & 0.19 & 87.05 & 0.125 & 0.187 & 2.229 & 2.61 & 1.09 &  98.65 & 0.801 & 1.081 & 1.819 \\ 
   & SAND & 0.26 & 0.19 & 89.10 & 0.141 & 0.187 & 1.773 & 2.61 & 1.09 &  88.65 & 0.540 & 1.081 & 4.013 \\ 
       \addlinespace
   			\multirow{7}{*}{ATO} 
& BOOT I & 0.02 & 0.09 & 94.75 & 0.085 & 0.086 & 1.014 & 0.02 & 0.09 &  95.80 & 0.091 & 0.088 & 0.943 \\ 
    & BOOT II & 0.02 & 0.09 & 94.85 & 0.085 & 0.086 & 1.010 & 0.02 & 0.09 & 100.00 & 0.530 & 0.088 & 0.028 \\ 
   & WBExp I & 0.02 & 0.09 & 93.90 & 0.083 & 0.086 & 1.078 & 0.02 & 0.09 & 100.00 & 0.513 & 0.088 & 0.029 \\ 
   & WBRad I & 0.02 & 0.09 & 93.75 & 0.084 & 0.086 & 1.053 & 0.02 & 0.09 & 100.00 & 0.525 & 0.088 & 0.028 \\ 
   & WBExp II & 0.02 & 0.09 & 93.75 & 0.083 & 0.086 & 1.064 & 0.02 & 0.09 & 100.00 & 0.505 & 0.088 & 0.030 \\ 
   & WBRad II & 0.02 & 0.09 & 94.05 & 0.084 & 0.086 & 1.042 & 0.02 & 0.09 & 100.00 & 0.520 & 0.088 & 0.029 \\ 
   & SAND & 0.02 & 0.09 & 94.60 & 0.085 & 0.086 & 1.033 & 0.02 & 0.09 &  95.00 & 0.088 & 0.088 & 0.990 \\ 
\addlinespace 
			\multirow{7}{*}{ATM}
  & BOOT I & 0.01 & 0.09 & 94.95 & 0.087 & 0.087 & 1.006 & 0.00 & 0.13 &  95.85 & 0.135 & 0.128 & 0.903 \\ 
    & BOOT II & 0.01 & 0.09 & 95.05 & 0.087 & 0.087 & 1.014 & 0.00 & 0.13 & 100.00 & 0.544 & 0.128 & 0.055 \\ 
   & WBExp I & 0.01 & 0.09 & 93.70 & 0.084 & 0.087 & 1.073 & 0.00 & 0.13 & 100.00 & 0.526 & 0.128 & 0.059 \\ 
   & WBRad I & 0.01 & 0.09 & 94.15 & 0.085 & 0.087 & 1.055 & 0.00 & 0.13 & 100.00 & 0.541 & 0.128 & 0.056 \\ 
   & WBExp II & 0.01 & 0.09 & 94.20 & 0.085 & 0.087 & 1.054 & 0.00 & 0.13 & 100.00 & 0.524 & 0.128 & 0.060 \\ 
   & WBRad II & 0.01 & 0.09 & 93.95 & 0.086 & 0.087 & 1.038 & 0.00 & 0.13 & 100.00 & 0.539 & 0.128 & 0.056 \\ 
   & SAND & 0.01 & 0.09 & 94.65 & 0.086 & 0.087 & 1.023 & 0.00 & 0.13 &  97.15 & 0.143 & 0.128 & 0.806 \\ 
			\addlinespace 
			\multirow{7}{*}{ATEN} 
& BOOT I & 0.01 & 0.09 & 94.40 & 0.086 & 0.087 & 1.015 & 0.07 & 0.11 &  95.40 & 0.110 & 0.109 & 0.980 \\ 
    & BOOT II & 0.01 & 0.09 & 94.65 & 0.086 & 0.087 & 1.002 & 0.07 & 0.11 & 100.00 & 0.524 & 0.109 & 0.044 \\ 
   & WBExp I & 0.01 & 0.09 & 93.75 & 0.083 & 0.087 & 1.082 & 0.07 & 0.11 & 100.00 & 0.506 & 0.109 & 0.047 \\ 
   & WBRad I & 0.01 & 0.09 & 93.90 & 0.084 & 0.087 & 1.052 & 0.07 & 0.11 & 100.00 & 0.521 & 0.109 & 0.044 \\ 
   & WBExp II & 0.01 & 0.09 & 93.80 & 0.084 & 0.087 & 1.072 & 0.07 & 0.11 & 100.00 & 0.498 & 0.109 & 0.048 \\ 
   & WBRad II & 0.01 & 0.09 & 93.75 & 0.085 & 0.087 & 1.047 & 0.07 & 0.11 & 100.00 & 0.513 & 0.109 & 0.045 \\ 
   & SAND & 0.01 & 0.09 & 94.45 & 0.085 & 0.087 & 1.040 & 0.07 & 0.11 &  94.80 & 0.107 & 0.109 & 1.045 \\ 
			\addlinespace 
			& &\multicolumn{6}{c}{Augmented: only OR correctly specified} & \multicolumn{6}{c}{Augmented: PS and OR misspecified}\\ 
			\cmidrule(lr){1-2}\cmidrule(lr){3-8}\cmidrule(lr){9-14}
	\multirow{7}{*}{ATE} 
& BOOT I & 0.02 & 0.19 & 94.65 & 0.135 & 0.186 & 1.913 & 27.52 & 3.35 & 92.30 & 1.280 & 3.161 & 6.099 \\ 
    & BOOT II & 0.02 & 0.19 & 93.55 & 0.130 & 0.186 & 2.069 & 27.52 & 3.35 & 90.45 & 1.201 & 3.161 & 6.924 \\ 
   & WBExp I & 0.02 & 0.19 & 95.20 & 0.141 & 0.186 & 1.747 & 27.52 & 3.35 & 95.90 & 1.430 & 3.161 & 4.887 \\ 
   & WBRad I & 0.02 & 0.19 & 96.90 & 0.173 & 0.186 & 1.167 & 27.52 & 3.35 & 97.50 & 1.956 & 3.161 & 2.612 \\ 
   & WBExp II & 0.02 & 0.19 & 95.20 & 0.141 & 0.186 & 1.747 & 27.52 & 3.35 & 95.90 & 1.430 & 3.161 & 4.887 \\ 
   & WBRad II & 0.02 & 0.19 & 96.90 & 0.173 & 0.186 & 1.167 & 27.52 & 3.35 & 97.50 & 1.956 & 3.161 & 2.612 \\ 
   & SAND & 0.02 & 0.19 & 91.40 & 0.127 & 0.186 & 2.165 & 27.52 & 3.35 & 87.35 & 1.205 & 3.161 & 6.879 \\
			\addlinespace 
   \multirow{7}{*}{ATT} 
& BOOT I & 0.15 & 0.26 & 92.50 & 0.184 & 0.257 & 1.937 & 77.99 & 5.24 & 81.75 & 1.694 & 4.208 & 6.174 \\ 
    & BOOT II & 0.15 & 0.26 & 92.00 & 0.178 & 0.257 & 2.071 & 77.99 & 5.24 & 79.20 & 1.613 & 4.208 & 6.808 \\ 
   & WBExp I & 0.15 & 0.26 & 94.40 & 0.228 & 0.257 & 1.270 & 77.99 & 5.24 & 97.55 & 2.462 & 4.208 & 2.921 \\ 
   & WBRad I & 0.15 & 0.26 & 96.15 & 0.307 & 0.257 & 0.697 & 77.99 & 5.24 & 98.50 & 3.611 & 4.208 & 1.358 \\ 
   & WBExp II & 0.15 & 0.26 & 94.40 & 0.229 & 0.257 & 1.256 & 77.99 & 5.24 & 97.55 & 2.482 & 4.208 & 2.874 \\ 
   & WBRad II & 0.15 & 0.26 & 96.20 & 0.306 & 0.257 & 0.701 & 77.99 & 5.24 & 98.50 & 3.625 & 4.208 & 1.348 \\ 
   & SAND & 0.15 & 0.26 & 87.10 & 0.167 & 0.257 & 2.349 & 77.99 & 5.24 & 73.60 & 1.558 & 4.208 & 7.293 \\ 
			\addlinespace 
   
   \multirow{7}{*}{ATO} 
& BOOT I & 0.02 & 0.09 & 94.80 & 0.086 & 0.086 & 1.003 & 13.50 & 0.78 & 84.75 & 0.556 & 0.567 & 1.038 \\ 
    & BOOT II & 0.02 & 0.09 & 94.90 & 0.086 & 0.086 & 1.001 & 13.50 & 0.78 & 83.70 & 0.546 & 0.567 & 1.079 \\ 
   & WBExp I & 0.02 & 0.09 & 92.40 & 0.080 & 0.086 & 1.163 & 13.50 & 0.78 & 81.65 & 0.525 & 0.567 & 1.166 \\ 
   & WBRad I & 0.02 & 0.09 & 92.35 & 0.081 & 0.086 & 1.142 & 13.50 & 0.78 & 83.30 & 0.543 & 0.567 & 1.089 \\ 
   & WBExp II & 0.02 & 0.09 & 93.05 & 0.081 & 0.086 & 1.142 & 13.50 & 0.78 & 81.65 & 0.531 & 0.567 & 1.140 \\ 
   & WBRad II & 0.02 & 0.09 & 92.75 & 0.081 & 0.086 & 1.124 & 13.50 & 0.78 & 84.25 & 0.553 & 0.567 & 1.051 \\ 
   & SAND & 0.02 & 0.09 & 94.15 & 0.085 & 0.086 & 1.027 & 13.50 & 0.78 & 84.65 & 0.554 & 0.567 & 1.048 \\ 
			\addlinespace 
			\multirow{7}{*}{ATM} 
  & BOOT I & 0.02 & 0.09 & 94.55 & 0.088 & 0.088 & 1.000 & 14.15 & 0.80 & 83.95 & 0.559 & 0.567 & 1.031 \\ 
   & BOOT II & 0.02 & 0.09 & 94.50 & 0.087 & 0.088 & 1.007 & 14.15 & 0.80 & 81.85 & 0.539 & 0.567 & 1.109 \\ 
   & WBExp I & 0.02 & 0.09 & 92.35 & 0.081 & 0.088 & 1.166 & 14.15 & 0.80 & 78.50 & 0.514 & 0.567 & 1.216 \\ 
   & WBRad I & 0.02 & 0.09 & 92.65 & 0.082 & 0.088 & 1.148 & 14.15 & 0.80 & 81.00 & 0.531 & 0.567 & 1.140 \\ 
   & WBExp II & 0.02 & 0.09 & 92.60 & 0.082 & 0.088 & 1.152 & 14.15 & 0.80 & 79.15 & 0.519 & 0.567 & 1.193 \\ 
   & WBRad II & 0.02 & 0.09 & 92.90 & 0.082 & 0.088 & 1.137 & 14.15 & 0.80 & 81.45 & 0.538 & 0.567 & 1.112 \\ 
   & SAND & 0.02 & 0.09 & 94.15 & 0.087 & 0.088 & 1.018 & 14.15 & 0.80 & 83.80 & 0.556 & 0.567 & 1.041 \\ 
			\addlinespace 
			\multirow{7}{*}{ATEN} 
& BOOT I & 0.02 & 0.09 & 94.75 & 0.086 & 0.086 & 0.998 & 11.51 & 0.76 & 87.55 & 0.581 & 0.600 & 1.066 \\ 
    & BOOT II & 0.02 & 0.09 & 94.75 & 0.087 & 0.086 & 0.994 & 11.51 & 0.76 & 87.45 & 0.580 & 0.600 & 1.070 \\ 
   & WBExp I & 0.02 & 0.09 & 93.00 & 0.081 & 0.086 & 1.124 & 11.51 & 0.76 & 86.00 & 0.572 & 0.600 & 1.099 \\ 
   & WBRad I & 0.02 & 0.09 & 93.40 & 0.083 & 0.086 & 1.090 & 11.51 & 0.76 & 87.40 & 0.600 & 0.600 & 1.000 \\ 
   & WBExp II & 0.02 & 0.09 & 93.40 & 0.082 & 0.086 & 1.106 & 11.51 & 0.76 & 86.55 & 0.583 & 0.600 & 1.058 \\ 
   & WBRad II & 0.02 & 0.09 & 93.55 & 0.083 & 0.086 & 1.074 & 11.51 & 0.76 & 88.30 & 0.614 & 0.600 & 0.955 \\ 
   & SAND & 0.02 & 0.09 & 94.30 & 0.085 & 0.086 & 1.030 & 11.51 & 0.76 & 87.25 & 0.577 & 0.600 & 1.082 \\ 
			\bottomrule
		\end{tabular}
		\begin{tablenotes}
			\scriptsize
			\item SE: median of standard errors by proposed method; ESD: empirical standard deviation; RE: relative efficiency; CP: coverage probability; SAND: sandwich variance; WBRad (resp. Exp) : wild bootstrap via Rademacher (resp. exponential) distribution; BOOT I: standard bootstrap; BOOT II: post-weighting bootstrap; SAND: sandwich variance estimator;  PS: propensity score; OR: outcome regression.
		\end{tablenotes}
	\end{threeparttable}
\end{table}

\begin{table}[H]\tiny
	\begin{threeparttable}
		\centering
		\caption{Results of Model 5 (SAND does not work, $p=0.80$), $N=50$, under heterogeneous treatment effect}
		\label{tab:sandnot-n50-hets-results}
		\begin{tabular}{rrcccccccccccc}
			\toprule
   Estimand & Method & ARBias\%& RMSE & CP\%  & SE & ESD & RE &ARBias\%& RMSE  & CP\% & SE & ESD & RE \\
			\cmidrule(lr){3-8}\cmidrule(lr){9-14}
			\addlinespace 
			& &\multicolumn{6}{c}{Augmented: PS and OR correctly specified} & \multicolumn{6}{c}{Augmented: Only PS correctly specified} \\\cmidrule(lr){3-8}\cmidrule(lr){9-14}
			\cmidrule(lr){1-2}\cmidrule(lr){3-8}\cmidrule(lr){9-14}
			\multirow{7}{*}{ATE} 
 & BOOT I & 27.13 & 186.89 & 98.60 & 31.590 & 186.874 &   34.995 &  3.00 & 14.16 & 98.95 & 12.766 & 14.154 &  1.229 \\ 
    & BOOT II & 27.13 & 186.89 & 98.60 & 31.584 & 186.874 &   35.008 &  3.00 & 14.16 & 98.75 & 12.509 & 14.154 &  1.280 \\ 
   & WBExp I & 27.13 & 186.89 & 78.30 &  2.590 & 186.874 & 5204.923 &  3.00 & 14.16 & 85.10 &  3.833 & 14.154 & 13.633 \\ 
   & WBRad I & 27.13 & 186.89 & 80.50 &  2.828 & 186.874 & 4365.618 &  3.00 & 14.16 & 87.55 &  4.179 & 14.154 & 11.472 \\ 
    & WBExp II & 27.13 & 186.89 & 78.30 &  2.590 & 186.874 & 5204.923 &  3.00 & 14.16 & 85.10 &  3.833 & 14.154 & 13.633 \\ 
   & WBRad II & 27.13 & 186.89 & 80.50 &  2.828 & 186.874 & 4365.618 &  3.00 & 14.16 & 87.55 &  4.179 & 14.154 & 11.472 \\ 
    & SAND & 34.11 & 221.80 & 85.23 & 2.692 & 221.803 & 6786.679 &  4.00 & 12.42 & 86.71 & 3.706 & 12.406 & 11.207 \\ 
    \addlinespace
   			\multirow{7}{*}{ATT} 
 & BOOT I & 33.68 & 229.83 & 98.70 & 39.133 & 229.816 &   34.489 & 15.06 & 16.35 & 98.85 & 14.570 & 16.151 &  1.229 \\ 
    & BOOT II & 33.68 & 229.83 & 98.50 & 37.315 & 229.816 &   37.930 & 15.06 & 16.35 & 98.50 & 14.092 & 16.151 &  1.314 \\ 
    & WBExp I & 33.68 & 229.83 & 73.55 &  2.608 & 229.816 & 7762.727 & 15.06 & 16.35 & 77.30 &  3.789 & 16.151 & 18.173 \\ 
   & WBRad I & 33.68 & 229.83 & 76.85 &  2.882 & 229.816 & 6357.181 & 15.06 & 16.35 & 81.60 &  4.138 & 16.151 & 15.232 \\ 
   & WBExp II & 33.68 & 229.83 & 75.95 &  2.824 & 229.816 & 6621.424 & 15.06 & 16.35 & 78.25 &  3.871 & 16.151 & 17.406 \\ 
   & WBRad II & 33.68 & 229.83 & 79.00 &  3.104 & 229.816 & 5480.131 & 15.06 & 16.35 & 82.10 &  4.247 & 16.151 & 14.462 \\ 
    & SAND & 42.92 & 273.24 & 84.86 & 2.998 & 273.244 & 8309.652 & 16.61 & 14.73 & 83.25 & 4.165 & 14.468 & 12.068 \\ 
       \addlinespace
   			\multirow{7}{*}{ATO} 
& BOOT I &  5.79 &   7.26 & 96.85 &  7.797 &   7.202 &    0.853 & 10.99 &  6.22 & 95.25 &  6.619 &  5.982 &  0.817 \\ 
    & BOOT II &  5.79 &   7.26 & 97.70 & 16.892 &   7.202 &    0.182 & 10.99 &  6.22 & 97.30 &  9.286 &  5.982 &  0.415 \\
   & WBExp I &  5.79 &   7.26 & 72.75 &  2.267 &   7.202 &   10.090 & 10.99 &  6.22 & 84.80 &  3.980 &  5.982 &  2.259 \\ 
   & WBRad I &  5.79 &   7.26 & 76.35 &  2.566 &   7.202 &    7.876 & 10.99 &  6.22 & 89.00 &  4.675 &  5.982 &  1.637 \\ 
   & WBExp II &  5.79 &   7.26 & 76.50 &  2.654 &   7.202 &    7.365 & 10.99 &  6.22 & 89.95 &  4.345 &  5.982 &  1.895 \\ 
   & WBRad II &  5.79 &   7.26 & 81.25 &  3.180 &   7.202 &    5.128 & 10.99 &  6.22 & 92.60 &  5.024 &  5.982 &  1.418 \\ 
   & SAND &  5.97 &   6.78 & 82.24 & 2.913 &   6.722 &    5.325 & 11.07 &  6.19 & 74.97 & 2.849 &  5.953 &  4.365 \\ 
			\addlinespace 
			\multirow{7}{*}{ATM}
  & BOOT I &  6.27 &  16.34 & 95.40 &  6.980 &  16.309 &    5.460 & 13.61 &  6.44 & 93.90 &  6.541 &  6.069 &  0.861 \\ 
    & BOOT II &  6.27 &  16.34 & 96.80 & 15.970 &  16.309 &    1.043 & 13.61 &  6.44 & 96.35 &  9.223 &  6.069 &  0.433 \\ 
   & WBExp I &  6.27 &  16.34 & 68.20 &  2.285 &  16.309 &   50.930 & 13.61 &  6.44 & 82.90 &  4.158 &  6.069 &  2.131 \\ 
   & WBRad I &  6.27 &  16.34 & 73.25 &  2.650 &  16.309 &   37.869 & 13.61 &  6.44 & 86.90 &  4.956 &  6.069 &  1.500 \\ 
   & WBExp II &  6.27 &  16.34 & 76.10 &  3.114 &  16.309 &   27.425 & 13.61 &  6.44 & 91.10 &  5.132 &  6.069 &  1.399 \\ 
   & WBRad II &  6.27 &  16.34 & 81.85 &  3.896 &  16.309 &   17.526 & 13.61 &  6.44 & 93.75 &  5.957 &  6.069 &  1.038 \\ 
   & SAND &  5.71 &  18.96 & 80.57 & 3.235 &  18.945 &   34.301 & 13.68 &  6.41 & 77.29 & 3.548 &  6.032 &  2.890 \\ 
			\addlinespace 
			\multirow{7}{*}{ATEN} 
& BOOT I &  8.12 &  13.83 & 97.55 &  8.785 &  13.777 &    2.460 & 10.74 &  6.38 & 96.15 &  6.749 &  6.163 &  0.834 \\ 
    & BOOT II &  8.12 &  13.83 & 97.80 & 17.885 &  13.777 &    0.593 & 10.74 &  6.38 & 97.40 &  9.506 &  6.163 &  0.420 \\ 
   & WBExp I &  8.12 &  13.83 & 73.70 &  2.240 &  13.777 &   37.833 & 10.74 &  6.38 & 85.45 &  3.860 &  6.163 &  2.549 \\ 
   & WBRad I &  8.12 &  13.83 & 76.75 &  2.526 &  13.777 &   29.759 & 10.74 &  6.38 & 88.95 &  4.469 &  6.163 &  1.902 \\ 
   & WBExp II &  8.12 &  13.83 & 74.75 &  2.449 &  13.777 &   31.648 & 10.74 &  6.38 & 88.15 &  4.051 &  6.163 &  2.315 \\ 
   & WBRad II &  8.12 &  13.83 & 79.75 &  2.894 &  13.777 &   22.661 & 10.74 &  6.38 & 91.70 &  4.681 &  6.163 &  1.733 \\ 
   & SAND &  9.02 &  15.33 & 82.39 & 2.741 &  15.268 &   31.026 & 10.63 &  6.40 & 75.69 & 2.799 &  6.180 &  4.877 \\ 
			\addlinespace 
			& &\multicolumn{6}{c}{Augmented: only OR correctly specified} & \multicolumn{6}{c}{Augmented: PS and OR misspecified}\\ 
			\cmidrule(lr){1-2}\cmidrule(lr){3-8}\cmidrule(lr){9-14}
	\multirow{7}{*}{ATE} 
 & BOOT I & 27.13 & 186.89 & 98.60 & 31.584 & 186.874 &   35.009 &  2.21 & 14.41 & 98.45 & 12.736 & 14.408 &  1.280 \\ 
    & BOOT II & 27.13 & 186.89 & 98.60 & 31.584 & 186.874 &   35.008 &  2.21 & 14.41 & 98.35 & 12.502 & 14.408 &  1.328 \\ 
   & WBExp I & 27.13 & 186.89 & 78.00 &  2.598 & 186.874 & 5174.969 &  2.21 & 14.41 & 80.90 &  3.857 & 14.408 & 13.953 \\ 
   & WBRad I & 27.13 & 186.89 & 80.40 &  2.831 & 186.874 & 4355.936 &  2.21 & 14.41 & 84.50 &  4.246 & 14.408 & 11.513 \\ 
   & WBExp II & 27.13 & 186.89 & 78.00 &  2.598 & 186.874 & 5174.969 &  2.21 & 14.41 & 80.90 &  3.857 & 14.408 & 13.953 \\ 
   & WBRad II & 27.13 & 186.89 & 80.40 &  2.831 & 186.874 & 4355.936 &  2.21 & 14.41 & 84.50 &  4.246 & 14.408 & 11.513 \\ 
    & SAND & 34.11 & 221.80 & 85.30 & 2.690 & 221.803 & 6799.534 &  3.01 & 12.72 & 84.33 & 4.035 & 12.718 &  9.933 \\ 
			\addlinespace 
   \multirow{7}{*}{ATT} 
& BOOT I & 30.60 & 218.75 & 98.65 & 37.652 & 218.745 &   33.752 & 11.50 & 16.20 & 98.70 & 14.453 & 16.086 &  1.239 \\ 
     & BOOT II & 30.60 & 218.75 & 98.50 & 36.476 & 218.745 &   35.963 & 11.50 & 16.20 & 98.45 & 13.921 & 16.086 &  1.335 \\ 
   & WBExp I & 30.60 & 218.75 & 75.20 &  2.641 & 218.745 & 6861.145 & 11.50 & 16.20 & 77.75 &  3.819 & 16.086 & 17.739 \\ 
    & WBRad I & 30.60 & 218.75 & 78.90 &  2.927 & 218.745 & 5585.757 & 11.50 & 16.20 & 81.65 &  4.184 & 16.086 & 14.779 \\ 
    & WBExp II & 30.60 & 218.75 & 76.95 &  2.809 & 218.745 & 6063.231 & 11.50 & 16.20 & 78.70 &  3.864 & 16.086 & 17.329 \\ 
   & WBRad II & 30.60 & 218.75 & 80.75 &  3.115 & 218.745 & 4930.913 & 11.50 & 16.20 & 82.40 &  4.307 & 16.086 & 13.949 \\ 
    & SAND & 38.62 & 260.23 & 84.72 & 2.837 & 260.247 & 8414.953 & 12.75 & 14.55 & 83.67 & 4.237 & 14.392 & 11.539 \\ 
			\addlinespace 
   
   \multirow{7}{*}{ATO} 
 & BOOT I & 12.88 &  58.03 & 97.00 & 12.231 &  58.009 &   22.494 & 12.68 &  7.10 & 94.75 &  7.061 &  6.827 &  0.935 \\ 
    & BOOT II & 12.88 &  58.03 & 98.65 & 20.132 &  58.009 &    8.303 & 12.68 &  7.10 & 97.20 &  9.595 &  6.827 &  0.506 \\ 
   & WBExp I & 12.88 &  58.03 & 77.55 &  2.259 &  58.009 &  659.614 & 12.68 &  7.10 & 78.65 &  3.761 &  6.827 &  3.294 \\ 
   & WBRad I & 12.88 &  58.03 & 80.85 &  2.545 &  58.009 &  519.720 & 12.68 &  7.10 & 84.05 &  4.455 &  6.827 &  2.348 \\ 
   & WBExp II & 12.88 &  58.03 & 83.90 &  3.093 &  58.009 &  351.722 & 12.68 &  7.10 & 87.25 &  4.615 &  6.827 &  2.188 \\ 
    & WBRad II & 12.88 &  58.03 & 87.90 &  3.893 &  58.009 &  222.041 & 12.68 &  7.10 & 91.15 &  5.404 &  6.827 &  1.596 \\ 
    & SAND & 18.63 &  64.65 & 87.55 & 2.804 &  64.607 &  530.981 & 12.72 &  7.07 & 83.67 & 4.181 &  6.794 &  2.640 \\ 
			\addlinespace 
			\multirow{7}{*}{ATM} 
 & BOOT I & 12.36 &  48.00 & 96.20 & 10.721 &  47.971 &   20.021 & 15.07 &  7.17 & 93.35 &  6.857 &  6.767 &  0.974 \\ 
    & BOOT II & 12.36 &  48.00 & 98.35 & 18.967 &  47.971 &    6.397 & 15.07 &  7.17 & 96.60 &  9.450 &  6.767 &  0.513 \\ 
   & WBExp I & 12.36 &  48.00 & 74.65 &  2.345 &  47.971 &  418.445 & 15.07 &  7.17 & 77.25 &  4.076 &  6.767 &  2.757 \\ 
   & WBRad I & 12.36 &  48.00 & 79.20 &  2.665 &  47.971 &  324.017 & 15.07 &  7.17 & 82.75 &  4.899 &  6.767 &  1.908 \\ 
   & WBExp II & 12.36 &  48.00 & 83.70 &  3.616 &  47.971 &  176.024 & 15.07 &  7.17 & 87.95 &  5.427 &  6.767 &  1.555 \\ 
   & WBRad II & 12.36 &  48.00 & 88.85 &  4.597 &  47.971 &  108.880 & 15.07 &  7.17 & 92.30 &  6.315 &  6.767 &  1.148 \\ 
    & SAND & 18.11 &  51.42 & 86.10 & 3.113 &  51.354 &  272.127 & 15.39 &  7.07 & 83.31 & 4.648 &  6.637 &  2.039 \\ 
			\addlinespace 
			\multirow{7}{*}{ATEN} 
& BOOT I & 15.41 &  71.83 & 97.60 & 13.950 &  71.809 &   26.497 & 11.53 &  7.42 & 95.95 &  7.440 &  7.199 &  0.936 \\ 
    & BOOT II & 15.41 &  71.83 & 98.75 & 21.317 &  71.809 &   11.348 & 11.53 &  7.42 & 97.65 &  9.819 &  7.199 &  0.538 \\ 
   & WBExp I & 15.41 &  71.83 & 77.75 &  2.229 &  71.809 & 1038.217 & 11.53 &  7.42 & 78.95 &  3.670 &  7.199 &  3.847 \\ 
    & WBRad I & 15.41 &  71.83 & 81.15 &  2.472 &  71.809 &  844.089 & 11.53 &  7.42 & 83.75 &  4.240 &  7.199 &  2.882 \\ 
   & WBExp II & 15.41 &  71.83 & 82.75 &  2.827 &  71.809 &  645.171 & 11.53 &  7.42 & 85.25 &  4.199 &  7.199 &  2.939 \\ 
   & WBRad II & 15.41 &  71.83 & 86.50 &  3.448 &  71.809 &  433.761 & 11.53 &  7.42 & 90.15 &  4.953 &  7.199 &  2.113 \\ 
    & SAND & 20.92 &  82.19 & 87.63 & 2.679 &  82.152 &  940.532 & 11.11 &  7.41 & 84.51 & 4.031 &  7.211 &  3.200 \\ 
			\bottomrule
		\end{tabular}
		\begin{tablenotes}
			\scriptsize
			\item ARBias\%: absolute percent relative bias; RMSE: root mean square error; CP\%: coverage probability (\%); SE: median of standard errors by proposed method; ESD: empirical standard deviation; RE: median relative efficiency; BOOT I: standard bootstrap; BOOT II: post-weighting bootstrap;  WBExp (resp. ExpRad): wild bootstrap via exponential (resp. Rademacher) distribution; SAND: sandwich variance estimator; PS: propensity score; OR: outcome regression.
		\end{tablenotes}
	\end{threeparttable}
\end{table}	

\begin{table}[H]\tiny
	\begin{threeparttable}
		\centering
		\caption{Results of Model 5 (SAND does not work, $p=0.80$), $N=50$, under homogeneous treatment effect}
		\label{tab:sandnot-n50-cons-results}
		\begin{tabular}{rrcccccccccccc}
			\toprule
   Estimand & Method & ARBias\%& RMSE & CP\%  & SE & ESD & RE &ARBias\%& RMSE  & CP\% & SE & ESD & RE \\
			\cmidrule(lr){3-8}\cmidrule(lr){9-14}
			\addlinespace 
			& &\multicolumn{6}{c}{Augmented: PS and OR correctly specified} & \multicolumn{6}{c}{Augmented: Only PS correctly specified} \\\cmidrule(lr){3-8}\cmidrule(lr){9-14}
			\cmidrule(lr){1-2}\cmidrule(lr){3-8}\cmidrule(lr){9-14}
			\multirow{7}{*}{ATE} 
  & BOOT I & 14.51 & 33.53 & 98.80 & 26.858 & 33.530 &    1.559 & 30.50 & 8.62 & 98.60 & 6.999 & 8.536 &  1.488 \\ 
    & BOOT II & 14.51 & 33.53 & 98.80 & 26.858 & 33.530 &    1.559 & 30.50 & 8.62 &  98.30 & 6.997 & 8.536 &  1.488 \\ 
    & WBExp I & 14.51 & 33.53 & 49.35 &  0.387 & 33.530 & 7490.192 & 30.50 & 8.62 &  64.60 & 1.574 & 8.536 & 29.400 \\ 
   & WBRad I & 14.51 & 33.53 & 53.50 &  0.420 & 33.530 & 6375.558 & 30.50 & 8.62 & 68.50 & 1.720 & 8.536 & 24.623 \\ 
    & WBExp II & 14.51 & 33.53 & 49.35 &  0.387 & 33.530 & 7490.192 & 30.50 & 8.62 &  64.60 & 1.574 & 8.536 & 29.400 \\ 
   & WBRad II & 14.51 & 33.53 & 53.50 &  0.420 & 33.530 & 6375.558 & 30.50 & 8.62 & 68.50 & 1.720 & 8.536 & 24.623 \\ 
    & SAND & 148.88 & 221.70 & 67.47 & 0.572 & 221.697 & 149997.297 & 44.97 & 12.05 & 65.26 & 1.620 & 11.920 & 54.171 \\ 
    \addlinespace
   			\multirow{7}{*}{ATT} 
& BOOT I & 16.49 & 39.08 & 98.95 & 33.099 & 39.081 &    1.394 & 50.07 & 9.61 & 97.75 & 8.211 & 9.401 &  1.311 \\ 
    & BOOT II & 16.49 & 39.08 & 98.90 & 31.312 & 39.081 &    1.558 & 50.07 & 9.61 &  97.35 & 7.876 & 9.401 &  1.425 \\ 
    & WBExp I & 16.49 & 39.08 & 44.65 &  0.426 & 39.081 & 8397.904 & 50.07 & 9.61 &  52.85 & 1.575 & 9.401 & 35.639 \\ 
    & WBRad I & 16.49 & 39.08 & 48.65 &  0.474 & 39.081 & 6797.943 & 50.07 & 9.61 &  57.65 & 1.741 & 9.401 & 29.160 \\ 
    & WBExp II & 16.49 & 39.08 & 46.80 &  0.452 & 39.081 & 7485.640 & 50.07 & 9.61 &  54.80 & 1.663 & 9.401 & 31.960 \\ 
    & WBRad II & 16.49 & 39.08 & 50.40 &  0.498 & 39.081 & 6147.979 & 50.07 & 9.61 &  59.50 & 1.830 & 9.401 & 26.392 \\ 
   & SAND & 183.31 & 273.13 & 66.16 & 0.699 & 273.132 & 152776.764 & 68.24 & 14.20 & 57.45 & 1.880 & 13.944 & 54.986 \\ 
       \addlinespace
   			\multirow{7}{*}{ATO} 
& BOOT I &  0.63 &  3.56 & 99.70 &  4.788 &  3.565 &    0.554 &  2.70 & 1.28 & 99.85 & 1.994 & 1.272 &  0.407 \\ 
    & BOOT II &  0.63 &  3.56 & 99.90 & 14.402 &  3.565 &    0.061 &  2.70 & 1.28 & 100.00 & 4.616 & 1.272 &  0.076 \\ 
    & WBExp I &  0.63 &  3.56 & 85.65 &  0.344 &  3.565 &  107.094 &  2.70 & 1.28 &  99.60 & 1.501 & 1.272 &  0.718 \\ 
   & WBRad I &  0.63 &  3.56 & 88.85 &  0.377 &  3.565 &   89.469 &  2.70 & 1.28 & 99.95 & 1.697 & 1.272 &  0.562 \\ 
   & WBExp II &  0.63 &  3.56 & 86.85 &  0.355 &  3.565 &  100.900 &  2.70 & 1.28 & 99.35 & 1.365 & 1.272 &  0.868 \\ 
   & WBRad II &  0.63 &  3.56 & 90.50 &  0.409 &  3.565 &   75.996 &  2.70 & 1.28 & 99.65 & 1.546 & 1.272 &  0.677 \\ 
   & SAND &   1.63 &   5.75 & 86.54 & 0.397 &   5.749 &    210.046 &  0.68 &  4.36 & 92.01 & 0.816 &  4.357 & 28.518 \\ 
			\addlinespace 
			\multirow{7}{*}{ATM}
& BOOT I &  0.83 &  3.25 & 99.80 &  3.765 &  3.252 &    0.746 &  3.90 & 1.23 & 99.80 & 1.880 & 1.224 &  0.424 \\ 
    & BOOT II &  0.83 &  3.25 & 99.90 & 13.550 &  3.252 &    0.058 &  3.90 & 1.23 & 100.00 & 4.463 & 1.224 &  0.075 \\
   & WBExp I &  0.83 &  3.25 & 86.45 &  0.354 &  3.252 &   84.421 &  3.90 & 1.23 & 99.60 & 1.515 & 1.224 &  0.653 \\ 
   & WBRad I &  0.83 &  3.25 & 89.70 &  0.391 &  3.252 &   69.215 &  3.90 & 1.23 & 99.90 & 1.730 & 1.224 &  0.501 \\ 
   & WBExp II &  0.83 &  3.25 & 88.85 &  0.387 &  3.252 &   70.792 &  3.90 & 1.23 & 99.55 & 1.615 & 1.224 &  0.574 \\ 
   & WBRad II &  0.83 &  3.25 & 93.25 &  0.450 &  3.252 &   52.236 &  3.90 & 1.23 & 99.75 & 1.836 & 1.224 &  0.444 \\ 
    & SAND &  10.62 &  18.52 & 89.16 & 0.435 &  18.524 &   1815.679 &  1.22 &  4.12 & 95.47 & 0.983 &  4.116 & 17.524 \\ 
			\addlinespace 
			\multirow{7}{*}{ATEN} 
& BOOT I &  0.47 &  4.73 & 99.55 &  5.838 &  4.728 &    0.656 &  0.86 & 1.43 & 99.95 & 2.183 & 1.425 &  0.426 \\ 
   & BOOT II &  0.47 &  4.73 & 99.85 & 15.276 &  4.728 &    0.096 &  0.86 & 1.43 & 100.00 & 4.805 & 1.425 &  0.088 \\ 
   & WBExp I &  0.47 &  4.73 & 84.85 &  0.340 &  4.728 &  193.027 &  0.86 & 1.43 & 99.55 & 1.489 & 1.425 &  0.916 \\ 
   & WBRad I &  0.47 &  4.73 & 87.05 &  0.370 &  4.728 &  162.979 &  0.86 & 1.43 & 99.95 & 1.697 & 1.425 &  0.705 \\ 
   & WBExp II &  0.47 &  4.73 & 83.15 &  0.339 &  4.728 &  194.357 &  0.86 & 1.43 & 99.05 & 1.308 & 1.425 &  1.187 \\ 
   & WBRad II &  0.47 &  4.73 & 87.70 &  0.391 &  4.728 &  146.557 &  0.86 & 1.43 & 99.35 & 1.485 & 1.425 &  0.921 \\ 
   & SAND &  13.73 &  14.94 & 83.33 & 0.387 &  14.935 &   1488.623 &  3.50 &  4.84 & 90.11 & 0.867 &  4.844 & 31.193 \\ 
			\addlinespace 
			& &\multicolumn{6}{c}{Augmented: only OR correctly specified} & \multicolumn{6}{c}{Augmented: PS and OR misspecified}\\ 
			\cmidrule(lr){1-2}\cmidrule(lr){3-8}\cmidrule(lr){9-14}
	\multirow{7}{*}{ATE} 
 & BOOT I & 14.52 & 33.53 & 98.80 & 26.857 & 33.531 &    1.559 & 15.62 & 8.88 & 98.00 & 7.140 & 8.861 &  1.540 \\ 
    & BOOT II & 14.52 & 33.53 & 98.75 & 26.856 & 33.531 &    1.559 & 15.62 & 8.88 & 97.75 & 7.070 & 8.861 &  1.571 \\ 
    & WBExp I & 14.52 & 33.53 & 49.20 &  0.384 & 33.531 & 7634.906 & 15.62 & 8.88 & 62.35 & 1.642 & 8.861 & 29.118 \\ 
    & WBRad I & 14.52 & 33.53 & 53.20 &  0.415 & 33.531 & 6523.337 & 15.62 & 8.88 & 67.60 & 1.856 & 8.861 & 22.795 \\ 
    & WBExp II & 14.52 & 33.53 & 49.20 &  0.384 & 33.531 & 7634.906 & 15.62 & 8.88 & 62.35 & 1.642 & 8.861 & 29.118 \\ 
   & WBRad II & 14.52 & 33.53 & 53.20 &  0.415 & 33.531 & 6523.337 & 15.62 & 8.88 & 67.60 & 1.856 & 8.861 & 22.795 \\ 
    & SAND & 148.90 & 221.70 & 67.83 & 0.567 & 221.697 & 152655.503 & 36.29 & 12.10 & 64.72 & 1.775 & 12.020 & 45.871 \\ 
			\addlinespace 
   \multirow{7}{*}{ATT} 
& BOOT I & 17.48 & 36.38 & 98.95 & 31.793 & 36.384 &    1.310 & 28.69 & 9.76 & 97.75 & 8.288 & 9.699 &  1.370 \\ 
    & BOOT II & 17.48 & 36.38 & 98.85 & 30.568 & 36.384 &    1.417 & 28.69 & 9.76 & 97.15 & 7.965 & 9.699 &  1.483 \\ 
    & WBExp I & 17.48 & 36.38 & 45.70 &  0.406 & 36.384 & 8015.563 & 28.69 & 9.76 & 54.95 & 1.641 & 9.699 & 34.951 \\ 
   & WBRad I & 17.48 & 36.38 & 49.20 &  0.448 & 36.384 & 6593.161 & 28.69 & 9.76 & 60.10 & 1.846 & 9.699 & 27.621 \\ 
    & WBExp II & 17.48 & 36.38 & 49.80 &  0.446 & 36.384 & 6649.142 & 28.69 & 9.76 & 56.75 & 1.720 & 9.699 & 31.797 \\ 
   & WBRad II & 17.48 & 36.38 & 53.60 &  0.490 & 36.384 & 5515.500 & 28.69 & 9.76 & 61.55 & 1.931 & 9.699 & 25.243 \\ 
    & SAND & 172.57 & 260.15 & 66.81 & 0.658 & 260.157 & 156305.367 & 53.28 & 14.00 & 60.13 & 1.934 & 13.845 & 51.249 \\ 
			\addlinespace 
   
   \multirow{7}{*}{ATO} 
& BOOT I &  1.74 & 22.11 & 97.75 &  9.449 & 22.117 &    5.478 & 15.95 & 2.62 & 97.30 & 2.588 & 2.540 &  0.963 \\ 
    & BOOT II &  1.74 & 22.11 & 99.20 & 17.423 & 22.117 &    1.611 & 15.95 & 2.62 & 99.05 & 4.785 & 2.540 &  0.282 \\ 
    & WBExp I &  1.74 & 22.11 & 68.90 &  0.370 & 22.117 & 3573.907 & 15.95 & 2.62 & 81.85 & 1.517 & 2.540 &  2.803 \\ 
   & WBRad I &  1.74 & 22.11 & 74.40 &  0.401 & 22.117 & 3049.583 & 15.95 & 2.62 & 87.55 & 1.710 & 2.540 &  2.208 \\ 
   & WBExp II &  1.74 & 22.11 & 66.05 &  0.383 & 22.117 & 3338.024 & 15.95 & 2.62 & 80.40 & 1.442 & 2.540 &  3.102 \\ 
    & WBRad II &  1.74 & 22.11 & 72.30 &  0.433 & 22.117 & 2612.204 & 15.95 & 2.62 & 86.35 & 1.648 & 2.540 &  2.377 \\ 
   & SAND &  56.95 &  64.45 & 81.88 & 0.500 &  64.430 &  16631.410 &  7.71 &  5.03 & 86.77 & 1.651 &  5.020 &  9.240 \\ 
			\addlinespace 
			\multirow{7}{*}{ATM} 
& BOOT I &  2.97 & 21.91 & 97.15 &  8.006 & 21.912 &    7.491 & 16.84 & 2.55 & 96.75 & 2.269 & 2.465 &  1.180 \\ 
   & BOOT II &  2.97 & 21.91 & 99.25 & 16.418 & 21.912 &    1.781 & 16.84 & 2.55 & 99.15 & 4.629 & 2.465 &  0.284 \\ 
   & WBExp I &  2.97 & 21.91 & 71.45 &  0.379 & 21.912 & 3340.168 & 16.84 & 2.55 & 83.45 & 1.553 & 2.465 &  2.519 \\ 
   & WBRad I &  2.97 & 21.91 & 77.65 &  0.419 & 21.912 & 2731.210 & 16.84 & 2.55 & 88.45 & 1.770 & 2.465 &  1.940 \\ 
   & WBExp II &  2.97 & 21.91 & 72.85 &  0.423 & 21.912 & 2685.489 & 16.84 & 2.55 & 87.50 & 1.698 & 2.465 &  2.108 \\ 
   & WBRad II &  2.97 & 21.91 & 77.70 &  0.484 & 21.912 & 2050.420 & 16.84 & 2.55 & 91.65 & 1.969 & 2.465 &  1.568 \\ 
   & SAND &  46.22 &  51.16 & 86.46 & 0.522 &  51.147 &   9596.265 & 10.06 &  4.40 & 90.58 & 1.794 &  4.382 &  5.963 \\
			\addlinespace 
			\multirow{7}{*}{ATEN} 
& BOOT I &  0.68 & 22.64 & 97.80 & 11.128 & 22.643 &    4.140 & 13.99 & 2.89 & 97.75 & 2.939 & 2.833 &  0.929 \\ 
    & BOOT II &  0.68 & 22.64 & 99.15 & 18.440 & 22.643 &    1.508 & 13.99 & 2.89 & 99.00 & 5.064 & 2.833 &  0.313 \\ 
    & WBExp I &  0.68 & 22.64 & 64.85 &  0.366 & 22.643 & 3832.086 & 13.99 & 2.89 & 79.90 & 1.507 & 2.833 &  3.537 \\ 
    & WBRad I &  0.68 & 22.64 & 70.65 &  0.398 & 22.643 & 3233.189 & 13.99 & 2.89 & 86.40 & 1.709 & 2.833 &  2.749 \\ 
   & WBExp II &  0.68 & 22.64 & 61.15 &  0.369 & 22.643 & 3771.210 & 13.99 & 2.89 & 76.35 & 1.359 & 2.833 &  4.348 \\ 
   & WBRad II &  0.68 & 22.64 & 67.50 &  0.407 & 22.643 & 3099.994 & 13.99 & 2.89 & 82.95 & 1.550 & 2.833 &  3.342 \\ 
    & SAND &  68.53 &  82.01 & 78.46 & 0.498 &  81.995 &  27073.645 &  3.52 &  5.73 & 82.84 & 1.619 &  5.727 & 12.517 \\ 
			\bottomrule
		\end{tabular}
		\begin{tablenotes}
			\scriptsize
			\item ARBias\%: absolute percent relative bias; RMSE: root mean square error; CP\%: coverage probability (\%); SE: median of standard errors by proposed method; ESD: empirical standard deviation; RE: median relative efficiency; BOOT I: standard bootstrap; BOOT II: post-weighting bootstrap;  WBExp (resp. ExpRad): wild bootstrap via exponential (resp. Rademacher) distribution; SAND: sandwich variance estimator; PS: propensity score; OR: outcome regression.
		\end{tablenotes}
	\end{threeparttable}
\end{table}	

\section{Additional Data Analysis Results}\label{apx:data}

\begin{figure}[H]
     \centering
     \begin{subfigure}[b]{0.45\textwidth}
         \centering
         \includegraphics[trim=5 5 5 5, clip, width=\textwidth]{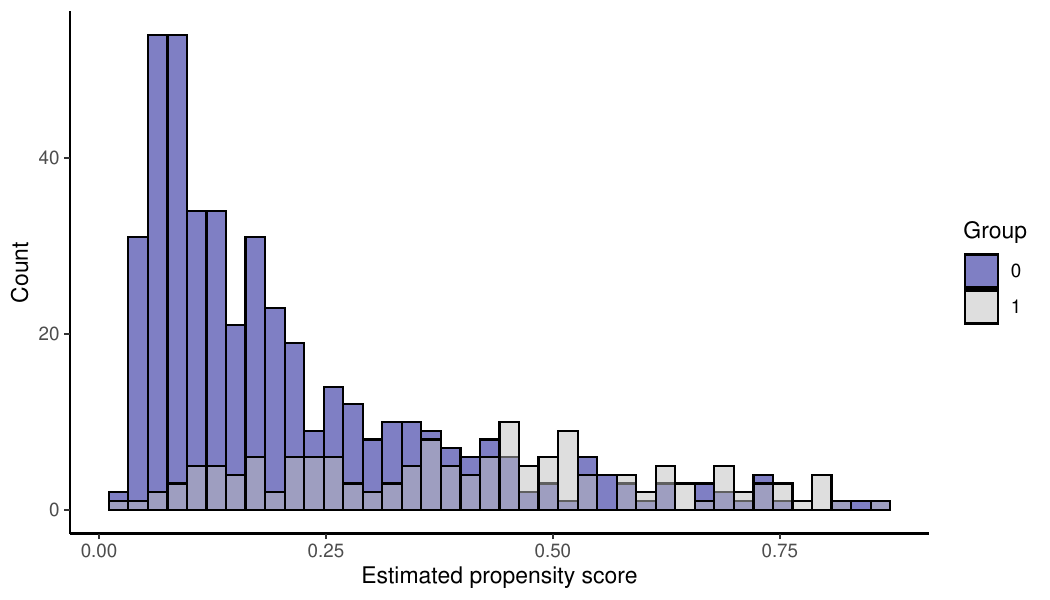}
         \caption{Estimated PSs by treatment group, where 0 = low fish consumption level, 1 = high fish consumption level}
         \label{fig:dat-ps2}
     \end{subfigure}
     \hspace{-.1cm}
     \begin{subfigure}[b]{0.50\textwidth}
         \centering
         \includegraphics[trim=5 5 5 5, clip, width=\textwidth]{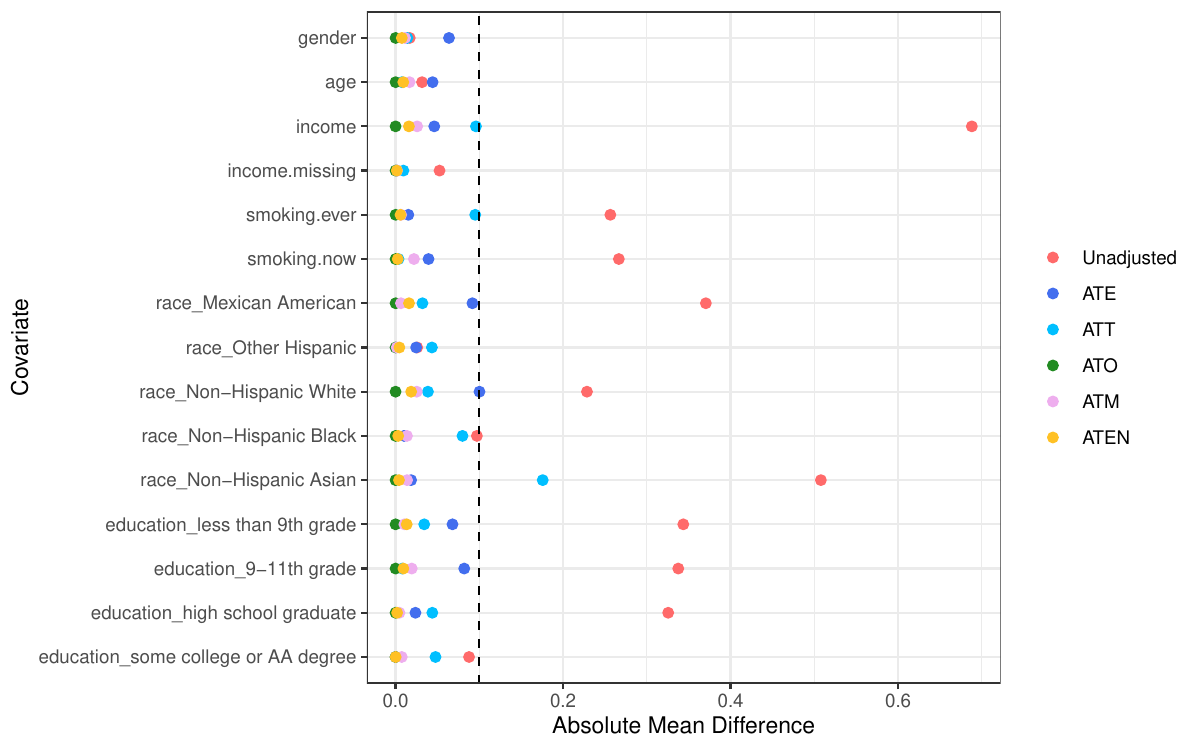}
         \caption{Covariates balance via different weights}
         \label{fig:dat-covbala2}
     \end{subfigure}
        \caption{Estimated PS distributions by treatment group and covariates balance of the NHANES 2013-2014 data for the secondary analysis in Section \ref{subsec:second}.}
        \label{fig:Data_EX2}
\end{figure}

\end{document}